\documentclass[preprint,3p, times, sort&compress]{elsarticle}
\usepackage{amsmath}      
\DeclareMathOperator*{\argmin}{arg\,min}
\usepackage{mathtools, cuted, cases}
\usepackage{amsfonts}
\usepackage{tikz}

\usepackage{caption}
\usepackage[labelformat=simple]{subcaption}    
\usepackage{algorithm}
\usepackage{algpseudocode}

\renewcommand{\algorithmiccomment}[1]{\bgroup\hfill\#~#1\egroup}

\usepackage{xfrac}
\usepackage{siunitx}

\usepackage{tcolorbox}
\newfloat{Box}{tbp}{lop}
\floatname{Box}{Box}

\usepackage[makeroom]{cancel}

\makeatletter
\renewcommand\p@subfigure{\thefigure\,}

\makeatother

\usepackage{amssymb}
\usepackage{mathrsfs}
\DeclareMathAlphabet\mathbfcal{OMS}{cmsy}{b}{n}

\journal{Computer Methods in Applied Mechanics and Engineering}

\begin{document}

\begin{frontmatter}

\title{Echocardiogram-based ventricular isogeometric cardiac analysis using multi-patch fitted NURBS}

\author[1]{Robin Willems\corref{cor1}}
\cortext[cor1]{Corresponding author}
\ead{R.Willems@tue.nl}
\author[2]{Lex Verberne}
\author[3,4]{Olaf van der Sluis}
\author[2]{Clemens V. Verhoosel}
\address[1]{Faculty of Biomedical Engineering, Cardiovascular Biomechanics, Eindhoven University of Technology, The Netherlands}
\address[2]{Faculty of Mechanical Engineering, Energy Technology and Fluid Dynamics, Eindhoven University of Technology, The Netherlands}
\address[3]{Faculty of Mechanical Engineering, Mechanics of Materials, Eindhoven University of Technology, The Netherlands}
\address[4]{Philips Research, AI, Data Science and Digital Twin Department, High Tech Campus 34 Eindhoven, The Netherlands}





\begin{abstract}
Monitoring the cardiac function often relies on non-invasive vital measurements, electrocardiograms, and low-resolution echocardiograms. The monitoring data can be augmented with numerical modeling results to support treatment-risk assessment. Often, medical images are not suitable for high-fidelity modeling due to their spatial sparsity and low resolution. In this work, we present a workflow that converts a limited number of two-dimensional (2D) echocardiogram images into analysis-suitable left-ventricle geometries using isogeometric analysis (IGA). The image data is fitted by reshaping a multi-patch NURBS template. This fitting procedure results in a smooth interpolation with a limited number of degrees of freedom. In addition, a coupled 3D-0D cardiac model is extended with a pre-stressing method to perform ventricular-cardiac-mechanic analyses based on \emph{in vivo} images. The workflow is benchmarked using population-averaged imaging data, which demonstrates that both the shape of the left ventricle and its mechanical response are well predicted in sparse-data scenarios. The case of a clinical patient is considered to demonstrate the suitability of the workflow in a real-data setting.
\end{abstract}





\end{frontmatter}


\section{Introduction}\label{sec:Intro}
For decades, there has been an increasing trend in the use of patient-specific computational models in the clinical practice \cite{antiga_image-based_2008, lopez-perez_personalized_2019, niederer_computational_2019, fumagalli_role_2024}.  Such models help to improve the understanding of biophysical phenomena and can provide decision support for clinicians \cite{niederer_computational_2019, fumagalli_role_2024}. In the context of cardiac modeling, patient-specific computational models have been used to study a broad range of phenomena and diseases, such as valvular diseases~\cite{bosmans_validated_2016, luraghi_modeling_2019}, dyssynchronous heart failure~\cite{niederer_length-dependent_2011, huntjens_influence_2014, kayvanpour_towards_2015}, atrial arrhythmias~\cite{dang_evaluation_2005, mcdowell_mechanistic_2013, trayanova_mathematical_2014}, and ventricular arrhythmias under various conditions~\cite{ten_tusscher_organization_2009, kazbanov_effect_2014,lopez-perez_personalized_2019, kruithof_influence_2021, willems_isogeometric-mechanics-driven_2023}. These models are most commonly based on high-quality image data (\emph{e.g.}, from magnetic resonance imaging (MRI) \cite{dang_evaluation_2005, niederer_length-dependent_2011, kruithof_influence_2021} or computer tomography (CT) \cite{bosmans_validated_2016, luraghi_modeling_2019}), but modeling based on sparse, irregular, or low-resolution image data (\emph{e.g.}, from echocardiograms \cite{alessandrini_realistic_2018}) is also possible. This work focuses on patient-specific analyses based on echocardiogram imaging data, which is typically only available in a limited number of cross-sectional planes (\emph{i.e.}, the data is sparse and irregular) in which only relatively large features can be identified properly (\emph{i.e.}, the resolution of the data is low).

The choice of a particular computational modeling technique is strongly influenced by the characteristics of the imaging data. Computational techniques to extract an analysis-suitable geometry from scan data can be classified as either \emph{direct segmentation techniques} or as \emph{template-fitting techniques}. In \emph{direct segmentation techniques}, a mesh for the object of interest is obtained by converting imaging cells directly into computational elements. This typically requires scan data with sufficient resolution throughout the complete domain. The marching cubes method \cite{lorensen_marching_1987,newman_survey_2006} -- in which voxels are meshed based on an interpolation of the grayscale intensity -- is the most prominent direct segmentation method in the context of the finite element method (FEM) \cite{frey_fully_1994,cebral_medical_2001,kim_construction_2019}. Scan-based immersed analysis methods \cite{verhoosel_scan-based_2023} -- in which the direct segmentation takes place on a background mesh that covers the scan domain -- are a common state-of-the-art alternative. In \emph{template-fitting techniques}, a reference geometry that captures the essential features of the object of interest is (gradually) deformed to match the scan data. This approach is less dependent on the resolution of the scan data, as the template can interpolate the geometry in regions where data is sparse. In the context of the FEM, typically the boundary of a template mesh (\emph{e.g.}, a surface triangulation (STL)) is deformed based on imaging data, after which the interior elements are updated in accordance with the boundary deformation to ensure sufficient mesh quality \cite{baghdadi_template-based_2005}. Alternatively, template geometries based on splines can be used in the context of isogeometric analysis \cite{urick_review_2019}. While both fitting paradigms have their own merits, the suitability of template-fitting techniques for sparse and irregular data scenarios makes it the favorable approach in the context of this manuscript.

\begin{figure*}[ht]
\centering
\begin{tikzpicture}
    \node[anchor=south west,inner sep=0] at (0,0) {\includegraphics[width=\textwidth]{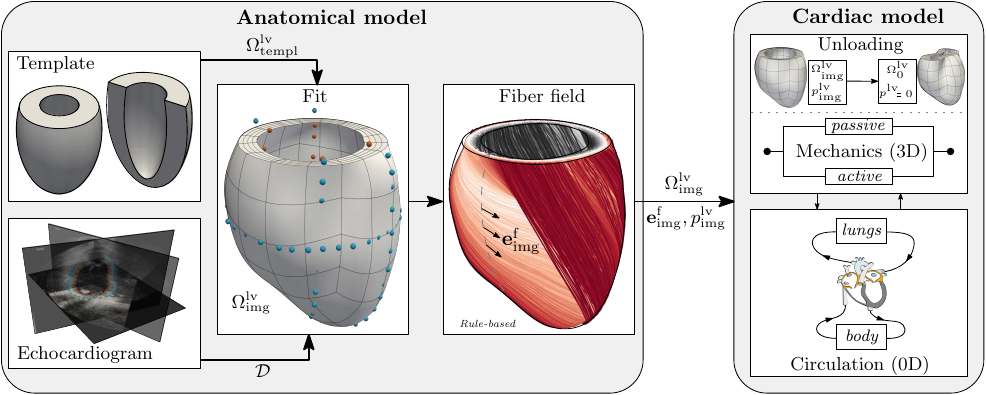}};
    \node at (5.8, -0.4) {\footnotesize (a)};
    \node at (14.5 , -0.4) {\footnotesize (b)};
\end{tikzpicture}
\caption{The IGA workflow proposed in this work consists of two main stages: the anatomical modeling stage and the cardiac modeling stage. In the anatomical modeling stage, a parametrized NURBS (Non-Uniform Rational B-Splines) template of the left ventricle, which is denoted by $\Omega^{\mathrm{lv}}_{\mathrm{templ}}$, is deformed according to the sparse echocardiogram data, $\mathcal{D}$. The resulting fitted geometry is deemed the image configuration, $\Omega^{\mathrm{lv}}_{\mathrm{img}}$, that is subject to an image cavity pressure, $p^{\mathrm{lv}}_{\mathrm{img}}$. A rule-based fiber field, with direction $\mathbf{e}^{\rm{f}}_{\mathrm{img}}$, is then constructed based on the image configuration. The geometrical properties, $\Omega^{\mathrm{lv}}_{\mathrm{templ}}$, $\mathbf{e}^{\rm{f}}_{\mathrm{img}}$, and $p^{\mathrm{lv}}_{\mathrm{img}}$, resulting from the anatomical modeling stage are then used as input for the cardiac modeling stage, which solves the relevant physics and consists of a three-dimensional (3D) mechanical component and a zero-dimensional (0D) circulatory component. An essential part of the cardiac modeling stage is the unloading of the image configuration -- gradually decrease of $p^{\mathrm{lv}}_{\mathrm{img}}$ to $0$ -- to retrieve the (stress-free) reference configuration, $\Omega^{\mathrm{lv}}_{0}$.} \label{fig:workflow}
\end{figure*}

The template-fitting technique proposed in this work builds on the mechanical cardiac modeling approach proposed by Willems \emph{et al.} \cite{willems_isogeometric_2023}, which performs an isogeometric analysis (IGA) based on a multi-patch NURBS \cite{cottrell_isogeometric_2009} spline model of the ventricles (illustrated in Fig.~\ref{fig:workflow}a). The main advantage of the proposed IGA approach -- in which both the reference geometry and the deformations are interpolated using the same spline basis -- is that the curved ventricles can be accurately represented using a minimal number of degrees of freedom (in the form of control points). Additional degrees of freedom are then only inserted in regions where this is needed to accurately mimic the mechanical response. The mechanical cardiac model of Ref.~\cite{willems_isogeometric_2023} has been benchmarked based on a state-of-the-art finite element solver and has been extended in Ref.~\cite{willems_isogeometric-mechanics-driven_2023} to incorporate the effect of infarct regions on the cardiac response, coupled with electrophysiology simulations. The IGA approach is demonstrated to make accurate predictions with a substantial reduction in the number of degrees of freedom compared to a traditional FE approach. Additionally, the absence of geometry clean-up and meshing steps in IGA makes it suitable for studying geometry variations.

Patient-specific analyses can be performed using the approach of Willems \emph{et al.} \cite{willems_isogeometric_2023}, in the sense that the parameters of the spline model (\emph{e.g.}, cavity volume, wall thickness) can be retrieved from imaging data. Although the obtained ventricle models will follow the general characteristics of the scanned object, geometric details are lost in the parameter-retrieval step. In this regard, the proposed IGA approach does not fully exploit the geometric information encoded in the scan data. However, the spline-based ventricle model is an ideal starting point for a template-fitting technique. The main idea of this work is to extend the approach of Ref.~\cite{willems_isogeometric_2023} to take into account geometric details, as illustrated by the workflow in Fig.~\ref{fig:workflow}. While in the original approach the mechanical analysis is directly performed on the template, in this work we perform an intermediate step in which we deform the spline-based ventricle template based on the scan data using a template-fitting technique. This additional step will match the scanned object more precisely, generating more accurate patient-specific analysis results.

We herein assume that the echocardiogram data has been segmented, resulting in a three-dimensional point cloud (or polyline) representation of the boundary of the ventricle in each echocardiogram plane. In combination, these planar point clouds result in a sparse and irregular point cloud representation of the ventricle boundaries. A template-fitting technique based on a spline-based ventricle model is anticipated to be powerful in this setting, as the template can be fitted to the data points in regions where data is available (\emph{i.e.}, in the echocardiogram slices), whereas the splines will provide a smooth interpolation based on the template in the data scant regions (\emph{i.e.}, in between the slices). Although there exist many algorithms to fit splines to point clouds (see, \emph{e.g.}, Refs.~\cite{piegl_nurbs_1997,weiss_advanced_2002,wang_fitting_2006} for overviews of various algorithms), these are often not directly applicable to the echocardiogram-based isogeometric analysis setting considered in this work. Specific challenges in this regard are the sparsity and irregularity of the point cloud and the use of a multi-patch NURBS object which is desired to retain smoothness across the patch interfaces. These challenges make techniques that approach the fitting as a point distance minimization problem promising in the context of the current work, since these allow for the simultaneous minimization of a distance measure and a smoothness measure \cite{weiss_advanced_2002}. Our work incorporates various aspects established in the literature on such techniques, most notably the use of an iterative procedure for updating the points on the parametrized spline surface closest to the data points \cite{hoschek_intrinsic_1988,saux_improved_2003}. Another aspect included in our work is the consideration of spline refinements to improve the quality of the fit \cite{greiner_interpolating_1998}. While the fundamental ingredients of the spline-fitting algorithm required in this work are established, an automatic algorithm capable of generating an analysis-suitable multi-patch NURBS object from the specific data set under consideration is not available.

A key contribution of the current work is the development of a fitting algorithm for multi-patch NURBS based on point distance minimization, suitable for sparse, irregular, and low-resolution echocardiogram imaging data. Novel features that improve the robustness and efficiency of the developed algorithm include the consideration of a convolution-based control point update scheme -- which draws inspiration from scan-based immersed isogeometric analysis procedures \cite{verhoosel_scan-based_2023} -- and the use of a Voronoi tessellation to quantify the importance of data points relative to the fitting process. The developed fitting algorithm extends the patient-specific workflow of Willems \emph{et al.} \cite{willems_isogeometric_2023}, allowing the IGA framework to exploit the full potential of the echocardiogram scan data. The extended workflow is tested using both dense and sparse data sets, considering both the geometry approximation and the mechanical cardiac response. With respect to the cardiac analysis, in this paper, we also extend the model of Ref.~\cite{willems_isogeometric_2023} to adjust for the fact that the imaging data is typically not obtained in an unloaded state. As part of this extension, the general pre-stressing algorithm proposed by Weisbecker \emph{et al.} \cite{weisbecker_generalized_2014} is adapted to the considered isogeometric setting. We demonstrate the proposed patient-specific isogeometric analysis workflow in a real-data setting using a clinical patient.

This paper is outlined as follows. Section~\ref{sec:Sec2} commences with the introduction of the spline-based anatomical model. Section~\ref{sec:Sec3} then introduces the fitting algorithm, including examples to demonstrate its performance. Section~\ref{sec:Sec4} then reviews the cardiac model of Ref.~\cite{willems_isogeometric_2023} and introduces the extension to accommodate the pressure-loaded image configuration. The IGA workflow is then validated using dense point cloud data in Section~\ref{sec:Sec5}, where the impact of data sparsity is investigated. An application scenario for a clinical patient is then studied in Section~\ref{sec:Sec6}, after which conclusions and recommendations are presented in Section~\ref{sec:ConcRec}.
\newpage

\section{The spline-based anatomical model}\label{sec:Sec2}
The patient-specific isogeometric analysis workflow considered in this work requires the definition of a suitable spline-based computational domain. The multi-patch anatomical model of the idealized left ventricle developed in Willems~\emph{et al.}~\cite{willems_isogeometric_2023} serves as the template geometry in our patient-specific analyses. While the reader is referred to Ref.~\cite{willems_isogeometric_2023} for details, in this section we will review the nomenclature essential to this work.

\subsection{The multi-patch NURBS template}\label{sec:multipatch}
\begin{figure*}[!t]
     \centering
     \begin{subfigure}[b]{0.48\textwidth}
         \centering
             \includegraphics[width=\textwidth]{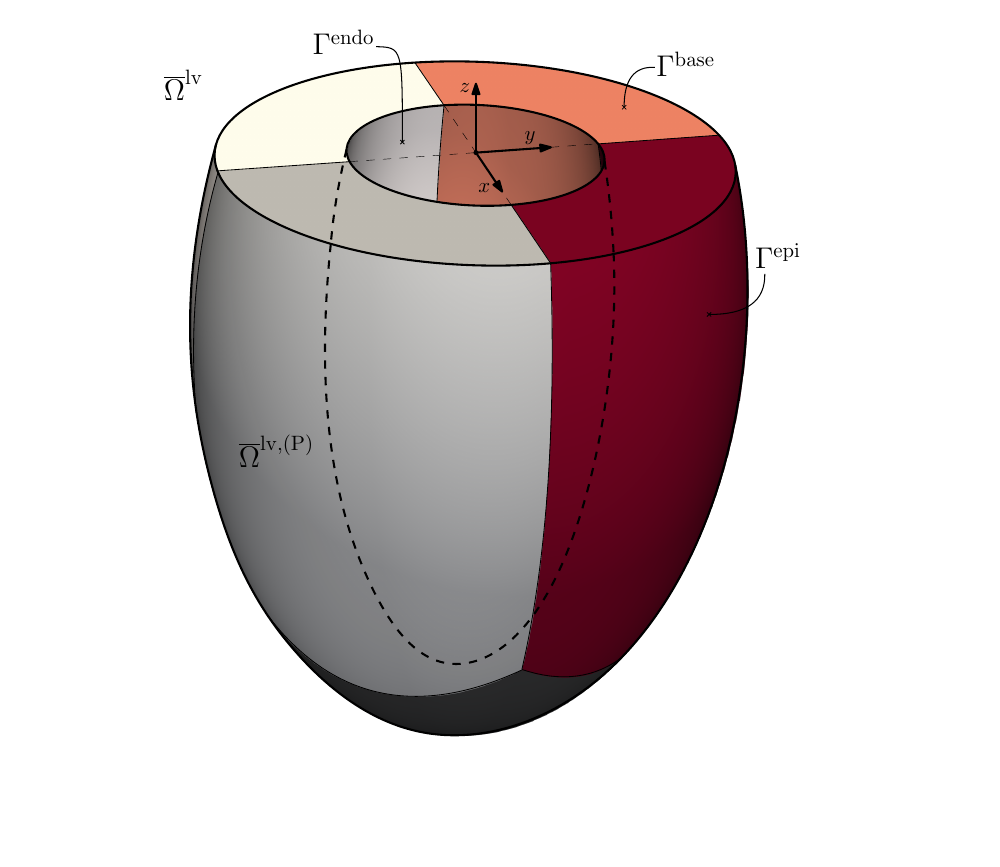}
         \caption{}
         \label{fig:TemplateSolid} 
     \end{subfigure}
     \hfill
     \begin{subfigure}[b]{0.48\textwidth}
         \centering
              \includegraphics[width=\textwidth]{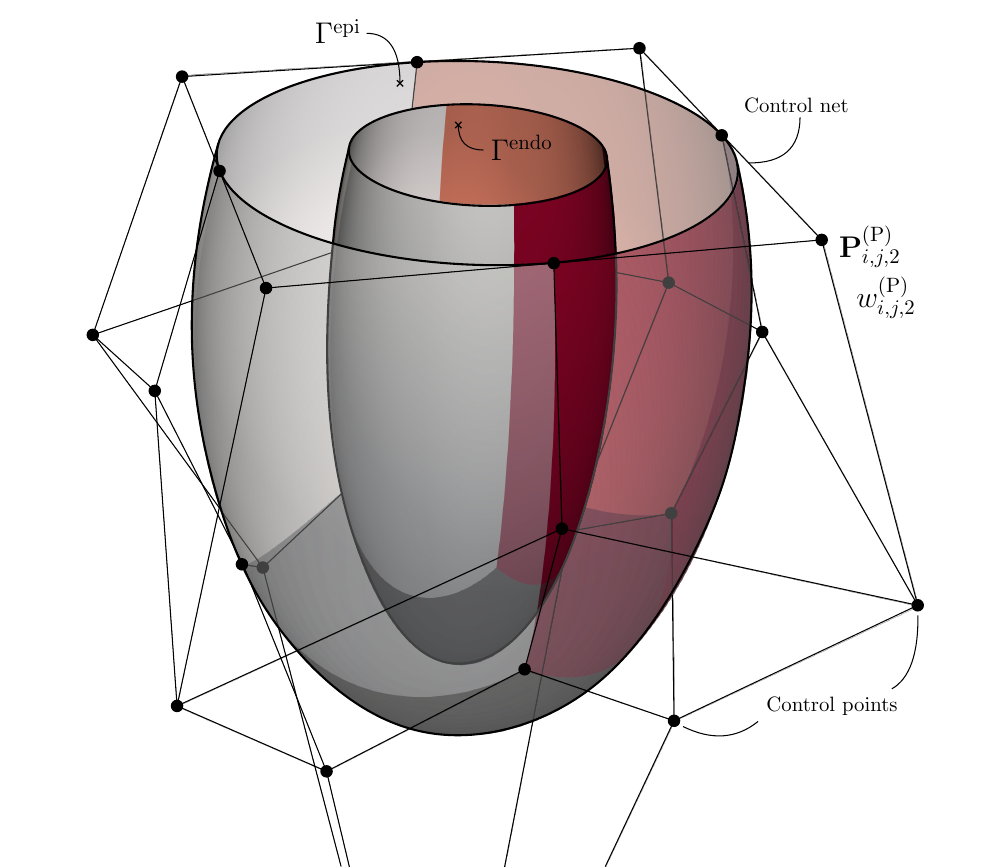}
         \caption{}
         \label{fig:TemplateSurface}
     \end{subfigure}
        \caption{(a) Visualization of the left ventricle multi-patch NURBS template, $\overline{\Omega}^{\mathrm{lv}}$, which is described by 5 (trivariate) quadratic NURBS patches, $\overline{\Omega}^{\mathrm{lv},(\mathrm{P})}$, indicated in different colors. The ventricle is truncated at the basal plane, $\Gamma^{\mathrm{base}}$, and bounded by the outer (epicardial) and inner (endocardial) surfaces, $\Gamma^{\mathrm{epi}}$ and $\Gamma^{\mathrm{endo}}$. (b) Epi- and endocardial boundaries, $\{ {\Gamma}^{\mathrm{epi}}, {\Gamma}^{\mathrm{endo}}\} \subset \overline{\Omega}^{\mathrm{lv}}$, of the ventricle template. Each multi-patch NURBS surface is described by a control net, consisting of $i\times j$ control points, $\mathbf{P}^{(\mathrm{P})}_{i,j,2}$, and corresponding weights, $w^{(\mathrm{P})}_{i,j,2}$. For visibility, we only show the epicardial control net.}
        \label{fig:Template}
\end{figure*}

We employ a common simplification of the left ventricle using a thick-walled ellipsoid that is truncated at the basal plane (Fig.~\ref{fig:TemplateSolid}). This idealized model assumes a smooth representation of the epi- and endocardium and neglects the papillary muscles and trabeculations inside the ventricles, which are responsible for opening the mitral and tricuspid valves. The idealized ventricle is represented by a multi-patch Non-Uniform Rational B-spline (NURBS) volume, which, in contrast to regular B-splines, enables the exact representation of conic sections. The physical geometry is obtained by mapping a multi-patch parameter domain to the physical domain, \emph{i.e.}, the truncated ellipsoid, using a geometrical map based on the NURBS basis functions, for which the control net provides the corresponding coefficients. This mapping is $C^0$-continuous at the patch interfaces, in principle allowing for kinks in the geometry at the patch interfaces, but the control net is selected such that the geometry is (almost) smooth, \emph{i.e.}, $G^1$-continuous.

We define the left ventricle template by the open bounded computational domain $\Omega^{\rm lv} \subset \mathbb{R}^{3}$. %
The boundary of this domain comprises the epicardial (outer) surface, $\Gamma^{\rm epi}$, the endocardial (inner) surfaces, $\Gamma^{\rm endo}$, and the basal plane surface, $\Gamma^{\rm base}$. The closure of the computational domain is denoted by $\overline{\Omega}^{\mathrm{lv}}$, which is partition by $n_{\rm{patch}}=5$ subdomains, $\Omega^{\mathrm{lv},\rm{(P)}}$, referred to as patches, where $\rm{P}=\{1,...,n_{\rm{patch}}\}$ is the patch index. The closed domain, $\overline{\Omega}^{\mathrm{lv}}$, is then defined as the union of the non-overlapping patches:
\begin{align}
\label{eq:MULTIPATCH}
    \overline{\Omega}^{\mathrm{lv}} &=\bigcup^{n_{\rm{patch}}}_{\rm{P}=1} \overline{\Omega}^{\mathrm{lv},\rm{(P)}},  &  \Omega^{\mathrm{lv},(i)} \cap \Omega^{\mathrm{lv},(j)} &= \varnothing \quad i \neq j.
\end{align}
The volumetric patches, $\overline{\Omega}^{\mathrm{lv},\rm{(P)}}$, are mapped from a parametric domain, $\widehat{\Omega} \subset \mathbb{R}^{3}$, with coordinates $(\xi,\eta,\zeta)$, in accordance with 
\begin{equation}
\label{eq:MAPP}
    \mathbf{S}_{\mathrm{lv}}^{\rm{(P)}}\left( \xi, \eta, \zeta \right) = \sum_{i} \sum_{j} \sum_{k} R^{(\mathrm{P})}_{i,j,k}(\xi,\eta,\zeta) \mathbf{P}^{\mathrm{(P)}}_{i,j,k},
\end{equation}
where $\mathbf{P}^{\mathrm{(P)}}_{i,j,k} \in \mathbb{R}^{3}$ are the patch control points. The NURBS basis functions are defined as
\begin{equation}
\label{eq:NURBS}
    R_{i,j,k}^{\rm (P)}\left( \xi,\eta,\zeta \right) = \frac{N_{i,j,k} \left(\xi,\eta,\zeta \right) w_{i,j,k}^{\rm (P)} }{\sum_{p} \sum_{q} \sum_{r} N_{p,q,r}\left(\xi,\eta,\zeta \right) w_{p,q,r}^{\rm (P)}},
\end{equation}
where $w^{\rm (P)}_{i,j,k}$ are the control point weights and 
\begin{equation}
\label{eq:NURBSBI}
    N_{i,j,k}(\xi,\eta,\zeta)= N_{i}(\xi)N_{j}(\eta)N_{k}(\zeta)
\end{equation}
are tensor-product trivariate B-splines. The coordinate $\zeta$ corresponds to the through-the-thickness direction, with $\zeta=0$ corresponding to the inner surface, and $\zeta=1$ to the outer surface. The coordinates $\xi$ and $\eta$ correspond to the directions tangential these surfaces. The reader is referred to Ref.~\cite{piegl_nurbs_1997} for further details on the construction of B-spline bases.

In this work, for the left ventricle geometry parametrization second-order splines are used for the tangential directions and first-order basis functions are used for the through-the-thickness direction. In this setting, the complete volumetric control net is specified through the control nets of the inner and outer surfaces in accordance with
\begin{align}
    \mathbf{S}_{\mathrm{endo}}^{\rm{(P)}}\left( \xi, \eta \right) &= \sum_{i} \sum_{j}  R^{(\mathrm{P})}_{i,j,1}(\xi,\eta,0) \mathbf{P}^{\mathrm{(P)}}_{i,j,1}, & \mathbf{S}_{\mathrm{epi}}^{\rm{(P)}}\left( \xi, \eta \right) &= \sum_{i} \sum_{j}  R^{(\mathrm{P})}_{i,j,2}(\xi,\eta,1) \mathbf{P}^{\mathrm{(P)}}_{i,j,2},
    \label{eq:surfacedefs}
\end{align}
where $\mathbf{P}^{\mathrm{(P)}}_{i,j,1}$ are the control point associated with the inner surface and $\mathbf{P}^{\mathrm{(P)}}_{i,j,2}$ those associated with the outer surface. From the perspective of the fitting algorithm this means that when the fitting is performed for the inner and outer surface individually, a complete parametrization of the ventricle is attained. In order to perform the isogeometric analysis, further refinement (knot insertion) and order elevation procedures may still be required.


\newpage

\section{The sparse data fitting algorithm}\label{sec:Sec3}
In this section, we present the developed multi-patch NURBS fitting algorithm suitable for sparse and irregular point clouds. Since the left ventricle geometry is completely defined through the inner and outer surfaces, we introduce the fitting procedure in the generic setting of fitting a multi-patch NURBS surface to a point cloud. To not over-complicate notation, the surface to be fitted is denoted by $\overline{\Omega}$. In the context of the left ventricle, this means that $\overline{\Omega} = \Gamma^{\rm endo}$ or $\overline{\Omega}= \Gamma^{\rm epi}$. The control net and NURBS basis functions corresponding to the surface $\overline{\Omega}$ are here denoted by $\mathbf{P}^{(\mathrm{P})}_{i,j}$ and $R_{i,j}^{\rm (P)}(\xi,\eta)$, respectively, simplifying the notation as used for the surface definitions \eqref{eq:surfacedefs}. In Section~\ref{sec:thefittingalgorithm} we introduce the point distance minimization problem underlying the fitting procedure, along with the developed iterative algorithm used to solve this problem. In Section~\ref{sec:example} we then analyse the developed fitting algorithm using a well-defined test case.

\subsection{The fitting algorithm} 
\label{sec:thefittingalgorithm}
\begin{figure*}[!t]
     \centering
     \begin{subfigure}[b]{0.49\textwidth}
         \centering
             \includegraphics[width=\textwidth]{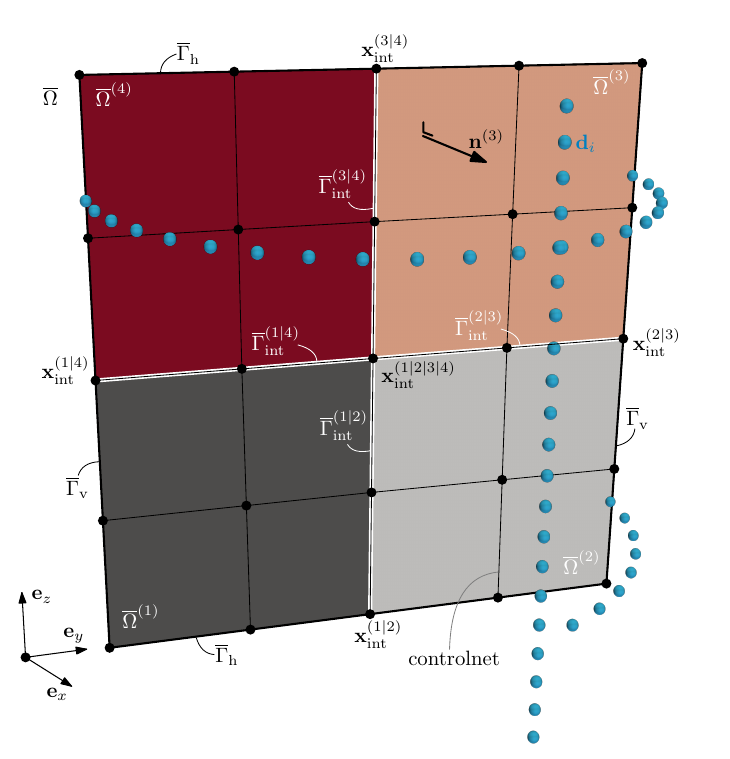}
         \caption{}
         \label{fig:example_graphical_a} 
     \end{subfigure}
     \hfill
     \begin{subfigure}[b]{0.49\textwidth}
         \centering
              \includegraphics[width=\textwidth]{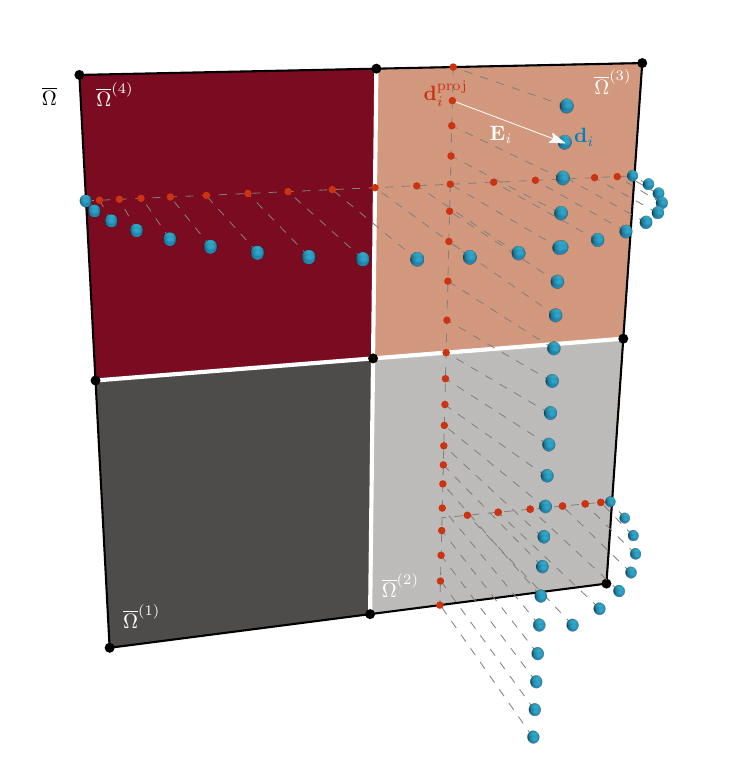}
         \caption{}
         \label{fig:example_graphical_b}
     \end{subfigure}
        \caption{(a) Graphical representation of the considered test-case with $\overline{\Omega}$ the closed initial multi-patch domain, \emph{i.e.,} the template, and $\mathbf{d}_i \in \mathcal{D}$ the data points to which the template is to be fitted, marked in blue. The multi-patch domain consists of four closed patches, $\overline{\Omega}^{(\rm{P})} \subset \overline{\Omega}$ for $\mathrm{P}=\{1,2,3,4\}$, connected via their interfaces, $\overline{\Gamma}^{(i|j)}_{\rm{int}} =\overline{\Omega}^{(i)} \cap \overline{\Omega}^{(j)}$ with $i \neq j$. The external closed boundaries are denoted by, $\{ \overline{\Gamma}_{\rm{v}} \cup \overline{\Gamma}_{\rm{h}} \} \subset  \overline{\Omega}$. The control net vertices located on the interfaces are denoted as, $\mathbf{x}_{\mathrm{int}}$, and represent the endpoints of a closed interface, $\overline{\Gamma}_{\mathrm{int}}$. (b) Visualization of the nearest-point-projected data points, $\mathbf{d}^{\mathrm{proj}}_i$, on the multi-patch domain, indicated in red. The difference is given by the error vector, $\mathbf{E}_i$.}
        \label{fig:example_graphical}
\end{figure*}

The fitting algorithm discussed in this section is developed to deform multi-patch surface topologies in 3D space, given a set of arbitrary (spatially sparse) data points. The data points can come from different sources, but in this manuscript we limit ourselves to 2D echocardiogram data (Figure~\ref{fig:workflow}). Such echocardiogram data relies on 2D slices or planes that, combined in 3D space, result in a sparse point cloud. We introduce the test case visualized in Figure~\ref{fig:example_graphical} as an exemplary sparse point cloud fitting problem, and use it throughout this section to explain each algorithm component. The data points (indicated in blue) are defined in 3 different planes and are sampled from a cylindrical shape with unit radius. The goal of the fitting algorithm is to interpolate these data points using a multi-patch domain that is analysis-suitable. We do this by defining an initial \emph{template} multi-patch geometry which is iteratively deformed such that it matches the data points and maintains smoothness, \emph{i.e.}, retains $G^1$-continuity.

Let $\overline{\Omega}$ be a closed multi-patch domain comprised of four patches as illustrated in Figure~\ref{fig:example_graphical_a}. The domain consists of external boundaries $\{ \overline{\Gamma}_{\rm{v}} \cup \overline{\Gamma}_{\rm{h}} \}  \subset  \overline{\Omega}$ and four internal boundaries, $\overline{\Gamma}_{\mathrm{int}}$, \emph{i.e.}, interfaces. Given a set of $n_{\mathrm{data}}$
 arbitrary spatial data points, $\mathcal{D}=\{\mathbf{d}_1, \mathbf{d}_2, ..., \mathbf{d}_{n_{\mathrm{data}}}\}$, where each item corresponds to the spatial coordinate of a data point, we require a solution to the minimization problem:
 \begin{equation}
\left\{ \begin{aligned}
 \multicolumn{2}{l}{\mbox{Find the closed multi-patch domain } $\overline{\Omega}_{\mathrm{fit}}$ \mbox{such that}}\\
 \overline{\Omega}_{\mathrm{fit}} &= \argmin_{\tilde{\Omega} \, \subset \, \mathbb{R}^3}\left( \mathcal{F}( \tilde{\Omega}, \mathcal{D})  \right),  && \label{eq:argmin} \\
 \multicolumn{2}{l}{\mbox{where}} \\
 \mathcal{F}( \tilde{\Omega}, \mathcal{D}) & = \underbrace{\varepsilon_{\mathcal{D}}(\tilde{\Omega}, \mathcal{D}) }_{\text{Data fit}} + \lambda \underbrace{ \varepsilon_{\mathcal{C}}(\tilde{\Omega})}_{\text{Continuity}}. &&
\end{aligned}
\right.
\end{equation}
The functional $\mathcal{F}(\cdot)$ to be minimized is the summation of the data fit error and the continuity/smoothness error across the interfaces. The constant $\lambda$ introduces dimensional consistency and balances the importance of the error contributions \cite{weiss_advanced_2002}.

The displacement error norm is an averaged discrete $L^2$-norm that quantifies how close the data points are to the multi-patch surface and is defined as
\begin{equation}\label{eq:error_displ}
    \varepsilon_{\mathcal{D}}(\tilde{\Omega}, \mathcal{D}) = \frac{1}{{n_{\mathrm{data}}}} \sum^{{n_{\mathrm{data}}}}_{i=1} \left| \, \mathbf{E}_i \, \right|,
\end{equation}
with the error vector, $\mathbf{E}_i$, visualized in Figure~\ref{fig:example_graphical_b}, given by
\begin{equation}
\label{eq:error_vec}
    \mathbf{E}_i =  \mathbf{d}_i - {\mathbf{d}}^{\mathrm{proj}}_i.
\end{equation}
The projected data points, ${\mathbf{d}}^{\mathrm{proj}}_i$, are defined as the nearest-point-projection between the multi-patch surface and the data points (indicated in red in Figure~\ref{fig:example_graphical_b}), such that
\begin{equation}
\label{eq:projection}
    \mathbf{d}^{\mathrm{proj}}_i = \argmin_{\tilde{\mathbf{d}}_i \, \in \, \tilde{\Omega}}  | \, \tilde{\mathbf{d}}_i - \mathbf{d}_i \, |, \quad \forall \, \mathbf{d}_i  \in \mathcal{D}.
\end{equation}
The continuity error norm, $\varepsilon_{\mathcal{C}}$, is quantified by the average angle between the normal vectors, $\mathbf{n}$, across the interfaces, \emph{i.e.},
\begin{equation}\label{eq:error_cont}
    \varepsilon_{\mathcal{C}} = \frac{1}{\int \text{d} \overline{\Gamma}_{\mathrm{int}}} \int \left|  \text{arccos} \left( \frac{\mathbf{n}^{(i)} \cdot \mathbf{n}^{(j)} }{ \left|\mathbf{n}^{(i)} \right| \left|\mathbf{n}^{(j)} \right| }   \right) \right| \text{d} \overline{\Gamma}_{\mathrm{int}}, \quad \text{with} \quad i \neq j; \ i,j \in \{1,2, \ldots , n_{\mathrm{patch}} \},
\end{equation}
where the superscripts refer to the patch indices that share the considered interface, $\overline{\Gamma}_{\mathrm{int}}$. Note that this error norm assumes that a maximum of two patches are allowed to share a single interface. Both error terms are chosen such that they represent a physically interpretable measure of the fitted surface, facilitating the user in choosing a suitable stopping criterion for the considered application.

\begin{algorithm}
\caption{Multi-patch template fitting algorithm.}\label{alg:fitting}
    \begin{algorithmic}[1]
        \Require \\ $\overline{\Omega}, \ \mathbf{d}_c \subset \mathcal{D}, \ \mathscr{C}$ \Comment{Initial template, data point set, user-defined constraint operator}
        \Ensure $\tilde{\Omega}^{0} \gets \overline{\Omega}$
        \While{$(t \leq t_{\rm{max}} )\lor (\varepsilon > \text{tol})$}
            \State $\mathbf{d}_c^{\mathrm{proj}} \gets \Call{project\_on\_surface}{\tilde{\Omega}^{t}, \mathbf{d}_c}$ \Comment{Nearest-point projection \emph{cf.}~\eqref{eq:projection}}\label{alg:pointproject}
            \State ${\mathbf{E}_c} \gets \Call{get\_error\_vector}{\mathbf{d}_c, \mathbf{d}_c^{\mathrm{proj}}}$ \Comment{Error vector between surface and data point, \emph{cf.}~\eqref{eq:error_vec}} 
                \For{$( \tilde{\Omega}^{(\mathrm{P})}, \mathbf{E}^{(\mathrm{P})}_k, \mathbf{d}^{\mathrm{proj},(\mathrm{P})}_k )$ \textbf{in} $( \tilde{\Omega}^{t}, {\mathbf{E}_c, \mathbf{d}_c^{\mathrm{proj}}} )$  } \Comment{Loop over the patches}
                    \State $\mathbf{d}_k^{\mathrm{local},(\mathrm{P})} \gets \Call{get\_local\_coordinates}{\tilde{\Omega}^{(\mathrm{P})}, \mathbf{d}_k^{\mathrm{proj},(\mathrm{P})}}$ \Comment{Convert to $(\xi,\eta)$-coordinates, \emph{cf.}~\eqref{eq:dlocal}}\label{alg:dlocal}
                    \State $M^{(\mathrm{P})}_{i,j}, m^{\mathrm{(P)}}_{k} \gets \Call{get\_voronoi\_mass}{\tilde{\Omega}^{(\rm{P})}, \, \mathbf{d}_k^{\mathrm{local},(\mathrm{P})}}$ \Comment{Data point mass, \emph{cf.}~\eqref{eq:Mij}}\label{alg:Mij}
                    \State $\mathbf{u}^{(\mathrm{P})}_{i,j} \gets \alpha_{\mathrm{relax}} \ \Call{get\_cps\_data\_fit}{\tilde{\Omega}^{(\rm{P})}, \, \mathbf{d}_k^{\mathrm{local},(\mathrm{P})}, \, {\mathbf{E}}_k^{(\mathrm{P})}, \, m^{\mathrm{(P)}}_{k}}$  \Comment{Control point data fit displacement, \emph{cf.}~\eqref{eq:displace}}\label{alg:cpsdisplace}
                \EndFor
            \State $\tilde{\Omega}^{t+0.5} \gets \Call{update\_surface}{\tilde{\Omega}^t, \, \mathbf{u}_{i,j}, \, \mathscr{C}}$ \Comment{Update surface control points, \emph{cf.}~\eqref{eq:Pupdate}}\label{alg:update1}
            \State $\mathbf{u}_{i,j} \gets \Call{correct\_for\_G0\_continuity}{\tilde{\Omega}^{t+0.5}, \, \mathbf{u}_{i,j}, \, M_{i,j}}$ \Comment{Connect patches, \emph{cf.}~\eqref{eq:connect}}\label{alg:G0cont}
            \State $\tilde{\Omega}^{t+0.75} \gets \Call{update\_surface}{\tilde{\Omega}^{t+0.5}, \, \mathbf{u}_{i,j} \, \mathscr{C}}$ \Comment{Update surface control points, \emph{cf.}~\eqref{eq:Pupdate}}\label{alg:update2}
            \State $\mathbf{u}_{i,j}\gets\Call{correct\_for\_G1\_continuity}{\tilde{\Omega}^{t+0.75}, \, \mathbf{u}_{i,j}, \, M_{i,j}}$ \Comment{Approximate inter-patch continuity, \emph{cf.}~\eqref{eq:G1displace}}\label{alg:G1cont}
            \State $\tilde{\Omega}^{t+1} \gets \Call{update\_surface}{\tilde{\Omega}^{t+0.75}, \,\mathbf{u}_{i,j}, \, \mathscr{C}}$ \Comment{Update surface control points, \emph{cf.}~\eqref{eq:Pupdate}}\label{alg:update3}
            \State $\varepsilon_{\mathcal{D}}  \gets\Call{get\_error}{ \mathbf{E}_c }$ \Comment{Displacement error, \emph{cf.}~\eqref{eq:error_displ}}
            \If{$\Delta\varepsilon_{\mathcal{D}} < \Delta\mathrm{tol}$} 
               \State $\tilde{\Omega}^{t+1} \gets \Call{refine}{\tilde{\Omega}^{t+1}}$ \Comment{Uniform knot refinement}\label{alg:refinement}
            \EndIf
            \State $t \gets t+1$
        \EndWhile 
    \end{algorithmic}
\end{algorithm}

\begin{figure*}[!t]
     \centering
     \begin{subfigure}[b]{0.33\textwidth}
         \centering
             \includegraphics[width=\textwidth]{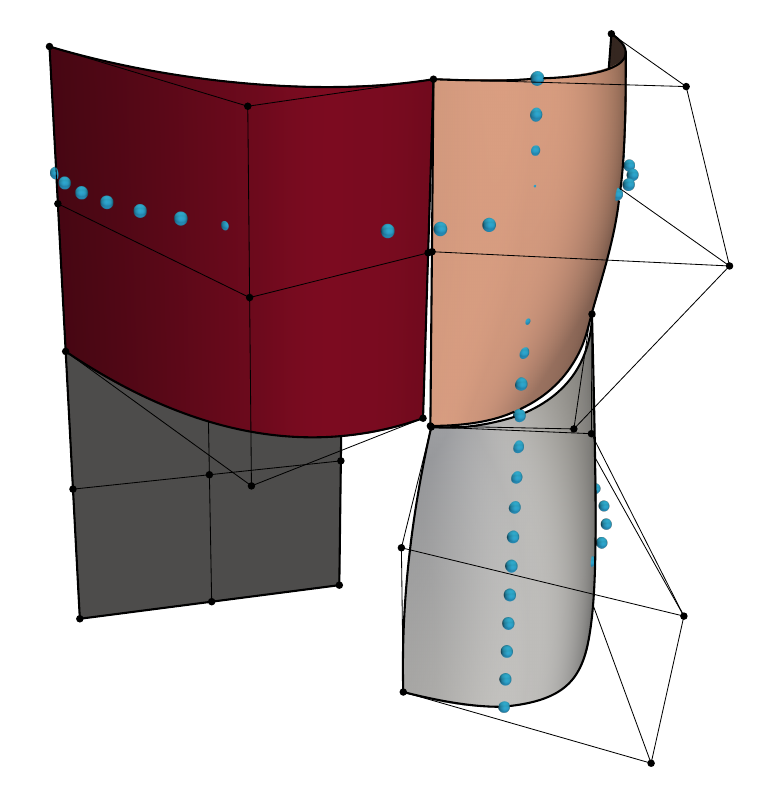}
         \caption{}
         \label{fig:fitting_graphical_a} 
     \end{subfigure}
     \hfill
     \begin{subfigure}[b]{0.33\textwidth}
         \centering
              \includegraphics[width=\textwidth]{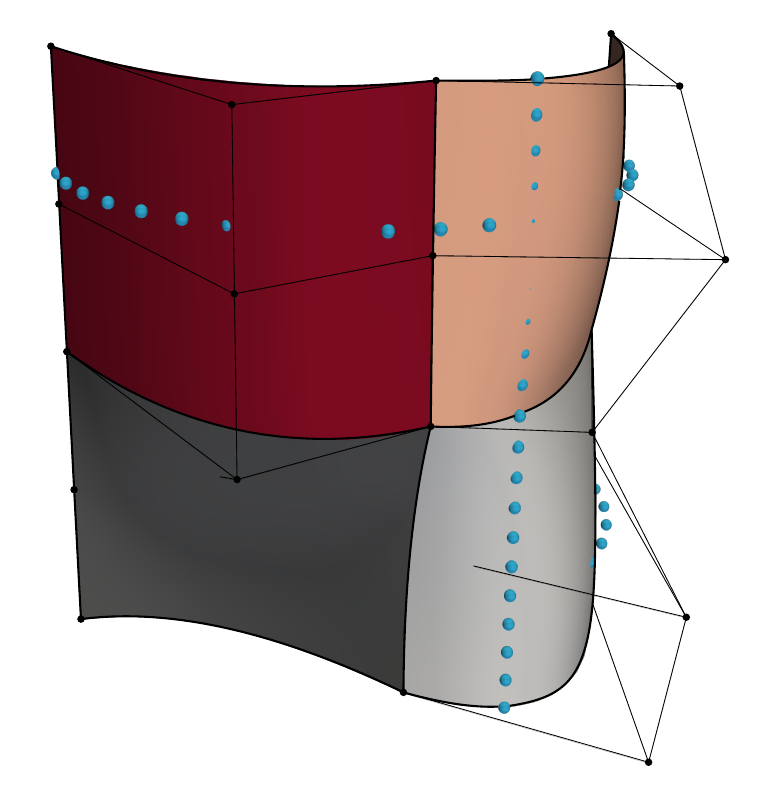}
         \caption{}
         \label{fig:fitting_graphical_b}
     \end{subfigure}
     \begin{subfigure}[b]{0.33\textwidth}
         \centering
              \includegraphics[width=\textwidth]{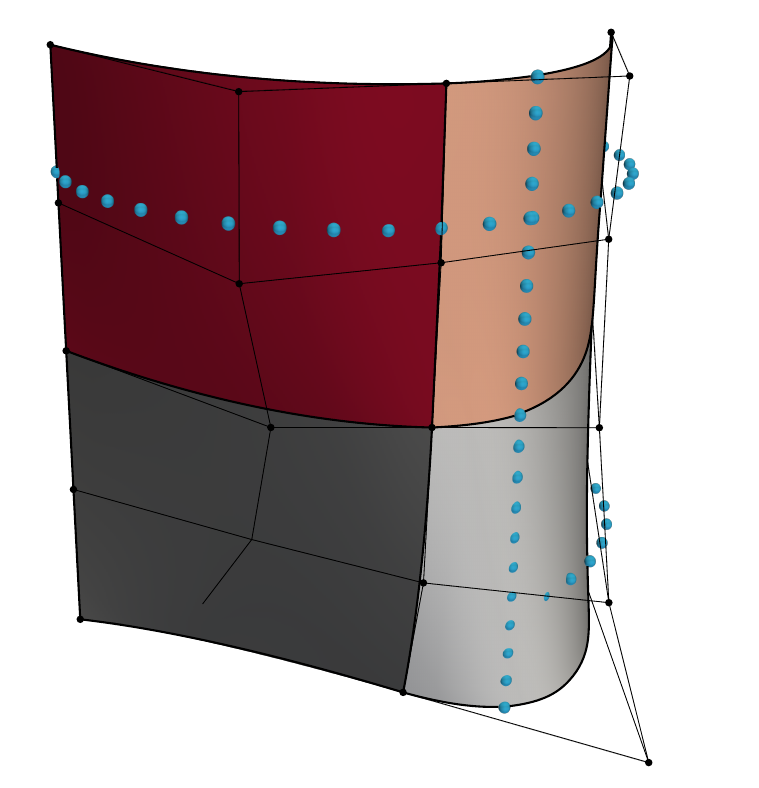}
         \caption{}
         \label{fig:fitting_graphical_c}
     \end{subfigure}
        \caption{Graphical overview of the three main steps of the multi-patch surface fitting algorithm: (a) Fitting the individual patches to the data points; (b) Correcting for $G^0$-patch-interface continuity, \emph{i.e.}, connecting the patches; (c) Correcting for $G^1$-patch-interface continuity, \emph{i.e.}, smoothing the multi-patch geometry at the patch interfaces.}
        \label{fig:fitting_graphical}
\end{figure*}

The algorithm aims to reduce both error contributions by changing the spatial coordinates of the control net, $\mathbf{P}_{i,j}$, following Equation~\eqref{eq:MAPP}, in an iterative sequential manner. This procedure, outlined in Algorithm~\ref{alg:fitting}, is graphically illustrated in Figure~\ref{fig:fitting_graphical}. The sequential algorithm comprises three main steps: (I) data point fitting of the individual patches (Figure~\ref{fig:fitting_graphical_a} and Lines~\ref{alg:dlocal}-\ref{alg:cpsdisplace} in Alg.~\ref{alg:fitting}); (II) connecting the individual patches (Figure~\ref{fig:fitting_graphical_b} and Line~\ref{alg:G0cont}); and (III) approximating $G^1$ inter-patch continuity (Figure~\ref{fig:fitting_graphical_c} and Line~\ref{alg:G1cont}). Steps~(I)-(III), which will be elaborated upon in the remainder of this section, are iterated, resulting in a gradual minimization of the functional in Equation~\eqref{eq:argmin}. Each step is proceeded by a control net and \emph{surface update}, (Lines~\ref{alg:update1}, \ref{alg:update2} and \ref{alg:update3}), where only the spatial coordinates of the control net, $\mathbf{P}_{i,j}$, are altered while the NURBS weights, $w_{i,j}$, remain unchanged (visualized in Figure~\ref{fig:fitting_graphical_a}). Therefore, the set of control points that defines the multi-patch domain, $\mathbf{P}_{i,j}=\{\mathbf{P}^{(\mathrm{1})}_{i,j}, ..., \mathbf{P}^{(\mathrm{P})}_{i,j} \}$, is updated every iteration (indexed by $t$) by a set of displacement vectors, $\mathbf{u}_{i,j}=\{\mathbf{u}^{(\mathrm{1})}_{i,j}, ..., \mathbf{u}^{(\mathrm{P})}_{i,j} \}$ for every patch $\mathrm{P}$, as
\begin{equation}
\label{eq:Pupdate}
    \mathbf{P}^{(\mathrm{P})^{t+1}}_{i,j} = {\mathbf{P}^{(\mathrm{P})^t}_{i,j} + \mathscr{C}(\mathbf{u}^{(\mathrm{P})}_{i,j}} ),  
\end{equation}
where the $\mathbf{P}^{(\mathrm{P})^{t+1}}_{i,j}$ is the updated control net of patch $\mathrm{P}$ and $\mathscr{C}(\cdot)$ is a user-defined constraint operator. This operator ensures that any computed displacement, $\mathbf{u}^{(\mathrm{P})}_{i,j}$, adheres to user-defined constraints, \emph{e.g.}, is only allowed to move within a plane, along a line, or is completely fixed. Note that a relaxation factor, $\alpha_{\mathrm{relax}}$, is used in Line~\ref{alg:cpsdisplace} to control the computed displacement, $\mathbf{u}^{(\mathbf{P})}_{i,j}$, before updating the control net, \emph{cf.} Equation~\eqref{eq:Pupdate}.


\subsubsection{Step I: Single patch data point fit (Algorithm~\ref{alg:fitting}, Lines~\ref{alg:dlocal}-\ref{alg:cpsdisplace})}\label{sec:displace}
\begin{figure*}[!t]
     \centering
     \begin{subfigure}[b]{0.455\textwidth}
         \centering
             \includegraphics[width=\textwidth]{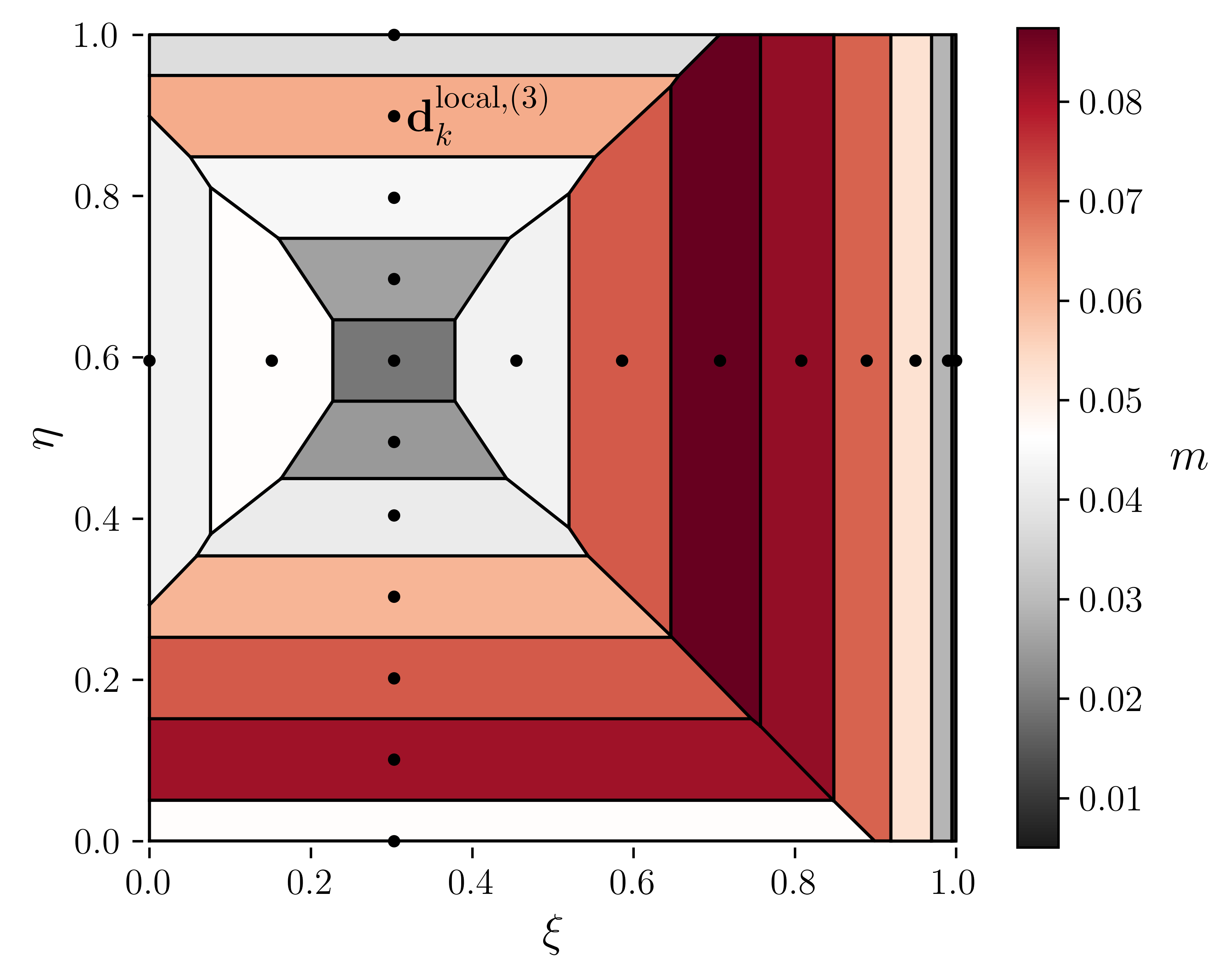}
         \caption{}
         \label{fig:voronoi_diagram} 
     \end{subfigure}
     \hfill
     \begin{subfigure}[b]{0.48\textwidth}
         \centering
              \includegraphics[width=\textwidth]{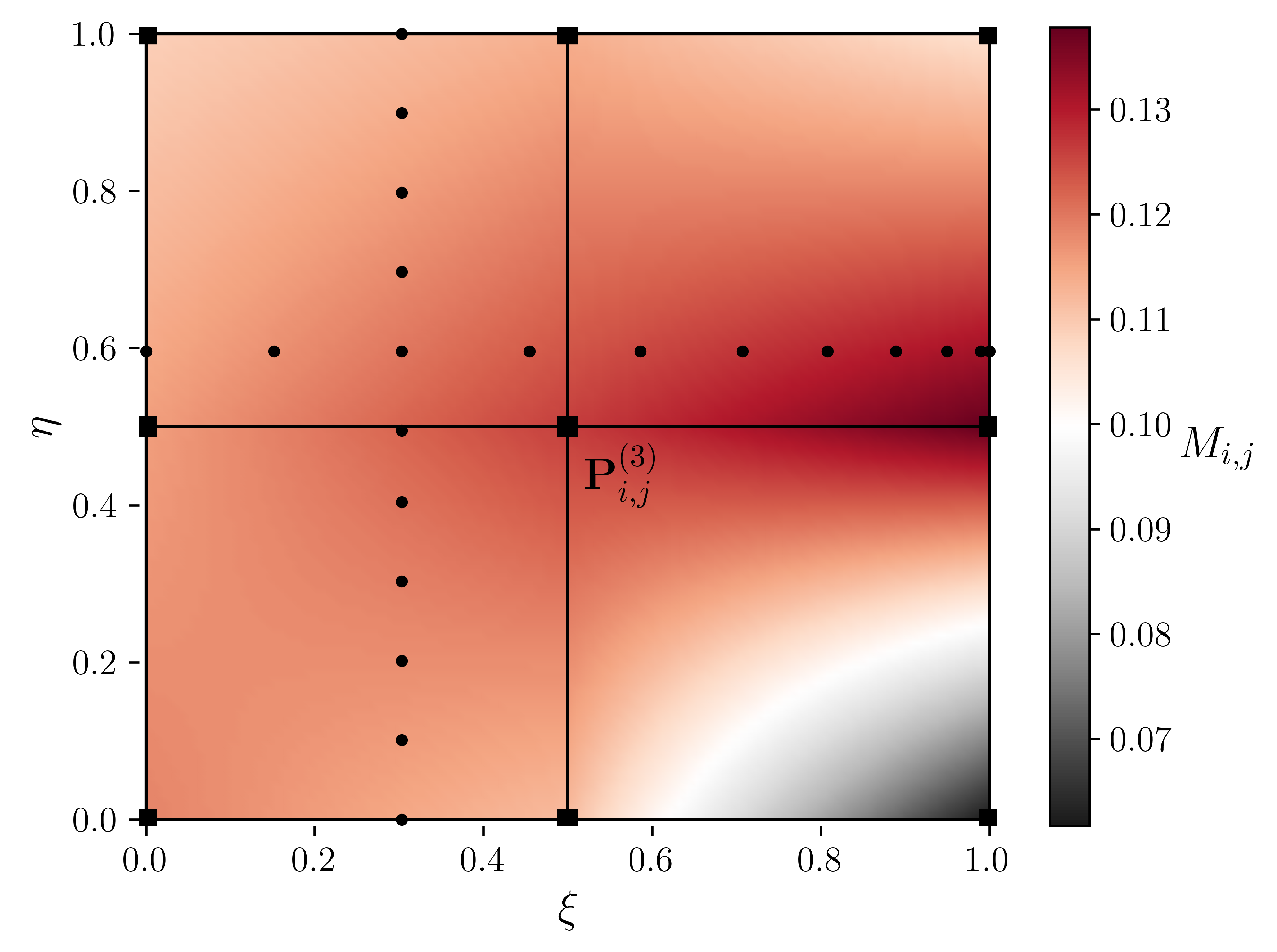}
         \caption{}
         \label{fig:weight_matrix}
     \end{subfigure}
        \caption{(a) Graphical representation of the Voronoi segments on the closed parametric patch domain, $\{\xi,\eta\} \in [0,1]\times [0,1]$, as a result of the local data points, $\mathbf{d}^{\mathrm{local},(3)}_k$, that correspond to the third patch, $\overline{\Omega}^{(3)}$ in Figure~\ref{fig:example_graphical_b}. The local data points are visualized as black dots with corresponding colors indicating the surface area of each segment, denoted by $m$. (b) Heat map showing the relative importance of each control point with respect to the data fit, quantified by $M_{i,j}$ as defined in Equation~\eqref{eq:Mij}. The control net $\mathbf{P}^{(3)}_{i,j}$ is indicated by the squares.}
        \label{fig:datapoint_mapp}
\end{figure*}

The first step of the fitting algorithm involves displacing the multi-patch surface toward the data points. We do this for each patch separately, where we treat each patch as a separate surface that is not connected to adjacent patches, \emph{i.e.}, the patches are discontinuous. To approximate the control net displacement vectors $\mathbf{u}^{(\mathrm{P})}_{i,j}$ of a single patch (Line~\ref{alg:cpsdisplace} in Algorithm~\ref{alg:fitting}), we draw inspiration from the B-spline convolution operation considered in Ref.~\cite{verhoosel_image-based_2015}, which has favorable properties in terms of, \emph{e.g.}, the occurrence of oscillations. In the context of the setting considered here, the NURBS-based convolution operation is given by
\begin{equation}
\label{eq:displace}
    \mathbf{u}^{(\mathrm{P})}_{i,j} =  \frac{\sum^{n^{(\mathrm{P})}_{\mathrm{data}}}_{k=1} R^{(\mathrm{P})}_{i,j}( \mathbf{d}^{\mathrm{local},(\mathrm{P})}_k ) \, m^{(\mathrm{P})}_k \, \mathbf{E}^{(\mathrm{P})}_k }{\sum^{n^{(\mathrm{P})}_{\mathrm{data}}}_{k=1} R^{(\mathrm{P})}_{i,j}( \mathbf{d}^{\mathrm{local},(\mathrm{P})}_k ) \, m^{(\mathrm{P})}_k},
\end{equation}
where the mismatch in \textit{data-point-to-surface} is represented by the error vector, $\mathbf{E}^{(\mathrm{P})}_k = \mathbf{d}^{(\mathrm{P})}_k - \mathbf{d}^{\mathrm{proj},(\mathrm{P})}_k $ (Equation~\eqref{eq:error_vec}). The error vector dictates the direction and distance the projected surface point, $\mathbf{d}^{\mathrm{proj},(\mathrm{P})}_k$, should be displaced. The NURBS basis functions $R^{(\mathrm{P})}_{i,j}$ are evaluated in the local data points, $\mathbf{d}^{\mathrm{local},(\mathrm{\mathrm{P}})}_k$. These local data points are defined in the $(\xi,\eta)$-parameter domain, $\mathbf{d}^{\mathrm{local},(\mathrm{P})}_k \subset \mathbb{R}^{2}$, and are obtained by solving
\begin{equation}\label{eq:dlocal}
    \mathbf{d}^{\mathrm{local}, (\mathrm{P})}_k =  \argmin_{\tilde{\mathbf{d}}^{\mathrm{local}} \, \in \, [0,1]^2} {   \left\| \mathbf{S}^{\rm{(P)}}(\tilde{\mathbf{d}}^{\mathrm{local}}) - \mathbf{d}^{\mathrm{proj},(\mathrm{P})}_k \right\|_{L^2} }.
\end{equation}
The projected data points, $\mathbf{d}^{\mathrm{proj},(\mathrm{P})}_k \subset \mathbb{R}^{3}$, are related to the local data points by $\mathbf{d}^{\mathrm{proj},(\mathrm{P})}_k = \mathbf{S}^{\rm{(P)}}(\mathbf{d}^{\mathrm{local},(\mathrm{P})}_k )$, following Equation~\eqref{eq:MAPP}. The rational basis functions in Equation~\eqref{eq:displace} are weighted by a mass $m^{(\mathrm{P})}_k$. This mass represents the importance of the specific data point, $\mathbf{d}_k$, in relation to the fitting process and relative to other data points associated with the same patch. The data point mass is related to the surface area of Voronoi segments corresponding to the local projected data point, $\mathbf{d}^{\mathrm{local},(\mathrm{P})}_k$, visualized in Figure~\ref{fig:voronoi_diagram}. These segments partition the parameter domain in regions that are closest to each of the local projected data points. In the proposed algorithm, the Voronoi segments are evaluated for each patch individually and are bounded by the local minima and maxima, \emph{i.e.}, $\xi,\eta \in [0,1]^2$. As a result, the Voronoi segments are not continuous across patch interfaces. Satisfying interface continuity of the Voronoi segments is conceptually possible, but constitutes a non-essential extension of the algorithm which is beyond the scope of this manuscript. Nonetheless, the projected data points that are located on a patch interface are still shared or seen by the patches connected to it. The Voronoi segments are defined in the parametric space, $\{\xi, \eta\} \subset [0,1]^2$, resulting in a total unit area $A^{(\mathrm{P})}=\sum^{(\mathrm{P})}_k  m_k^{(P)}= 1$. Using the partition of unity property of NURBS, the area can also be expressed as a sum of control point masses, $M^{(\mathrm{P})}_{i,j}$, as
\begin{equation}
    A^{(\mathrm{P})}=\sum^{n}_{i=1} \sum^{m}_{j=1} M^{(\mathrm{P})}_{i,j} = 1,
\end{equation}
provided that $n^{(\mathrm{P})}_{\mathrm{data}}\neq0$ and with
\begin{equation}
\label{eq:Mij}
    M^{(\mathrm{P})}_{i,j} = \sum^{n^{(\mathrm{P})}_{\mathrm{data}}}_k R^{(\mathrm{P})}_{i,j}( \mathbf{d}^{\mathrm{local},(\mathrm{P})}_k ) \, m^{(\mathrm{P})}_{k}.
\end{equation}
These masses indicate how important the control points are to the fitting process relative to one another and will be used in the continuity correction steps discussed below. In the case of an empty set of data points, $n^{(\mathrm{P})}_{\mathrm{data}}=0$, Equation~\eqref{eq:Mij} becomes a zero-valued matrix $M^{(\mathrm{P})}_{i,j} = 0_{i,j}$. A graphical representation of Equation~\eqref{eq:Mij} is visualized in Figure~\ref{fig:weight_matrix}, where the control point values are visualized as a heat map at the first iteration indicated for the third patch, $\overline{\Omega}^{\mathrm{(P)}}$.

\subsubsection{Step II: $G^0$-patch continuity (Algorithm~\ref{alg:fitting}, Line~\ref{alg:G0cont})}\label{sec:g0cont}
After displacing the individual patches following Equation~\eqref{eq:displace}, $G^0$-patch continuity is not ensured and should be corrected for as is visualized in Figure~\ref{fig:fitting_graphical_a}. Provided that we assume the patch interfaces to remain conforming, \emph{i.e.}, the knot vector at each interface side is identical, $G^0$-continuity at the closed interface between two patches, $\overline{\Gamma}^{(i|j)}_{\mathrm{int}} = \overline{\Omega}^{(i)} \cap \overline{\Omega}^{(j)}$, is satisfied when
\begin{equation}
\label{eq:G0}
    \mathbf{P}^{(i)}= \mathbf{P}^{(j)}, \quad w^{(i)} = K w^{(j)}
\end{equation}
for all control points on the interface, where $K$ is a positive real number \cite{che_g1_2005}. It requires the spatial coordinates of the control points, $\mathbf{P}$, at the shared interface to match identically, while the weights associated with these control points, $w$, may differ by a factor of $K$. Given that we keep the weights constant during the fitting process, we merely have to match the control points at the interfaces. However, one should consider the importance that the individual patches have on the data point fit, \emph{i.e.}, a simple average of the interface and vertex control points would hamper the fitting procedure if one of the patches does not contribute to the data fitting. This is the case for the first patch, $\overline{\Omega}^{\mathrm{(1)}}$, visualized in Figure~\ref{fig:example_graphical_b}, where no projected data points are attributed to this patch (indicated by the red dots). As a result, we perform a weighted average of the control points located at the internal interfaces and shared vertices \emph{cf.} Equation~\eqref{eq:Mij}, to obtain a $G^0$-continuous multi-patch domain (visualized in Figure~\ref{fig:fitting_graphical_b}).

To perform the weighted average we first make a distinction between \emph{open interfaces} and \emph{interface vertices}, \emph{i.e.}, the corner points of patches on interfaces. Let $\chi_{\mathrm{int}}=\{\mathbf{x}^{k}_{\mathrm{int}} | \, k \in \mathcal{I}_{\mathbf{x}_{\mathrm{int}}} \}$ be the unique set of interface vertices, where the set $\mathcal{I}_{\mathbf{x}_{\mathrm{int}}}$ comprises the set of patch indices adjacent to this vertex. For the example case in Figure~\ref{fig:example_graphical_a} we have $\mathcal{I}_{\mathbf{x}_{\mathbf{int}}}=\{ (1|2), (2|3), (3|4), (1|4), (1|2|3|4)\}$. Similarly, we define $\Sigma_{\mathrm{int}}=\{\Gamma^{k}_{\mathrm{int}} | \, k \in \mathcal{I}_{\Gamma_{\mathrm{int}}} \}$ as the unique set of open multi-patch interfaces, where the set $\mathcal{I}_{\Gamma_{\mathrm{int}}}$ comprises the set of patch indices adjacent to this interface. For the example case in Figure~\ref{fig:example_graphical_a} we have $\mathcal{I}_{\Gamma_{\mathrm{int}}}=\{ (1|2), (2|3), (3|4), (1|4)\}$. We define open interfaces such that they do not include their endpoints, which are interface vertices by definition, meaning $\Sigma_{\mathrm{int}} \cap \chi_{\mathrm{int}} = \emptyset$. By evaluating the displacement and distributed mass, Equation~\eqref{eq:displace} and \eqref{eq:Mij} respectively, for each patch that shares the control points located on the unique interfaces and vertices, $\overline{\Sigma}_{\mathrm{int}}$, we compute the weighted average displacement such that, 
\begin{equation}
\label{eq:connect}
    \mathbf{u}^{\phantom{}}_{\lambda} = \frac{ \sum_{\mathrm{P} \in \mathcal{I}_{\lambda}} M^{(\mathrm{P})}_{\lambda} \, \mathbf{u}^{(\mathrm{P})}_{\lambda} }{  \sum_{\mathrm{P} \in \mathcal{I}_{\lambda}} M^{(\mathrm{P})}_{\lambda} }, \quad \text{for} \quad \lambda \in \{\Gamma_{\mathrm{int}}, \mathbf{x}_{\mathrm{int}}\},
\end{equation}
where the subscript $\lambda$ indicates whether the quantity is evaluated at an open interface or interface vertex. This distinction between an open interface and an interface vertex is due to the difference in \emph{valency}. The valency of an interface is typically limited to $2$, with some exceptions, while the valency of interface vertices is usually higher~\cite{hughes_chapter_2021}. Because of this, open interfaces and interface vertices should be treated separately to ensure a correct weighted average is obtained following Equation~\eqref{eq:connect}.

\subsubsection{Step III: $G^1$-patch continuity (Algorithm~\ref{alg:fitting}, Line~\ref{alg:G1cont})}\label{sec:g1cont}
Following the correction on $G^0$-patch continuity is the correction for $G^1$-continuity, Figure~\ref{fig:fitting_graphical_c}. This step aims to reduce the sharp $G^0$ edges at interfaces of interest and obtain a \emph{smooth} $G^1$ representation of the geometry. These sharp edges are inherent to the multi-patch configuration due to the $C^0$-continuity of the rational basis functions at the patch interfaces. Even if the parametrization is limited to $C^0$-continuity at the patch edges, one can still obtain $G^1$-continuity (or higher) as this is a geometric property rather than a parametrization argument. The geometric continuity, specifically $G^1$, is quantified by the angle of the normal vectors (tangent planes) of each patch at the interface. While angle values should ideally be zero, angles with an absolute value below a specific threshold can be considered almost \emph{smooth} (following Ref.~\cite{karciauskas_can_2015} we use a threshold of $0.1$~[deg]). We note that, in light of the developed fitting algorithm, the $G^1$-continuity is not merely aesthetic, but plays an important role in maintaining the template shape during the fitting process, as will be discussed in the example below.

Satisfying $G^1$-continuity is an active field of research in which the control points adjacent to and at the interface are positioned such that non-linear constraints ensuring the $G^1$-continuity criterion are satisfied~\cite{che_g1_2005, mourrain_dimension_2016, blidia_g1-smooth_2017, kiciak_geometric_2017}. For arbitrary (complex) topologies, these non-linear constraints cannot always be satisfied, and as such often require degree elevation~\cite{peters_complexity_2010}. In view of our analysis-suitable geometry, we do not desire excessive degree elevation or knot insertion since this would both defeat the purpose of fitting a coarse template on sparse data and increase computational load following the IGA paradigm. As a result, we propose a $G^1$ approximation procedure, which approximates $G^1$-continuity at specific landmarks (set to be the Greville points). The number of landmarks is proportional to the number of control points, resulting in convergent $G^1$-continuity behavior across the interface upon knot refinement, similar to Refs.~\cite{shi_reconstruction_2004,luong_approach_2022}. It should be mentioned that these geometry-based continuity methods are different from the continuity methods developed in the IGA framework: the Analysis-Suitable $G^1$~\cite{farahat_isogeometric_2023}, the Approximate $C^1$ constructions~\cite{weinmuller_construction_2021}, the D-patch~\cite{toshniwal_smooth_2017}, and the Almost-$C^1$~\cite{takacs_almost-c1_2023}. Such continuity methods focus on the isogeometric function space rather than on the geometric property; see Ref.~\cite{verhelst_comparison_2024} for a comparison. Such alterations of the function space are not considered in this manuscript.

\begin{figure*}[!t]
     \centering
     \begin{subfigure}[b]{0.48\textwidth}
         \centering
             \includegraphics[width=\textwidth]{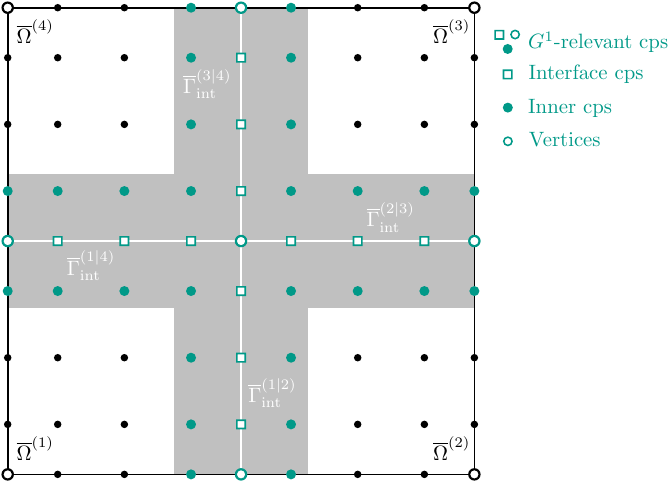}
         \caption{}
         \label{fig:G1_schematic_a} 
     \end{subfigure}
     \hfill
     \begin{subfigure}[b]{0.48\textwidth}
         \centering
              \includegraphics[width=\textwidth]{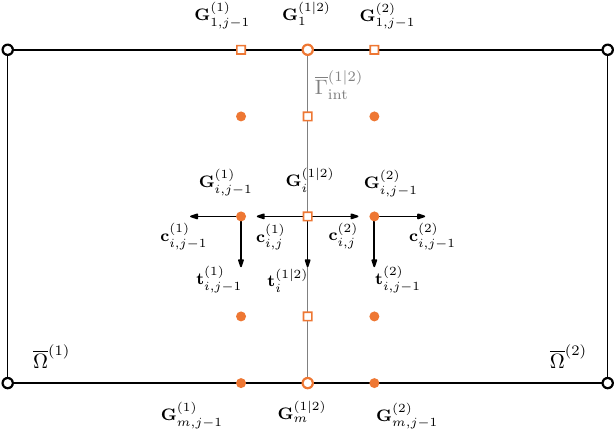}
         \caption{}
         \label{fig:G1_schematic_b}
     \end{subfigure}
        \caption{(a) Schematic overview of the multi-patch domain, $\overline{\Omega}$, in which 16 control points (cps) per patch are visualized for illustration purposes only. Control points located within the grey area are considered $G^1$-relevant cps and comprise vertices, interface cps, and inner cps (indicated in teal). (b) Schematic illustration of the relevant quantities of interest at the interface, $\overline{\Gamma}^{(1|2)}_{\mathrm{int}} = \overline{\Omega}^{(1)} \cap \overline{\Omega}^{(2)}$. The markers indicate the Greville points, $\mathbf{G}$, associated with the $G^1$-relevant control points in (a). For an interface with $m$ rows of Greville points, indicated by the row index, we evaluate the cross- and tangent-boundary vectors, $\mathbf{c}$ and $\mathbf{t}$ respectively, at each Greville point $\mathbf{G}$. These quantities are required to approximate $G^1$-continuity across the interface.}
        \label{fig:G1_schematic}
\end{figure*}

We consider again the multi-patch domain as shown in Figure~\ref{fig:example_graphical}, and provide a schematic overview of it in Figure~\ref{fig:G1_schematic_a} with additional control points introduced for clarification. The control points that affect $G^1$-continuity are located in the grey zone, which are either at the interface or adjacent to the interface (inner control points). The goal is to modify only the marked control points such that the multi-patch surface obtains globally approximate $G^1$-continuity. The procedure of obtaining approximate $G^1$-continuity across patch interfaces is explained using patches $\overline{\Omega}^{(1)}$ and $\overline{\Omega}^{(2)}$ as an example, Figure~\ref{fig:G1_schematic_b}. Unlike the open interfaces defined in the previous section, we consider only closed interfaces which are interfaces that include their endpoints (vertices), \emph{e.g.}, $\overline{\Gamma}^{(1|2)}_{\mathrm{int}} = {\Gamma}^{(1|2)}_{\mathrm{int}} \cup \mathbf{x}^{(1|2)}_{\mathrm{int}} \cup \mathbf{x}^{(1|2|3|4)}_{\mathrm{int}}$, visualized in Figure~\ref{fig:example_graphical_a}. Let $\mathbf{t}^{(1)}$ and $\mathbf{t}^{(2)}$ be the tangent-boundary vectors and $\mathbf{c}^{(1)}$ and $\mathbf{c}^{(2)}$ be the cross-boundary vectors defined at the interface $\overline{\Gamma}^{(1|2)}_{\mathrm{int}}$ of patches (1) and (2), respectively. Following the definition of $G^0$-continuity in Equation~\eqref{eq:G0}, it holds that $\mathbf{t}^{(1|2)}=\mathbf{t}^{(1)}=\mathbf{t}^{(2)}$. $G^1$-continuity is then guaranteed when the vectors, $\mathbf{t}^{(1|2)}$, $\mathbf{c}^{(1)}$, and $\mathbf{c}^{(2)}$, are linearly dependent across the entire closed interface, $\overline{\Gamma}^{(1|2)}_{\mathrm{int}}$. This means that the tangent space formed by, $\{\mathbf{c}^{(1)}$, $\mathbf{t}^{(1|2)}\}$, and, $\{\mathbf{c}^{(2)}$, $\mathbf{t}^{(1|2)}\}$, are both the same, \emph{i.e.}, the vectors are coplanar: $\text{det}(\mathbf{c}^{(1)}$, $\mathbf{t}^{(1|2)},\mathbf{c}^{(2)})=0$ \cite{che_g1_2005}. The procedure outlined in the following paragraph will therefore focus on modifying the specific control points such that this coplanarity condition is approximated. 

Satisfying the coplanarity condition is achieved by first defining a common plane ($\mathcal{C}\text{-plane}$) on the interface onto which the cross- and tangent-boundary vectors should be projected. In the general setting, this plane should be continuous along the interface, but satisfying continuity would put additional constraints on the spline degree~\cite{peters_complexity_2010}. We therefore evaluate the $G^1$-continuity at specific landmarks on the surface at which we minimize the $G^1$-continuity jump in an iterative manner. We use the Greville abscissae~\cite{bajaj_curves_1990} for the local coordinates of these landmarks because of their correlation to the curve. Let $\tilde{\Xi}_i$, $i=1,...,n$, and $\tilde{H}_j$, $j=1,...,m$, be the Greville abscissae related to the knot vector $\Xi$ and $H$. Following Equation~\eqref{eq:NURBS}, we define the global Greville points on the surface $\mathbf{S}^{(\mathrm{P})}$ as
\begin{equation}
\label{eq:Greville}
    \mathbf{G}^{(\mathrm{P})}_{i,j} = \mathbf{S}^{(\mathrm{P})}( \tilde{\Xi}_i, \tilde{H}_j ).
\end{equation}
In the remainder, $\mathbf{G}_{i,j}$, will be referred to as the Greville points, which are visualized in Figure~\ref{fig:G1_schematic_b}. For each \emph{row} of Greville points that are located along the interface, denoted by the subscript $i$ in Figure~\ref{fig:G1_schematic_b}, the same $G^1$ correction procedure is performed as explained in the next paragraph. Therefore, in the remainder, we only consider a single row and omit the row index $i$ for notational brevity. Furthermore, Greville points located inside a patch, denoted by subscript $j-1$, are called \emph{inner} Greville points. Greville points located on the interface $\overline{\Gamma}^{(1|2)}_{\mathrm{int}}$ are denoted by, $\mathbf{G}^{(1|2)}$, and are related to the individual patch Greville points as, $\mathbf{G}^{(1|2)}=\mathbf{G}^{(1)}_j=\mathbf{G}^{(2)}_j$, following the definition of $G^0$-continuity in Equation~\eqref{eq:G0}.

\begin{figure*}[!t]
     \centering
     \begin{subfigure}[b]{0.48\textwidth}
         \centering
             \includegraphics[width=\textwidth]{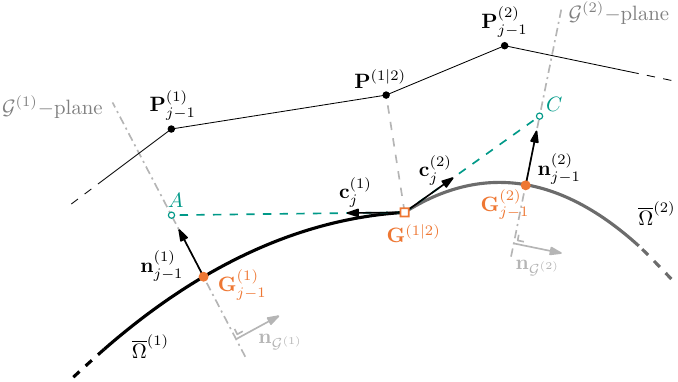}
         \caption{}
         \label{fig:G1_schematic_cross_a} 
     \end{subfigure}
     \hfill
     \begin{subfigure}[b]{0.48\textwidth}
         \centering
              \includegraphics[width=\textwidth]{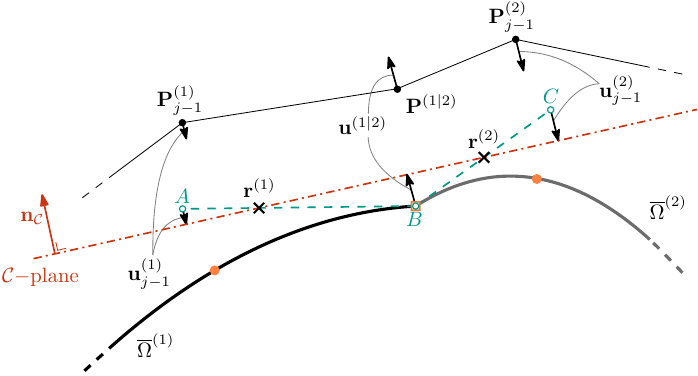}
         \caption{}
         \label{fig:G1_schematic_cross_b}
     \end{subfigure}
        \caption{Cross-sectional representation of two arbitrary NURBS patch surface, $\overline{\Omega}^{(1)}$ and $\overline{\Omega}^{(2)}$, at an interface Greville point, $\mathbf{G}^{(1|2)}$, which can be, but is not a patch vertex in this case. (a) The first step involves the construction of two $\mathcal{G}^{(\mathrm{P})}\text{-planes}$ at the inner Greville points, subscript $j-1$, based on their normal vector, $\mathbf{n}^{(\mathrm{P})}_{j-1}$, and tangent-boundary vector, $\mathbf{t}^{(\mathrm{P})}_{j-1}$ (out of plane component and not shown here). Two intersection points, $A$ and $C$, are calculated based on the cross-boundary vectors at the interface, $\mathbf{c}^{(\mathrm{P})}_j$. (b) The next step involves two line segments, $\overline{BA}$ and $\overline{BC}$, on which two individual points, $\mathbf{r}^{(\mathrm{P})}$, are positioned based on the fitting relevance. The common $\mathcal{C}\text{-plane}$ is constructed based on  Equation~\eqref{eq:Cplane}, of which the difference with the line segments' endpoints serves as the control point displacement correction, $\mathbf{u}^{(\mathrm{P})}$, \emph{cf.} Equation~\eqref{eq:G1displace}.}
        \label{fig:G1_schematic_cross}
\end{figure*}

\begin{figure*}[!t]
     \centering
     \begin{subfigure}[b]{0.33\textwidth}
         \centering
             \includegraphics[width=\textwidth]{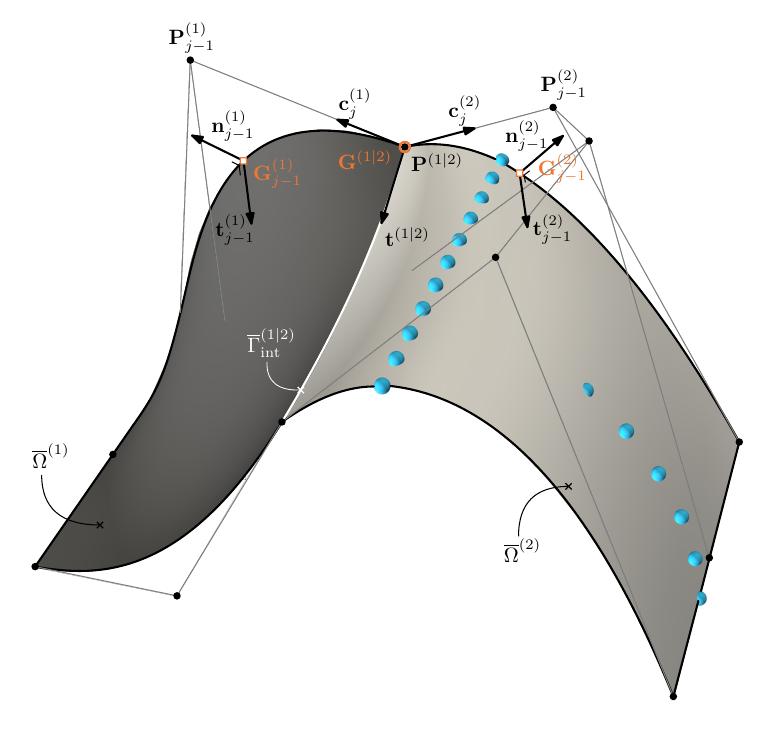}
         \caption{}
         \label{fig:G1_graphical_a} 
     \end{subfigure}
     \hfill
     \begin{subfigure}[b]{0.33\textwidth}
         \centering
              \includegraphics[width=\textwidth]{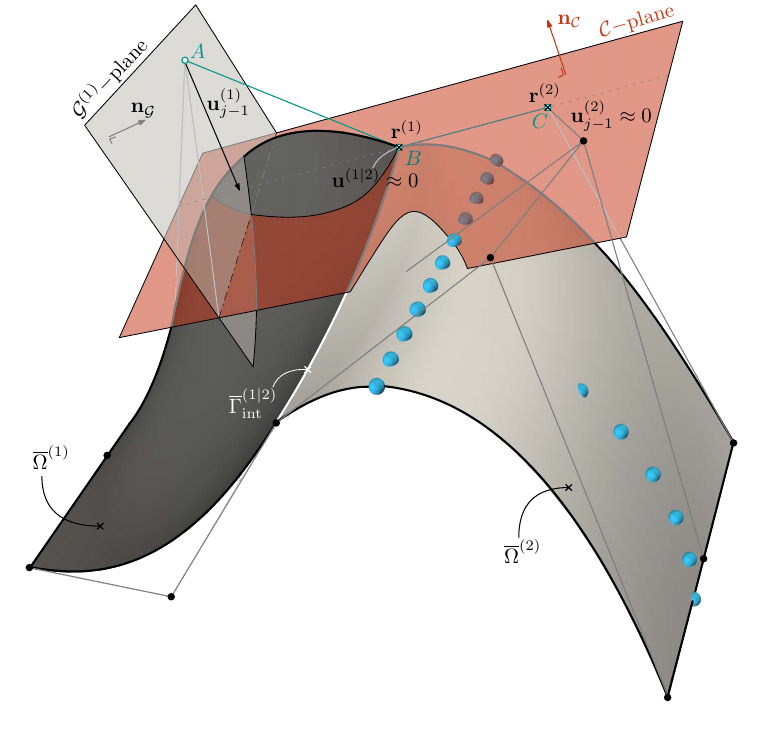}
         \caption{}
         \label{fig:G1_graphical_b}
     \end{subfigure}
     \begin{subfigure}[b]{0.33\textwidth}
         \centering
              \includegraphics[width=\textwidth]{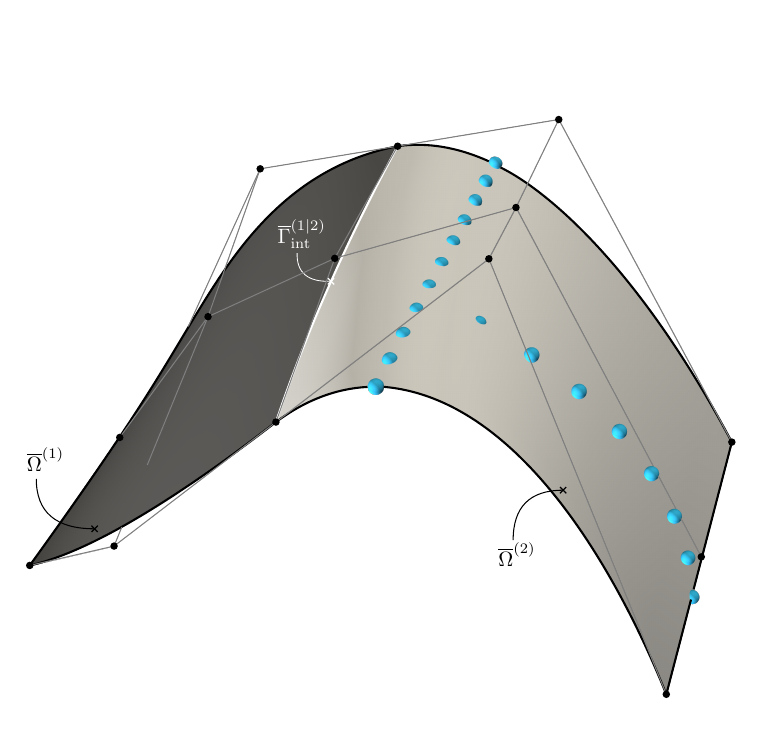}
         \caption{}
         \label{fig:G1_graphical_c}
     \end{subfigure}
        \caption{Graphical overview of the $G^1$-continuity correction steps visualized for patch $\overline{\Omega}^{(1)}$ and  $\overline{\Omega}^{(2)}$ at $\overline{\Gamma}^{(1|2)}_{\mathrm{int}}$ of the example case visualized in Figure~\ref{fig:example_graphical}. (a) The quantities of interest are first defined on the Greville points associated with the $G^1$-relevant control points for a single interface-Greville point. (b) A combination of three planes is consecutively constructed, first the $\mathcal{G}^{(i)}\text{-planes}$, \emph{cf.} Equation~\eqref{eq:Gplane}, and based on those the $\mathcal{C}\text{-plane}$, \emph{cf.} Equation~\eqref{eq:Cplane}. The $\mathcal{C}\text{-plane}$ serves as the common plane onto which the end-points of the line segments, $\overline{BA}$ and $\overline{BC}$, are projected resulting in the control point displacement, $\mathbf{u}^{(\mathrm{P})}$, \emph{cf.} Equation~\eqref{eq:G1displace}. Note that the result given in (c) is the final $G^1$-correction result of the entire multi-patch surface, \emph{i.e.}, the visualized patches have also been influenced by the remaining patches, $\overline{\Omega}^{(3)}$ and  $\overline{\Omega}^{(4)}$, but are not visualized here.}
        \label{fig:G1_graphical}
\end{figure*}

The first step of the correction procedure is to construct two $\mathcal{G}^{(\mathrm{P})}\text{-planes}$ at the inner Greville points of each patch, schematically visualized for the cross-section across the interface in Figure~\ref{fig:G1_schematic_cross_a} and graphically shown in Figure~\ref{fig:G1_graphical_b}. The normal vector of these planes, $\mathbf{n}_{\mathcal{G}^{(\mathrm{P})}}$, is perpendicular to the normal, $\mathbf{n}^{(\mathrm{P})}_{j-1}$, and tangent-boundary vectors, $\mathbf{t}^{(\mathrm{P})}_{j-1}$, visualized in Figure~\ref{fig:G1_graphical_a} and \ref{fig:G1_graphical_b}. The $\mathcal{G}^{(\mathrm{P})}\text{-plane}$ of patch $\mathrm{P}$ is then given as
\begin{equation}\label{eq:Gplane}
    \mathcal{G}^{(\mathrm{P})}\text{-plane}:= \quad \left\{ \left. \mathbf{n}_{\mathcal{G}^{(\mathrm{P})}} \cdot \mathbf{x} - \mathbf{n}_{\mathcal{G}^{(\mathrm{P})}} \cdot \mathbf{G}^{(\mathrm{P})}_{j-1} = 0 \right| \mathbf{x} \in \mathbb{R}^3 \right\},
\end{equation}
where
\begin{equation}
    \mathbf{n}_{\mathcal{G}^{(\mathrm{P})}}  =  \frac{\mathbf{t}^{(\mathrm{P})}_{j-1}\times \mathbf{n}^{(\mathrm{P})}_{j-1}}{\left| \mathbf{t}^{(\mathrm{P})}_{j-1}\times \mathbf{n}^{(\mathrm{P})}_{j-1} \right|}.
\end{equation}
Next, we define two line segments, $\overline{BA}$ and $\overline{BC}$, such that
\begin{subequations}
\label{eq:segments}
\begin{equation}
    \overline{BA} = \left\{ \, \left. \theta ( \mathcal{G}^{(1)} \cap \overrightarrow{BA} ) + \left( 1 - \theta \right) \mathbf{G}^{(1|2)} \ \right|  \ \theta \in [0,1] \, \right\},
\end{equation}
\begin{equation}
    \overline{BC} = \left\{ \, \left. \theta ( \mathcal{G}^{(2)} \cap \overrightarrow{BC} ) + \left( 1 - \theta \right) \mathbf{G}^{(1|2)} \ \right| \ \theta \in [0,1]\, \right\},
\end{equation}
\end{subequations}
with the lines, $\overrightarrow{BA}$ and $\overrightarrow{BC}$, defined as
\begin{subequations}
\begin{equation}
    \overrightarrow{BA} = \left\{ \left. \mathbf{G}^{(1|2)} + \lambda \mathbf{c}^{(1)}_j \, \right| \,  \lambda \in \mathbb{R} \right\},
\end{equation}
\begin{equation}
    \overrightarrow{BC} = \left\{ \left. \mathbf{G}^{(1|2)} + \lambda \mathbf{c}^{(2)}_j \, \right| \,  \lambda \in \mathbb{R} \right\}.
\end{equation}
\end{subequations}
The continuity plane, $\mathcal{C}\text{-plane}$, is then defined as
\begin{equation}
\label{eq:Cplane}
    \mathcal{C}\text{-plane}: \quad \{ \left. \mathbf{n}_{\mathcal{C}} \cdot \mathbf{x} - \mathbf{n}_{\mathcal{C}} \cdot \mathbf{r}^{(\mathrm{1})} = 0 \right| \mathbf{x} \in \mathbb{R}^3 \},
\end{equation}
where
\begin{equation}
\label{eq:nc}
    \mathbf{n}_{\mathcal{C}} =  \frac{\mathbf{t}^{(1|2)}\times (\mathbf{r}^{(1)} - \mathbf{r}^{(2)}) }{\left| \mathbf{t}^{(1|2)}\times (\mathbf{r}^{(1)} - \mathbf{r}^{(2)}) \right|}.
\end{equation}
The points $\mathbf{r}^{(1)}$ and $\mathbf{r}^{(2)}$ are located on the line segments, $\mathbf{r}^{(1)} \subset \overline{BA}$ and $\mathbf{r}^{(2)} \subset \overline{BC}$, and their location on these segments depends on the data fitting information as defined in Equation~\eqref{eq:Mij}, as follows
\begin{subequations}
\begin{equation}
    \mathbf{r}^{(1)} = \overline{BA}_{\theta=\theta^{(1)}}, \quad \text{with} \quad \theta^{(1)} = \frac{M^{(1)}_{{j-1}}}{M^{(1)}_{{j-1}} + M^{(1)}_{{j}} },
\end{equation}
\begin{equation}
    \mathbf{r}^{(2)} = \overline{BC}_{\theta=\theta^{(2)}}, \quad \text{with} \quad \theta^{(2)} = \frac{M^{(2)}_{{j-1}}}{M^{(2)}_{{j-1}} + M^{(2)}_{{j}} }.
\end{equation}
\end{subequations}
Considering the graphical representation in Figure~\ref{fig:G1_graphical}, it is apparent that $M^{(1)}_{i,j}=0_{i,j}$, since no data points are associated to patch $\overline{\Omega}^{(1)}$. We then require that the control points that have the smallest influence on the data fit (low values of $M_{i,j}$) should be displaced more than control points that have a higher influence on the data fit. This ensures that the surface is minimally changed in areas that are important to the data fit during the continuity correction. We normalize the $M^{(\mathrm{P})}_{i,j}$ weights for each line segment, $\overline{BA}$ and $\overline{BC}$. We standardize the weights $M^{(\mathrm{P})}_{i,j}$ individually for each line segment, namely $\overline{BA}$ and $\overline{BC}$. This normalization is performed such that, in Equation~\eqref{eq:segments}, a value of $\theta=0$ signifies the highest significance for the interface (point $B$), while $\theta=1$ designates the inner control point (point $A$ or $C$). For the special case that both segments yield $\theta=0$, resulting in $\mathbf{r}^{(1)}=\mathbf{r}^{(2)}$, we equate Equation~\eqref{eq:nc} to the average of the normal vectors, $0.5(\mathbf{n}^{(1)}_j+\mathbf{n}^{(2)}_{j})$. In the general setting, visualized in Figure~\ref{fig:G1_schematic_cross} for two arbitrary NURBS surfaces, where both surfaces have equal importance for the data fitting, the points $\mathbf{r}^{(1)}$ and $\mathbf{r}^{(1)}$ are typically positioned close to the center of the line segments, Figure~\ref{fig:G1_schematic_cross_b}.

The final step in satisfying $G^1$-continuity is to displace the endpoints of the line segments $\overline{BA}$ and $\overline{BC}$ onto the constructed $\mathcal{C}\text{-plane}$ along a specified direction
\begin{subequations}
\label{eq:G1displace}
    \begin{align}
        \mathbf{u}^{(1)}_{j-1} &= \left( \left\{ \left. \overline{BA}_{\theta\text{=1}} + \lambda ( \mathbf{n}_\mathcal{C} + \mathbf{n}^{(1)}_{j-1} )  \, \right| \,  \lambda \in \mathbb{R} \right\} \cap \mathcal{C}\right) - \overline{BA}_{\theta\text{=1}}, \\
        \mathbf{u}^{(1|2)} &=\left( \left\{ \left. \overline{BA}_{\theta\text{=0}} + \lambda \mathbf{n}_\mathcal{C} \, \right| \,  \lambda \in \mathbb{R} \right\} \cap \mathcal{C}\right) - \overline{BA}_{\theta\text{=0}}, \\
        \mathbf{u}^{(2)}_{j-1} &= \left( \left\{ \left. \overline{BC}_{\theta\text{=1}} + \lambda ( \mathbf{n}_\mathcal{C} + \mathbf{n}^{(2)}_{j-1} )  \, \right| \,  \lambda \in \mathbb{R} \right\} \cap \mathcal{C}\right) - \overline{BC}_{\theta\text{=1}}, 
    \end{align}
\end{subequations}
where subscript $\theta\text{=0}$ and $\theta\text{=1}$ are the parametric values corresponding to the endpoints of the line segments, \emph{cf.} Equation~\eqref{eq:segments}. By applying the displacements of Equation~\eqref{eq:G1displace} to the corresponding control points, $\mathbf{P}^{(1)}_{j-1}$, $\mathbf{P}^{(1|2)}$, and $\mathbf{P}^{(2)}_{j-1}$, the surfaces are displaced and $G^1$-continuity is approximated. Equation~\eqref{eq:G1displace} is best explained schematically by Figure~\ref{fig:G1_schematic_cross_b} and graphically by Figure~\ref{fig:G1_graphical_b}. Here, the common $\mathcal{C}\text{-plane}$ is visualized onto which the endpoints of $\overline{BA}$ and $\overline{BC}$ are displaced. The resulting displacement of the control points yields an updated surface that approximates the $G^1$-continuity across the patch interface, visualized in Figure~\ref{fig:G1_graphical_c}. However, the resulting multi-patch domain does not guarantee a continuous $G^1$-continuity across all interfaces after a single $G^1$ correction update. Several iterations may be required before reaching a converged $G^1$ approximation due to the spline bases' (degree, knot refinement) inability to satisfy exact continuous $G^1$-continuity at the interfaces. Nonetheless, the continuity is further improved upon knot insertion or spline degree elevation (the latter will not be considered in this manuscript). The proposed correction scheme can therefore be considered as an iterative and convergent scheme to achieve $G^1$-continuity. Furthermore, the correction scheme is not unconditionally stable, often requiring constraints (Equation~\eqref{eq:Pupdate}) and relaxation on the computed displacement \emph{cf.} Equation~\eqref{eq:G1displace}. A qualitative analysis of the continuity correction in combination with the data fit is examined in the next section.

\subsection{Analysis of the fitting algorithm}\label{sec:example}
The fitting algorithm outlined in Algorithm~\ref{alg:fitting} is applied to the example case visualized in Figure~\ref{fig:example_graphical}. We apply constraints to the computed displacement during each fitting iteration to ensure stable convergence toward a physically valid final shape. Constraints ensure that the final multi-patch fit is regular, \emph{i.e.}, no singular control points~\cite{hughes_chapter_2021}, and that the continuity correction behaves stable for any arbitrary problem setting. To ensure this, we constrain all patch edges, \emph{i.e.}, the external boundaries and interfaces of the multi-patch domain in Figure~\ref{fig:example_graphical}, using the constraint operator, $\mathscr{C}(\cdot)$ \emph{cf.} Equation~\eqref{eq:Pupdate}, such that
\begin{equation}
    \label{eq:constraints}
     \mathscr{C}\left\{\begin{matrix}
        \mathbf{u} &= \mathbf{0} \quad \text{at} \quad & \overline{\Gamma}_{\mathrm{v}}, \\
        \mathbf{u}\cdot \mathbf{e}_z &= 0 \quad \text{at} \quad & \{ \overline{\Gamma}_{\mathrm{h}}, \, \overline{\Gamma}^{(1|4)}_{\mathrm{int}}, \, \overline{\Gamma}^{(2|3)}_{\mathrm{int}} \},\\
        \mathbf{u}\cdot \mathbf{e}_y &= 0 \quad \text{at} \quad & \{ \overline{\Gamma}^{(1|2)}_{\mathrm{int}}, \, \overline{\Gamma}^{(3|4)}_{\mathrm{int}} \}.
    \end{matrix}\right.
\end{equation}
These constraints ensure that the fitted geometry maintains its initial rectangular shape as good as possible. The inner control points of each patch are not constrained, which, for this test case, does not impede the stability of the fitting algorithm. In the analyses of the left ventricle in Section~\ref{sec:Sec5} we adopt a slightly different set of constraints due to the problem setting discussed there, which emphasizes the problem-specific nature of the constraints.

\begin{figure*}[!t]
     \centering
     \begin{subfigure}[b]{0.48\textwidth}
         \centering
             \includegraphics[width=\textwidth]{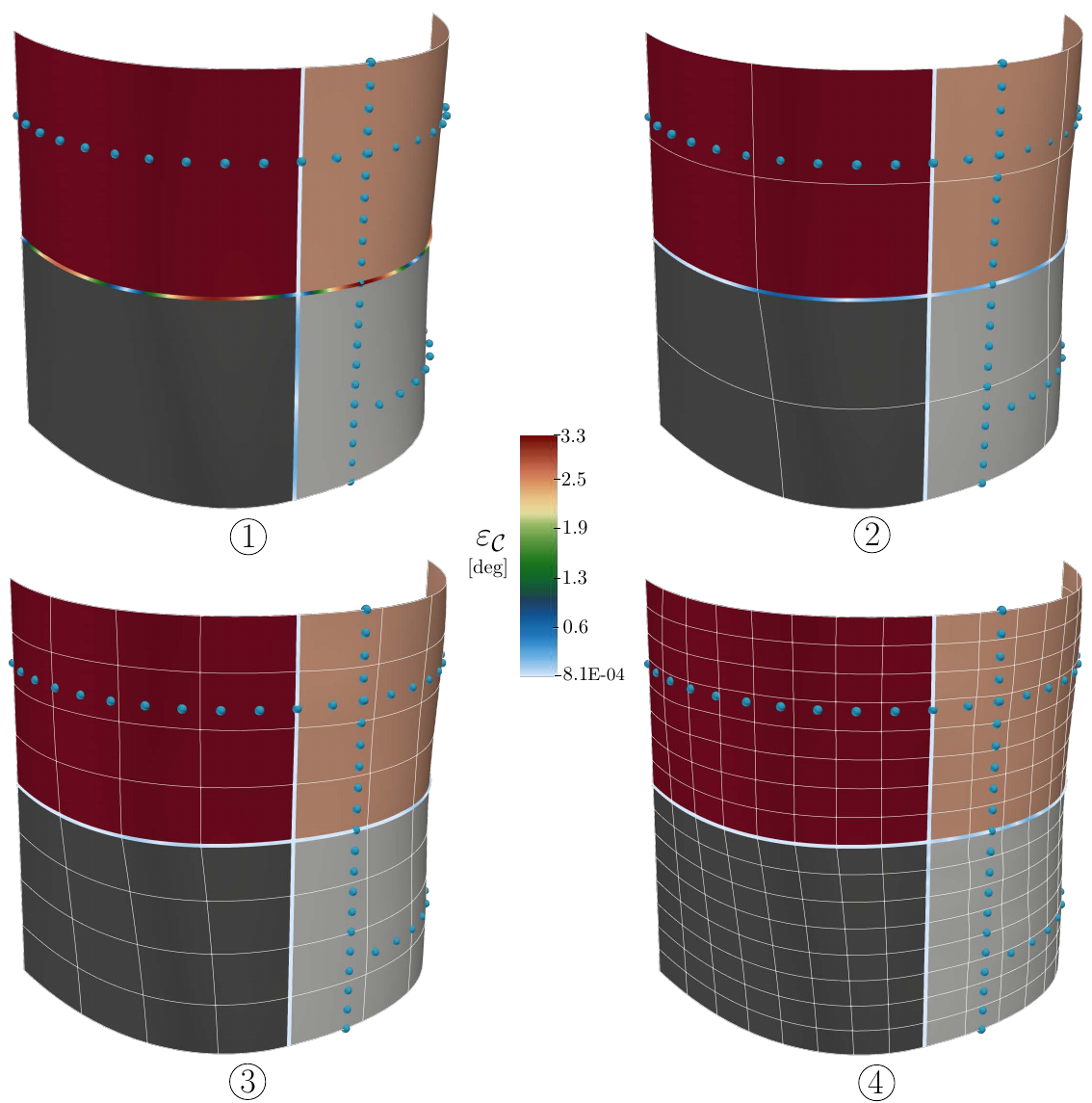}
         \caption{}
         \label{fig:Example_results_overview_a} 
     \end{subfigure}
     \hfill
     \begin{subfigure}[b]{0.48\textwidth}
         \centering
              \includegraphics[width=\textwidth]{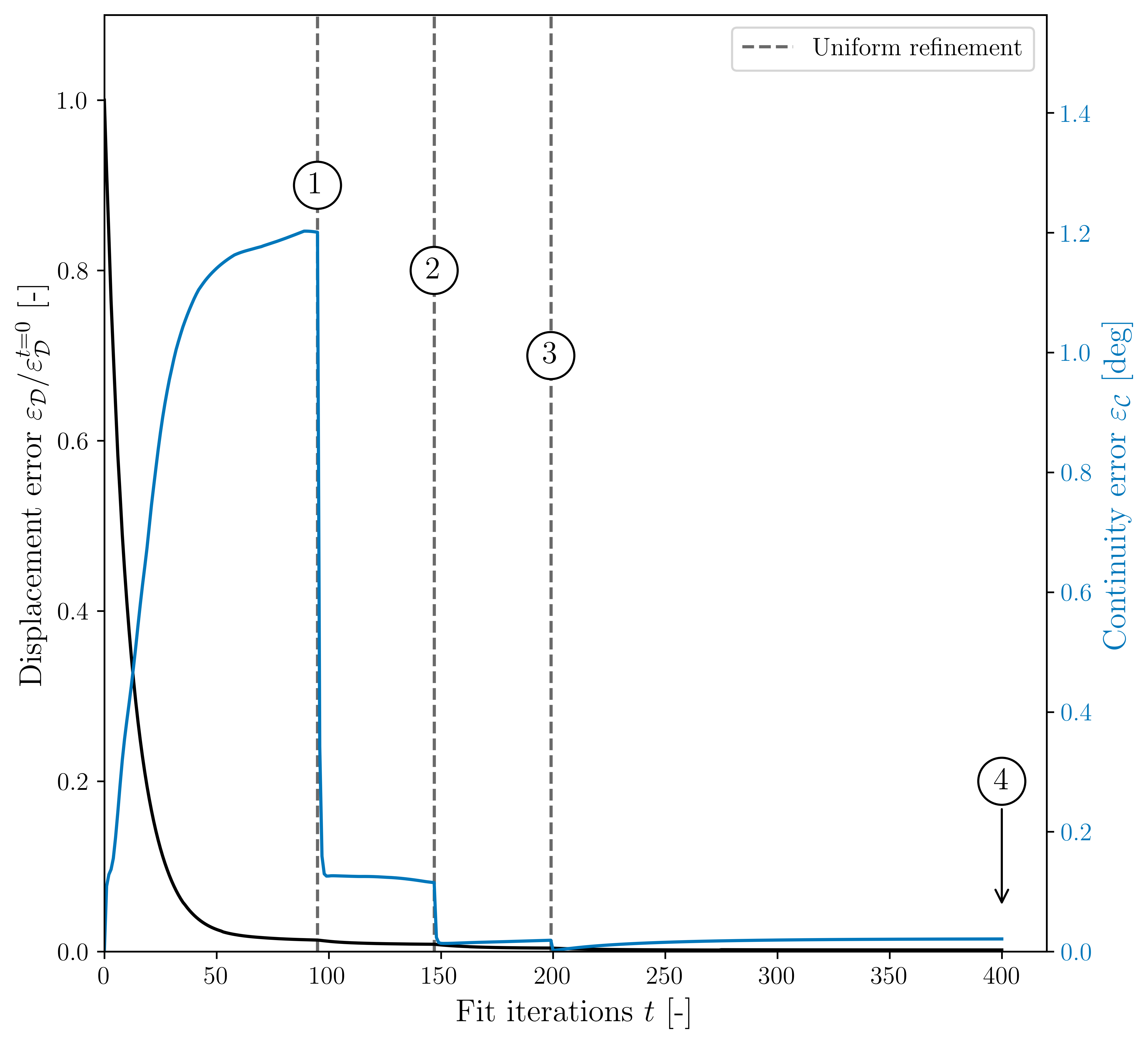}
         \caption{}
         \label{fig:Example_results_overview_b}
     \end{subfigure}
        \caption{(a) Overview of the different multi-patch configurations at different fitting iterations with different knot vectors or refinement levels. The color bar represents the angular jump, continuity error $\varepsilon_{\mathcal{C}}$, at the interfaces. (b) Trace of the displacement error $\varepsilon_{\mathcal{D}}$ and continuity error $\varepsilon_{\mathcal{C}}$ for a fixed number of iterations. The vertical dashed lines indicate the instance at which a uniform knot refinement is applied to the multi-patch surface, with the encircled numbers, \textcircled{1} at $t=80$, \textcircled{2} at $t=140$, \textcircled{3} at $t=192$, \textcircled{4} at $t=400$, corresponding to the configurations in (a). Uniform knot insertion ensures both a decrease in displacement and continuity error.}
        \label{fig:Example_results_overview}
\end{figure*}

The algorithm is run for $t_{\rm{max}}=400$ \emph{fit} iterations, during which a total of $3$ uniform knot refinements are allowed once the gradient of the displacement error is below a specified tolerance, $\Delta\varepsilon_{\mathcal{D}}=(\varepsilon^{\mathrm{t}}_{\mathcal{D}} - \varepsilon^{\mathrm{t+1}}_{\mathcal{D}})/\varepsilon^{\mathrm{t}}_{\mathcal{D}} < 10^{-2}$. The point-to-surface projection, Line~\ref{alg:pointproject} in Algorithm~\ref{alg:fitting}, is calculated each fit iteration while the continuity-correction step is iterated $5$ times during a single fit iteration. The computed displacement vectors following the data fit, Line~\ref{alg:cpsdisplace} in Algorithm~\ref{alg:fitting}, are relaxed by $\alpha_{\mathrm{relax}}=0.1$. The results are presented in Figure~\ref{fig:Example_results_overview}, where we aim to minimize the displacement error, $\varepsilon_{\mathcal{D}}$, and the $G^1$-continuity error, $\varepsilon_{\mathcal{C}}$, according to Equation~\eqref{eq:argmin}.

In Figure~\ref{fig:Example_results_overview_b} we plot the normalized displacement error with respect to the first iteration and the absolute average continuity error. The trace of the displacement error shows an exponential decay before reaching a plateau for each uniform refinement level. The plateau is caused by the inability of the initial non-rational B-spline multi-patch surface to precisely describe the cylindrical shape that the data points are based on~\cite{piegl_nurbs_1997}. Only after uniformly inserting additional knots (Line~\ref{alg:refinement}), the initial surface can achieve $<0.1\%$ displacement error. The continuity error of the multi-patch surface shows a different behavior, varying between refinement levels. We attribute this difference to the addition of knots: once the surface is uniformly refined it can locally adjust the shape of the geometry according to the data points. This local change in geometry introduces geometric gradients that impede the continuity correction as it becomes more difficult to satisfy $G^1$-continuity globally. Additional knot refinements can relax the conditions needed to satisfy the global continuity conditions using our continuity correction scheme, see Section~\ref{sec:g1cont}. Both error quantities show a converged solution upon reaching $400$ iterations, at which the displacement error has a value of $1.91\cdot10^{-3}$ and the continuity error $0.02$ [deg]. The latter is expected to be slightly above $0$ [deg] since the presented correction scheme in Section~\ref{sec:g1cont} requires further refinements to make the error vanish.

\begin{figure*}[!t]
     \centering
     \begin{subfigure}[b]{0.35\textwidth}
         \centering
             \includegraphics[width=\textwidth]{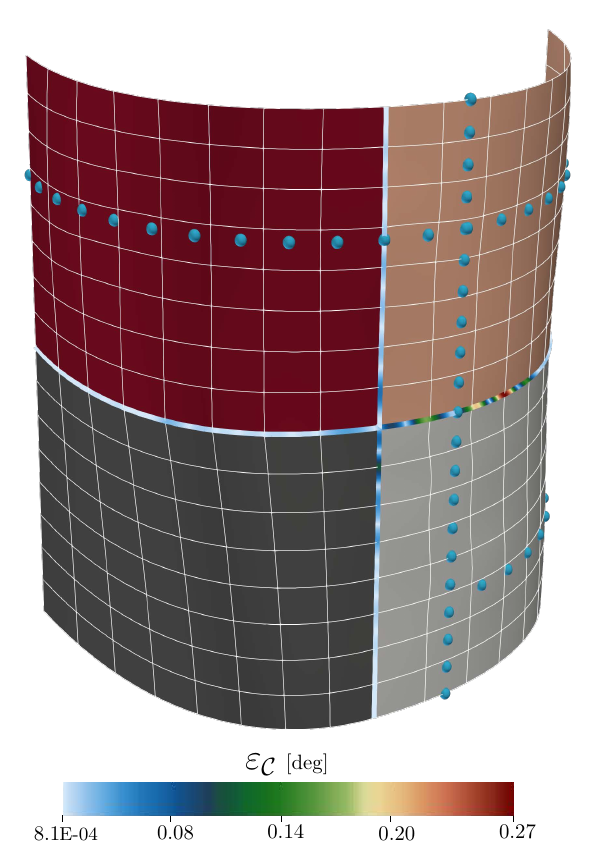}
         \caption{}
         \label{fig:Example_results_comp_a} 
     \end{subfigure}
     \hfill
     \begin{subfigure}[b]{0.35\textwidth}
         \centering
              \includegraphics[width=\textwidth]{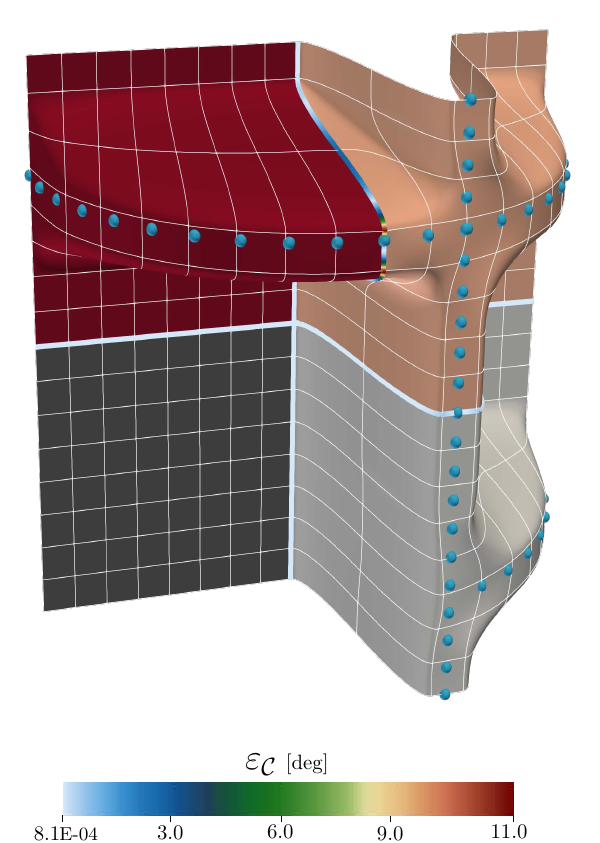}
         \caption{}
         \label{fig:Example_results_comp_b}
     \end{subfigure}
        \caption{Comparison of two fitted multi-patch surfaces after $400$ fit iterations with different initial knot vectors. (a) Multi-patch surface with initial $0\times0$ inner knots per patch and a final $7\times7$ inner knots per patch after uniform knot insertion during the fitting process. (b) Multi-patch surface with an initial and final $7\times7$ inner knots per patch, without any uniform knot insertion during the fitting process.}
        \label{fig:Example_results_comp}
\end{figure*}

The results presented in Figure~\ref{fig:Example_results_overview} are based on an initial template with quadratic B-splines and no internal knots, \emph{i.e.}, a single element per patch, as visualized in Figure~\ref{fig:example_graphical_a}. The initial knot vector of the multi-patch surface influences the final shape of the fitted geometry. To illustrate this, we use the knot vector from the fourth (final) surface in Figure~\ref{fig:Example_results_overview_a} as the starting knot vector for the initial template before running the fitting algorithm. The results are visualized in Figure~\ref{fig:Example_results_comp}, where a distinct difference is observed. Based on Equation~\eqref{eq:displace} only the control points that have support on the projected data points are displaced, and the remaining control points are unchanged. As a result, only a small portion of the multi-patch domain is displaced, while the rest adheres to the initial template shape. This comparison shows the importance of the initial number of control points of the chosen template. A large number of control points leads to a minimal change in the geometry in areas where there is no available data to be fitted, while a small number of control points ensures a higher degree of change. In the case of the left ventricle, we minimize the initial number of control points for the template. This can only be done if the minimal number of control points can describe the desired template shape, \emph{i.e.}, an ellipsoid in this case, with sufficient accuracy. This emphasizes the strong benefit of splines to describe a geometry with a relatively low number of degrees of freedom (control points) over conventional polygonal techniques.
\newpage

\section{The isogeometric cardiac model}\label{sec:Sec4}
The echocardiogram-based electromechanical cardiac model considered in this work builds on the isogeometric analysis (IGA) approach proposed by Willems \emph{et al.} \cite{willems_isogeometric_2023}. For self-containedness, in Section \ref{sec:cardiacmodel} we first briefly review the model of Ref.~\cite{willems_isogeometric_2023}. In Section~\ref{sec:gpa} we then discuss the model extensions required to accommodate the pressure-loaded image configuration.

\subsection{The isogeometric cardiac model}
\label{sec:cardiacmodel}
In Section~\ref{sec:governingequations} we first introduce the governing equations for the cardiac model, after which the solution techniques -- in particular the spatial discretization using isogeometric analysis -- are discussed in Section~\ref{sec:solutiontechniques}. We refer the reader to Willems \emph{et al.} \cite{willems_isogeometric_2023} for a detailed exposition of the model and the required solution techniques.

\subsubsection{Governing equations}
\label{sec:governingequations}
Our cardiac model consists of two main components, \emph{viz.} a three-dimensional continuum model mimicking the electromechanical behavior of the left ventricle and a zero-dimensional lumped parameter model describing the circulatory system. These two model components are coupled through the left ventricle cavity volume and pressure. Below we discuss the most important aspects of the cardiac model, a comprehensive overview of which is presented in Box~\ref{box:overview}. 

\begin{figure}
    \centering
    \includegraphics[width=\textwidth]{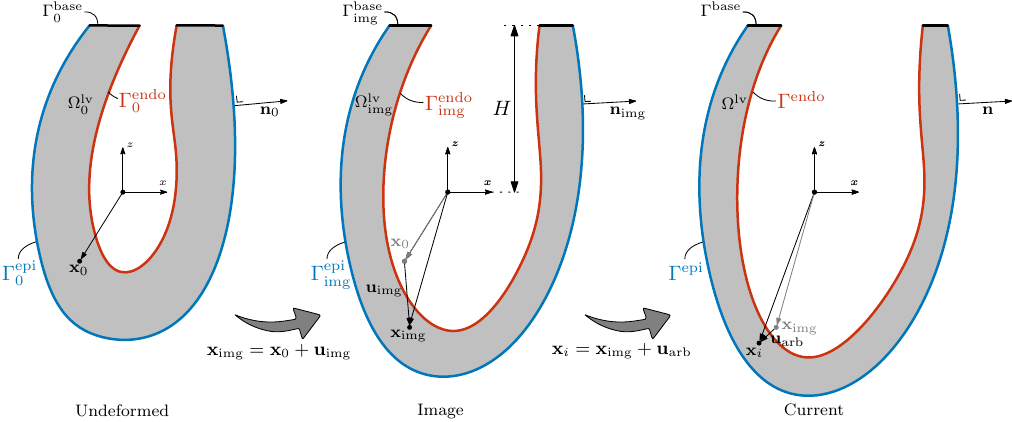}
    \caption{Schematic overview of the computational domain in (a) the undeformed configuration, $\Omega^{\mathrm{lv}}_0$,(b) the image configuration, $\Omega^{\mathrm{lv}}_{\rm img}$, and (c) the current configuration $\Omega^{\mathrm{lv}}$, corresponding to an arbitrary deformation.}
    \label{fig:configurations}
\end{figure}

To accommodate the large deformations and deformation gradients that are physically essential, a total Lagrangian formulation is used for the three-dimensional electromechanical model. To define the total Lagrange formulation, a distinction is made between the undeformed configuration, $\Omega^{\mathrm{lv}}_0$, and the current configuration, $\Omega^{\mathrm{lv}}$ (Figure~\ref{fig:configurations}). The position of a material point in the undeformed configuration is denoted by $x_{0_i}$ and that of a material point in the deformed configuration by $x_i$. The total displacement of a material point is denoted by $u_i= x_i-x_{0_i}$. In the discussion in this section, we assume the undeformed configuration to be known, which allows us to present the cardiac model without the need to refer to the intermediate image configuration, $\Omega^{\mathrm{lv}}_{\rm img}$ (Figure~\ref{fig:configurations}). The incorporation of the image configuration will be discussed in Section~\ref{sec:gpa}.

The mechanical equilibrium equations are presented in \ref{eq:B11a}--\ref{eq:B11g}. Ignoring inertia effects and body forces, the mechanical equilibrium is described in the current configuration, $\Omega^{\mathrm{lv}}$, by means of the Cauchy stress tensor $\sigma_{ij}$ (\ref{eq:B11a}). On the endocardium, $\Gamma^{\rm endo}$, a pressure load $p^{\rm lv}$ is applied (\ref{eq:B11b}). On the basal plane, $\Gamma^{\rm base}$, the deformation of the ventricle is constrained in the normal direction (\ref{eq:B11c}). The two remaining rigid body translations are constrained by prescribing the average of the in-plane components of the displacement field in the basal plane to be zero (\ref{eq:B11d}--\ref{eq:B11e}) and the rigid body rotation is constrained by prescribing the average curl of the displacement field to be zero (\ref{eq:B11f}). These rigid body constraints are defined in the current configuration (see \ref{app:rigidbodyconstraints}). For the initial condition of the mechanical equilibrium equations a zero displacement field is considered (\ref{eq:B11g}).

In the total Lagrangian formulation it is most convenient to express the constitutive behavior in terms of the Green-Lagrange strain tensor $E_{ij}$ (\ref{eq:B13}) -- which is directly related to the deformation gradient $F_{ij}$ (\ref{eq:B12}) -- and the second Piola-Kirchhoff stress tensor $S_{ij}$. The Cauchy stress $\sigma_{ij}$ is related to the second Piola-Kirchhoff stress by \ref{eq:B15a}. The second Piola-Kirchhoff stress in our model consists of two components, a passive component, $S^{\rm pas}_{ij}$, corresponding to the hyperelastic behavior of the myocardium, and an active component, $S^{\rm act}_{ij}$, corresponding to the force generated by the sarcomeres.

The passive component of the stress is expressed in terms of the invariants of the Green-Lagrange strain tensor $I_1$, $I_2$ and $I_3$ (\ref{eq:B14a}--\ref{eq:B14c}) and the quasi-invariant $I_4$ (\ref{eq:B14d}) representing the Green-Lagrange fiber strain. The fiber direction in the image configuration is determined using the model proposed in Ref.~\cite{rossi_thermodynamically_2014} and mapped toward the unloaded configuration as $\mathrm{e}_i^{\mathrm{f0}}$. The passive stress component is described by the Fung-type constitutive model proposed by Bovendeerd \emph{et al.}~\cite{bovendeerd_determinants_2009} (\ref{eq:B15b}). This model assumes a transverse isotropic material that is parametrized by the elastic modulus of the myocardium, $C$, and the bulk modulus, $\kappa$. The quadratic strain form $Q$ (\ref{eq:B15d}) depends on the strain invariants $I_1$, $I_2$ and $I_4$, with $a_1$, $a_2$ and $a_3$ being material constants. 

The active component of the stress is described through the phenomenological model of Refs.~\cite{bovendeerd_determinants_2009,kerckhoffs_homogeneity_2003}, which assumes the contraction of the myocytes to be initiated simultaneously at time $T^{\rm act}$. In this model, a sarcomere of length $l^s$ is regarded as a serial connection of a contractile element with length $l^c$, and an elastic element with length $l^s-l^c$. The stretch of the sarcomere is related to the Green-Lagrange strain in the fiber direction (\ref{eq:B15e}). The length of the contractile element is governed by a contractile dynamics relation (\ref{eq:B16a}), where $v^0$ represents the unloaded shortening velocity. Before activation, the contractile length is equal to the length of the sarcomere (\ref{eq:B16b}). The stiffness of the elastic element of the sarcomere model is denoted by $E^a$, and the stress generated by this element acts in the direction of the fibers in the undeformed configuration (\ref{eq:B15c}). The rise and decay of the myocyte contraction are governed by the twitch function, $f^{\rm twitch}$, which depends on the time since activation $t^a = t - T^{\rm act}$. The influence of the contractile length on the active stress component is prescribed by the isometric contraction function, $f^{\rm iso}$. For details regarding $f^{\rm twitch}$ and $f^{\rm iso}$, including the used parameters, the reader is referred to Ref.~\cite{willems_isogeometric_2023}. 

\begin{figure}
    \centering
    \includegraphics[width = 0.6\linewidth]{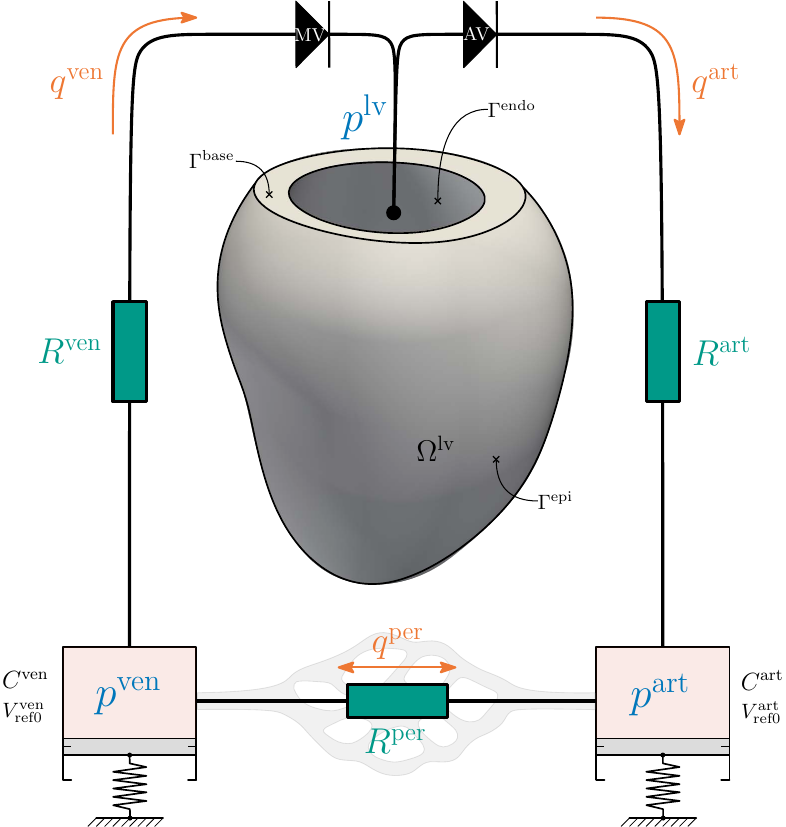}
    \caption{Schematic overview of the 0D circulatory system coupled to the 3D LV domain, $\Omega^{\mathrm{lv}}$. The closed-loop lumped parameter model represents a simplification of the human body's circulatory system. The circulatory system is modeled as two Windkessel compartments in series, characterized by a compliance, $C^i$, and a volume at zero reference pressure, $V^i_{\rm{ref0}}$, with $i\in\{ \mathrm{art}, \mathrm{ven} \}$ (art=arterial and ven=venous). The compartments and ventricles are linked via resistances, $R^i$ (green), through which arterial ($i=\text{art}$), venous ($i=\text{ven}$), and peripheral ($i=\text{per}$) volume flow occurs, $q^i$ (orange), caused by the compartment pressure, $p^i$ (blue). The two heart valves: the mitral valve (MV) and aortic valve (AV), are idealized to open and close instantaneously and prevent back-flow. The ventricle and compartment pressures, illustrated in blue, are the quantities that are solved in the circulatory model.}
    \label{fig:circulatoryScheme}
\end{figure}

The zero-dimensional model of the circulatory system is illustrated in Figure~\ref{fig:circulatoryScheme}. In essence, the circulatory model provides a (differential) relation between the pressure in the left ventricle, $p^{\rm lv}$, the volume of the left ventricle cavity, $V^{\rm lv}$, and the arterial and venous pressure $\tilde{p}=\{ p^{\rm art}, p^{\rm ven}\}$. The circulatory model is parametrized by the compliances, $C^{\rm art}$ and $C^{\rm ven}$, the reference volumes, $V_{\rm ref0}^{\rm art}$ and $V_{\rm ref0}^{\rm ven}$, and the resistances $R^{\rm art}$, $R^{\rm ven}$ and $R^{\rm per}$. The volume of the left ventricle depends on the solution to the mechanical model (\ref{eq:B17b}) and the mechanical model, in turn, depends on the pressure in the ventricle through the boundary condition (\ref{eq:B11b}). This makes the system coupled in two directions.

\begin{Box}
\begin{tcolorbox}[colback=white, boxrule=.3mm, boxsep=1pt,left=4pt,right=2pt,top=4pt,bottom=0pt]
\begin{subequations}
Equilibrium equation:
\begin{align}
  \frac{\partial}{\partial x_i} \sigma^{\phantom{}}_{ij}( l^{\rm{c}},u_k, t )  &= 0_j & \text{in } &\Omega^{\mathrm{lv}} & \times & (0, T] \tag{Box1.1a} \label{eq:B11a}\\
  n_i \sigma^{\phantom{}}_{ij} &= -p^{\mathrm{lv}}\ n_j & \text{on } & \Gamma^{\mathrm{endo}} & \times & (0, T] \tag{Box1.1b} \label{eq:B11b}\\
  n_i u_i  &= 0  & \text{on } &\Gamma^{\mathrm{base}} & & \tag{Box1.1c} \label{eq:B11c}\\
   \int u_1 \ \mathrm{d}\Gamma^{\mathrm{base}} &= 0 & \text{on } &\Gamma^{\mathrm{base}} & & \tag{Box1.1d} \label{eq:B11d}\\
   \int u_2 \ \mathrm{d}\Gamma^{\mathrm{base}} &= 0 & \text{on } &\Gamma^{\mathrm{base}} & & \tag{Box1.1e} \label{eq:B11e}\\
   \int \left( \frac{\partial u_2}{\partial x_1} - \frac{\partial u_1}{\partial x_2} \right) \ \mathrm{d}\Gamma^{\mathrm{base}} &= 0 & \text{on } &\Gamma^{\mathrm{base}} & & \tag{Box1.1f} \label{eq:B11f}\\
   u_i & = 0_i & \text{in } &\Omega^{\mathrm{lv}} & \times & 0 \tag{Box1.1g} \label{eq:B11g}\\  
\intertext{Deformation gradient:}
  F^{\phantom{}}_{ij} &= \frac{\partial x_{i}}{\partial {X}_{j}} = \delta_{ij} + \frac{\partial u_{i}}{\partial {X}_{j}} & \text{in } &\Omega^{\mathrm{lv}} & & \tag{Box1.2} \label{eq:B12}\\
\intertext{Green-Lagrange strain tensor:}
  E^{\phantom{}}_{ij} &= \frac{1}{2} \left( F^{\phantom{}}_{ki} F^{\phantom{}}_{kj} - \delta^{\phantom{}}_{ij} \right) & \text{in } &\Omega^{\mathrm{lv}} & & \tag{Box1.3} \label{eq:B13}\\
\intertext{Invariants of the strain tensor:}
  I_1 &= E^{\phantom{}}_{ii} & \text{in } &\Omega^{\mathrm{lv}} & & \tag{Box1.4a} \label{eq:B14a}\\
  I_2 & = \frac{1}{2} \left( E^{\phantom{}}_{ii} E^{\phantom{}}_{jj} -  E^{\phantom{}}_{ji} E^{\phantom{}}_{ij}\right) & \text{in } &\Omega^{\mathrm{lv}} & & \tag{Box1.4b} \label{eq:B14b}\\
  I_3 &= J = \text{det}\left( \mathbf{F} \right) & \text{in } &\Omega^{\mathrm{lv}} & & \tag{Box1.4c} \label{eq:B14c}\\
  I_4 &= {\mathrm{e}}^{\mathrm{f0}}_i E^{\phantom{}}_{ij}  {\mathrm{e}}^{\mathrm{f0}}_j & \text{in } &\Omega^{\mathrm{lv}} & & \tag{Box1.4d} \label{eq:B14d}\\
\intertext{Constitutive relation:}
  \sigma^{\phantom{}}_{ij} &= \frac{1}{J} F^{\phantom{}}_{ik} \left( S^{\mathrm{pas}}_{kl} + S^{\mathrm{act}}_{kl} \right) F^{\phantom{}}_{jl} & \text{in } &\Omega^{\mathrm{lv}}  & \times & (0, T] \tag{Box1.5a} \label{eq:B15a}\\
  S^{\mathrm{pas}}_{ij} &= \frac{\partial }{\partial E^{\phantom{}}_{ij}} \left( C [ \mathrm{exp}\left( Q \right) - 1    ] + \frac{1}{2} \kappa [ J^2 - 1    ]^2 \right) & \text{in } &\Omega^{\mathrm{lv}}_0 & & \tag{Box1.5b} \label{eq:B15b}\\
  S^{\mathrm{act}}_{ij} &= \frac{l^{\mathrm{s0}}}{l^{\mathrm{s}}}f^{\mathrm{iso}}\left( l^{\mathrm{c}} \right) \ f^{\mathrm{twitch}}\left(t^{\mathrm{a}}, l^{\mathrm{s}}  \right) \ E^{\mathrm{a}} \left( l^{\mathrm{s}} - l^{\mathrm{c}} \right) \ {\mathrm{e}}^{\mathrm{f0}}_i {\mathrm{e}}^{\mathrm{f0}}_j & \text{in } &\Omega^{\mathrm{lv}}_0  & \times & (0, T]\tag{Box1.5c} \label{eq:B15c}\\
  Q &=  a_1 I_1^2 - a_2 I_2 + a_3  I_4^2  & \text{in } &\Omega^{\mathrm{lv}}_0 & & \tag{Box1.5d} \label{eq:B15d} \\
  l^{\mathrm{s}} &= l^{\mathrm{s0}} \sqrt{2 I_4 + 1}  & \text{in } &\Omega^{\mathrm{lv}}_0 & & \tag{Box1.5e} \label{eq:B15e} \\
\intertext{Contractile dynamics:}
  \frac{d l^{\mathrm{c}}}{dt} &= \left[ E^{\mathrm{a}} \left( l^{\mathrm{s}} - l^{\mathrm{c}} \right) - 1\right] v^{\mathrm{0}} & \text{in } & \Omega^{\mathrm{lv}}_0 &\times & (T^{\mathrm{act}}, T] \tag{Box1.6a} \label{eq:B16a}\\
  l^{\mathrm{c}} &= l^s & \text{in } & \Omega^{\mathrm{lv}}_0 &\times & (0, T^{\mathrm{act}}] \tag{Box1.6b} \label{eq:B16b}
\intertext{Circulatory model:}
  \mathcal{M}_{\mathrm{circ}} \left( p^{\mathrm{lv}}, V^{\mathrm{lv}}_{\mathrm{cav}} , \tilde{p}\right) &= 0 & \text{in } & & & (0, T] \tag{Box1.7a} \label{eq:B17a}\\
  V^{\mathrm{lv}}_{\mathrm{cav}} &= \int \frac{1}{3} \left( x_{i} - H \right) n_i \ \text{d}\Gamma^{\rm{endo}} & \text{on } & \Gamma^{\mathrm{endo}} & & \tag{Box1.7b} \label{eq:B17b}
\end{align}
\end{subequations}
\end{tcolorbox}
\caption{Overview of the electromechanical cardiac model based on Ref.~\cite{willems_isogeometric_2023}.}
\label{box:overview}
\end{Box}

\subsubsection{Solution techniques}
\label{sec:solutiontechniques}
The formulation summarized in Box~\ref{box:overview} is discretized in space using isogeometric analysis. In this discretization technique, the displacement field and contractile sarcomere length field are approximated by
\begin{align}
    u_j^h &=\sum_{i} N_{ij}^u \tilde{u}_i, &  l^{c^h} &=\sum_{i} N_{i}^{l^c} \tilde{l}_i^c,
    \label{eq:functionapproximation}
\end{align}
where $N_{ij}^u$ are vector-valued B-spline basis functions for the displacement field and $N_{i}^{l^c}$ are scalar-valued B-spline basis functions for the contractile sarcomere length. The coefficients corresponding to the basis functions are denoted by $\tilde{u}_i$ for the displacement field, and by $\tilde{l}_i^c$ for the contractile sarcomere length. In this work, we consider maximum regularity cubic ($p=3$) B-splines for the displacement the contractile length field. This choice of B-spline basis functions makes the discretized displacement and sarcomere field $C^2$-continuous in the interior of the patches and $C^0$-continuous across patch boundaries.

We consider a Bubnov-Galerkin discretization of the weak form of Box~\ref{box:overview}, which is expressed completely on the reference configuration. The pressure boundary condition is weakly enforced (through the weak form), while the normal displacement boundary condition on the basal plane is enforced strongly (by constraining the relevant control point displacements). The rigid body constraints are enforced through scalar Lagrange multipliers. The weak forms and their matrix representations resulting from this discretization are elaborated in Willems \emph{et al.} \cite{willems_isogeometric_2023}.

For the temporal discretization of the time-dependent terms emanating from the contractile dynamics model and the circulatory system model we consider a backward Euler scheme. This results in a nonlinear system of equations for each time step, which is solved using Newton-Raphson iterations. To optimize computational effort while retaining stability, an adaptive time stepper is used. This time stepper scales the time step based on the required number of Newton iterations.
\subsection{The pressure-loaded image configuration}
\label{sec:gpa}

A fundamental aspect in scan-based cardiac modeling is that the images are generally not obtained in an unloaded state. To take this effect into account, we extend the modeling approach of Ref.~\cite{willems_isogeometric_2023} so that it can accommodate a loaded scan configuration. The fundamental idea behind the proposed extension is that all computations are performed on the isogeometric anatomical model obtained by the fitting procedure introduced in Section~\ref{sec:Sec3}, \emph{i.e.}, the image configuration. The rationale behind this is that the quality of the isogeometric geometry resulting from the fitting procedure can be assured, contrasting the case in which computations are performed on a geometric model deflated toward the unloaded configuration. In Section~\ref{sec:decomposition} we detail how the loaded image configuration is accommodated in the cardiac model. In Section~\ref{sec:theigpa} an isogeometric analysis version of the general pre-stress algorithm of Weisbecker \emph{et al.}~\cite{weisbecker_generalized_2014} is proposed to evaluate the mapping between the unloaded configuration and the image configuration.

\subsubsection{The decomposition of displacements and displacement gradients}
\label{sec:decomposition}
In our model, we make the distinction between three configurations, as schematically shown in Figure~\ref{fig:configurations} for the ventricle. We denote the position of the material point in the unloaded configuration $\Omega^{\mathrm{lv}}_0$  by $x_{0_i}$. When loaded by a cavity pressure $p^{\mathrm{lv}}_{\rm img}$, the ventricle is deformed toward the image configuration, $\Omega^{\mathrm{lv}}_{\rm img}$. The displacement corresponding to this image configuration is denoted by $u_{{\rm img}_i} = x_{{\rm img}_i} - x_{0_i}$. When loaded with an arbitrary pressure, $p^{\mathrm{lv}}=p^{\mathrm{lv}}_{\rm img} + p^{\mathrm{lv}}_{\rm arb}$, the material is deformed toward the current configuration, in which the total displacement of the material points is given by
\begin{align}
u_i = u_{{\rm img}_i} + u_{{\rm arb}_i},
\label{eq:displacementdecomposition}
\end{align}
where $u_{{\rm arb}_i}$ is the arbitrary displacement relative to the image configuration as a result of the arbitrary loading $p^{\mathrm{lv}}_{\rm arb}$. In the remainder, we will omit the superscript `lv', \emph{e.g.}, $p=p^{\mathrm{lv}}$, for clarity but also to emphasize the generality of the GPA.

In our formulation, we assume that the image displacement, $u_{{\rm img}_i}$, has been determined (in our case using the algorithm discussed in Section~\ref{sec:theigpa}). The displacement field corresponding to an arbitrary pressure loading is then determined by the model elaborated in Section~\ref{sec:cardiacmodel}, but with the total displacement field in accordance with equation~\eqref{eq:displacementdecomposition}. To evaluate the model it is necessary to determine the total deformation gradient. From the above definition of the configurations it follows that  $x_i = x_{0_i} + u_{{\rm img}_i} + u_{{\rm arb}_i}$, from which the total deformation gradient can be determined as
\begin{equation}
\begin{aligned}
 F^{\phantom{}}_{ij} &= \frac{\partial x_i}{\partial x_{0_j}} =  \delta_{ij} + \frac{\partial u_{{\rm img}_i}}{\partial x_{0_j}} + \frac{\partial u_{{\rm arb}_i}}{\partial x_{{\rm img}_k}} \frac{\partial x_{{\rm img}_k}}{\partial x_{0_j}} = F_{{\rm img}_{ij}} + \frac{\partial u_{{\rm arb}_i}}{\partial x_{{\rm img}_k}}  F_{{\rm img}_{kj}}= \left( \delta_{ik} +  \frac{\partial u_{{\rm arb}_i}}{\partial x_{{\rm img}_k}}  \right) F_{{\rm img}_{kj}} ,\\
 &=  F_{{\rm arb}_{ik}} F_{{\rm img}_{kj}},
\end{aligned}
\label{eq:Fdecomposition}
\end{equation}
where
\begin{align}
  F_{{\rm img}_{ij}} &= \delta_{ij} + \frac{\partial u_{{\rm img}_{i}}}{\partial {x}_{0_j}}    & F_{{\rm arb}_{ij}} &= \delta_{ij} + \frac{\partial u_{{\rm arb}_{i}}}{\partial {x}_{{\rm img}_{j}}}.
  \label{eq:Fcomponents}
\end{align}
The decomposition \eqref{eq:Fdecomposition} conveys that when the image deformation, $u_{{\rm img}_i}$ (or the image deformation gradient, $F_{{\rm img}_{ij}}$) is known, the total deformation gradient $F_{ij}$ can be evaluated. Using the definitions in Box~\ref{box:overview}, the constitutive behavior can then be evaluated for an arbitrary loading condition.

In our formulation, we perform all operations on the image configuration. This implies that the gradient of $u_{{\rm arb}_i}$ with respect to the image configuration can directly be evaluated and that the corresponding deformation gradient $F_{{\rm arb}_{ij}}$ can be used to evaluate gradients and integrals in the deformed configuration in accordance with
\begin{align}
\frac{\partial f}{\partial x_i} &= \frac{\partial f}{\partial x_{{\rm img}_j}} F_{{\rm arb}_{ji}}^{-1},  &   \int_{\Omega} f \, \mathrm{d}V &= \int_{\Omega_{\rm img}} f | \det{ \mathbf{F}_{\rm arb}} |\, \mathrm{d}V_{\rm img},
\end{align}
where $f$ is an arbitrary function.

The image displacement, $u_{{\rm img}_i}$, is also defined on the image configuration, such that the image deformation gradient in \eqref{eq:Fcomponents} cannot be evaluated directly. To evaluate this expression, the chain rule is applied to yield
\begin{align}
  F_{{\rm img}_{ij}} &= \delta_{ij} + \frac{\partial u_{{\rm img}_{i}}}{\partial {x}_{0_j}} = \delta_{ij} + \frac{\partial u_{{\rm img}_{i}}}{\partial {x}_{{\rm img}_{k}}} F_{{\rm img}_{kj}},
\end{align}
from which it follows that
\begin{align}
 F_{{\rm img}_{ij}}^{-1} =  \delta_{ij} - \frac{\partial u_{{\rm img}_{i}}}{\partial {x}_{{\rm img}_{j}}}.
 \label{eq:Fimginverse}
\end{align}
This reformulated expression for the (inverse of) the image deformation gradient can be evaluated since the gradient of the image displacement is now defined with respect to the image configuration.

\subsubsection{The isogeometric general pre-stressing algorithm}
\label{sec:theigpa}
To evaluate the cardiac model outlined in Section~\ref{sec:cardiacmodel} in the case of a loaded image configuration, the image displacement $\mathbf{u}_{\rm img}$ (or the image deformation gradient field $\mathbf{F}_{\rm img}$) must be known. To determine this image displacement, we adapt the general pre-stressing algorithm (GPA) of Weisbecker \emph{et al.} \cite{weisbecker_generalized_2014} to our isogeometric analysis setting.

The concept behind the GPA is to find the image deformation gradient field, $\mathbf{F}_{\rm img}$, for which the passive response in the image configuration, $\mathbf{x}_{\rm img}$, is in equilibrium with the applied pressure, $p_{\rm img}$. To formalize this concept, we express the passive model, \emph{i.e.}, the mechanical model comprised by the equilibrium equations (\ref{eq:B11a}--\ref{eq:B11g}) and the passive component of the constitutive relation (\ref{eq:B15b}), in abstract form as
\begin{align}
\mathbf{F} = \mathcal{M}_{\rm passive}( \mathbf{F}_{\rm img}, p).
\label{eq:abstractmodel}
\end{align}
This abstract form expresses that, given an image deformation gradient field $\mathbf{F}_{\rm img}$ and a pressure load $p$, the total deformation gradient field $\mathbf{F}$ can be computed. In the case that the pressure is equal to that of the image state, \emph{i.e.}, $p=p_{\rm img}$, the total deformation gradient field should be equal to the deformation gradient field of the image itself, that is
\begin{align}
    \mathbf{F}_{\rm img} = \mathcal{M}_{\rm passive}( \mathbf{F}_{\rm img}, p_{\rm img}).
    \label{eq:nonlineargpa}
\end{align}
This equilibrium condition for the image configuration is equivalent to stating that the arbitrary deformation gradient, $\mathbf{F}_{\rm arb}$, in the decomposition \eqref{eq:Fdecomposition} is equal to the identity tensor, and, following the definition \eqref{eq:Fcomponents} that the arbitrary displacement field vanishes, \emph{i.e.}, $\mathbf{u}_{\rm arb} = \mathbf{u} - \mathbf{u}_{\rm img} = \mathbf{0}$. 

Conceptually, the nonlinear equation \eqref{eq:nonlineargpa} can be solved for the image deformation gradient field $\mathbf{F}_{\rm img}$, given a certain $p_{\rm img}$. The original algorithm proposed in Ref.~\cite{weisbecker_generalized_2014} solves equation \eqref{eq:nonlineargpa} by gradually increasing the image pressure up to $p_{\rm img}$, using \eqref{eq:nonlineargpa} as an update rule for the image deformation gradient field, \emph{i.e.}, $\mathbf{F}_{\rm img}^{k+1} = \mathcal{M}_{\rm passive}( \mathbf{F}^k_{\rm img}, p^{k})$, where $k$ is the iteration number. When the image pressure is reached, \emph{i.e.}, $p^{k} = p_{\rm img}$, equation \eqref{eq:nonlineargpa} is used to perform load-increment-free fixed point iterations, \emph{i.e.}, $\mathbf{F}_{\rm img}^{k+1} = \mathcal{M}_{\rm passive}( \mathbf{F}^k_{\rm img}, p_{\rm img})$. For many (non-linear) problems, including the cases considered in this work, the gradual increase of the pressure before performing the fixed point iterations is required to ensure convergence.

\begin{algorithm}
\caption{Isogeometric general pre-stressing algorithm}\label{alg:igpa}
    \begin{algorithmic}[1]
        \Require $p_{\rm img}$, $n_{\mathrm{fixed}}$  \Comment{Image pressure, max free-load iterations}
        \Ensure  $\mathbf{u}_{\rm img}^0 = \mathbf{0}$, $p^0=0$, $\Delta p^0 = \Delta p_{\rm max}$, $k=0$
        \While{$(k - n_{\mathrm{load}} ) \leq n_{\mathrm{fixed}} $}
            \State $\mathbf{F}_{\rm img}^{k} \gets \Call{get\_image\_deformation\_gradient}{\mathbf{u}_{\rm img}^k}$ \Comment{Determine the image deformation gradient \emph{cf.}~\eqref{eq:Fimginverse}}
            \State $\mathbf{F}_{\rm img}^{k+1} \gets \Call{solve\_passive\_model}{\mathbf{F}_{\rm img}^k, p^k}$\Comment{Update the total deformation gradient \emph{cf.}~\eqref{eq:abstractmodel}}
            \State $\mathbf{u}_{\rm img}^{k+1} \gets \Call{get\_image\_deformation}{\mathbf{F}_{\rm img}^{k+1}}$\Comment{Update the image deformation \emph{cf.}~\eqref{eq:uimgprojection}}\label{alg:ln:projection}
            \If{solver converged in $n^{\mathrm{Newton}} \leq n^{\mathrm{Newton}}_{\rm max}$ Newton iterations}
                \State $\Delta p_{\rm max} \gets \min{\left(\Delta p_{\rm max},  p_{\rm img}-p^k \right)}$\Comment{Limit the maximum allowable step size by the image pressure}\label{alg:ln:maxdp}
                \State $\Delta p^{k+1} \gets \min{\left( \frac{n^{\mathrm{Newton}}_{\rm target}}{n^{\mathrm{Newton}}} \Delta p^k, \Delta p_{\rm max}  \right)}$\Comment{Adjust the step size based on the number of iterations}\label{alg:ln:target}
                \State $p^{k+1} \gets p^k + \Delta p^{k+1}$
                \State $\varepsilon = \Call{get\_relative\_increment}{\mathbf{u}_{\rm img}^{k}, \mathbf{u}_{\rm img}^{k+1}}$\Comment{Compute the relative image deformation increment \emph{cf.} \eqref{eq:uimgincrement}}\label{alg:ln:increment}
                \State $k \gets k + 1$
                \If{$p^{k+1} < p_{\mathrm{img}}$}
                    \State $n_{\mathrm{load}} \gets n_{\mathrm{load}} + 1$
                \EndIf
            \Else
                \State $\Delta p^{k} \gets \frac{n^{\mathrm{Newton}}_{\rm target}}{n^{\mathrm{Newton}}_{\rm max}} \Delta p^{k}$\Comment{Reduce the pressure step size}
                \State $p^{k} \gets p^{k-1} + \Delta p^{k}$
            \EndIf

        \EndWhile
    \end{algorithmic}
\end{algorithm}

To employ the general pre-stressing algorithm in the context of the considered isogeometric cardiac model, the original algorithm of Ref.~\cite{weisbecker_generalized_2014} has been adapted. The proposed isogeometric analysis GPA is outlined in Algorithm~\ref{alg:igpa}. Compared to the original algorithm, the following modifications have been made:
\begin{itemize}
    \item In principle, in the GPA the image deformation field $\mathbf{u}_{\rm img}$ does not need to be determined explicitly, as the model can be evaluated directly based on the image deformation gradient field, $\mathbf{F}_{\rm img}$. We find it practical, however, to explicitly compute the image deformation field, $\mathbf{u}_{\rm img}$, on the same spline basis as used for the discretization of the displacement field, \emph{i.e.}, equation~\eqref{eq:functionapproximation}, as
    \begin{align}
        u_{{\rm img}_j}^h &=\sum_{i} N_{ij}^u \tilde{u}_{{\rm img},i},
        \label{eq:uimgspline}
    \end{align}
    where $\tilde{u}_{{\rm img},i}$ are the control point image displacements. This discrete displacement image field is then computed by the projection of equation \eqref{eq:Fimginverse} in the Frobenius norm as
    \begin{align}
        {u}_{{\rm img},i}^h = \argmin_{{v}^h_{{\rm img},i}} \left( \left\|  F_{{\rm img}_{ij}}^{-1} -  \delta_{ij} + \frac{\partial v^h_{{\rm img}_{i}}}{\partial {x}_{{\rm img}_{j}}} \right\|  + \text{rigid body constraints} \right)
        \label{eq:uimgprojection}
    \end{align}
    In this projection (Line~\ref{alg:ln:projection}), the constraints are applied in the same way as considered for the model discretization in Section~\ref{sec:cardiacmodel}.\\
    Our choice to explicitly compute the image displacement field on the spline basis is practically motivated. In isogeometric analysis, the control point displacements are the natural way of storing the displacement field, in the sense that also the image geometry is represented by a control net. This means that an isogeometric analysis code will have the ability to interpolate the image displacement field in the same way as that it interpolates the geometry. The spline-based interpolation allows for the computation of the image deformation gradient field in the interior integration points (as used for the evaluation of the internal forces), on the endocardium (as used for the external forces) and on the basal plane (where Lagrange multiplier constraints are applied). It would be possible to store the image deformation gradient field in different quadrature point sets for the various contributions to the system of equations, but such a representation would not be standard. The representation in terms of the control point displacements standardizes the data format, which is beneficial from a computational workflow point of view. Although the spline interpolation \eqref{eq:uimgspline} introduces an additional approximation step compared to the direct usage of the image deformation gradient in the quadrature points, this approximation places the image displacement in the same discrete function space as the arbitrary displacement field $\mathbf{u}_{\rm arb}$. This means that the image displacement field is approximated in the same way, and with the same approximation properties, as the increments with respect to the image configuration.
    \item Due to the large deformations and high degree of non-linearity in the constitutive behavior of the considered isogeometric cardiac model, gradually increasing the pressure toward the image pressure is a necessity. To optimize the number of load steps required to reach the image pressure, an adaptive pressure step size is employed. The pressure increment is adjusted according to the number of Newton iterations, $n^{\mathrm{Newton}}$, required to solve the isogeometric passive model up to a specified tolerance (Lines \ref{alg:ln:maxdp}-\ref{alg:ln:target}), where $n^{\mathrm{Newton}}_{\rm target}$ is the desired number of Newton iterations. This adaptive load step procedure is initialized by prescribing $\Delta p^0 = \Delta p_{\rm max}$, and a load increment is redone with an adapted step size if the Newton solver does not converge within $n^{\mathrm{Newton}}_{\rm max}$ iterations.
\end{itemize}
Convergence of the algorithm is measured via the increments in the image displacement (Line~\ref{alg:ln:increment}) as
\begin{equation}
    \varepsilon = \frac{\| \mathbf{u}_{\rm img}^{k+1} - \mathbf{u}_{\rm img}^{k} \|}{\| \mathbf{u}_{\rm img}^{k} \|},
    \label{eq:uimgincrement}
\end{equation}
where use is made of the $L^2$-norm over the image configuration. The image displacement is assumed to be converged when the full image pressure load is applied and the increment measure \eqref{eq:uimgincrement} is below a specified tolerance. The number of steps required to reach the pressure load is denoted by $n_{\rm load}$, and the number of load-increment-free iterations by $n_{\rm fixed}$. The influence of these numerical parameters is studied for benchmark problems in \ref{app:igpabenchmarks}.
\newpage

\section{Benchmarking the patient-specific workflow}\label{sec:Sec5}
\begin{figure*}
     \centering
     \begin{subfigure}[b]{0.3\textwidth}
         \centering
             \includegraphics[width=\textwidth]{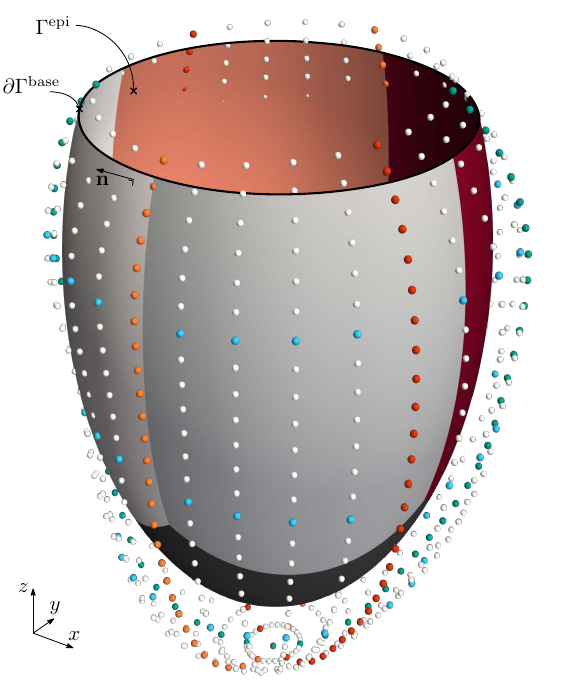}
         \caption{}
         \label{fig:Atlas_example_a} 
     \end{subfigure}
     \hfill
     \begin{subfigure}[b]{0.3\textwidth}
         \centering
              \includegraphics[width=\textwidth]{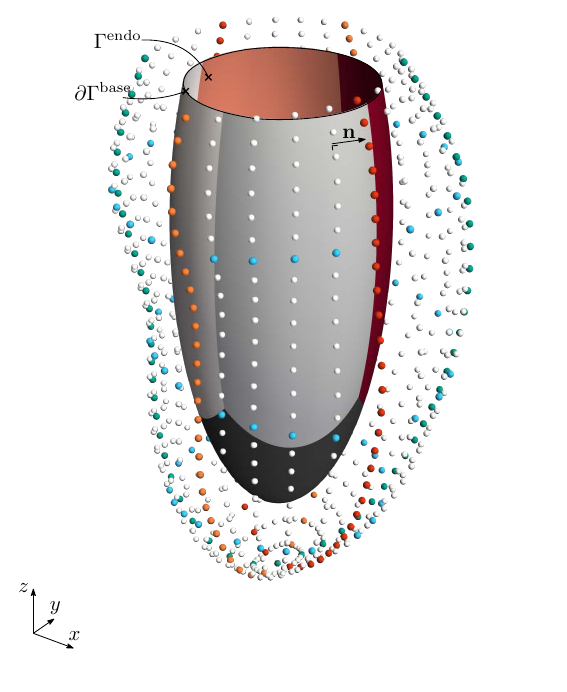}
         \caption{}
         \label{fig:Atlas_example_b}
     \end{subfigure}
     \hfill
     \begin{subfigure}[b]{0.3\textwidth}
         \centering
              \includegraphics[width=\textwidth]{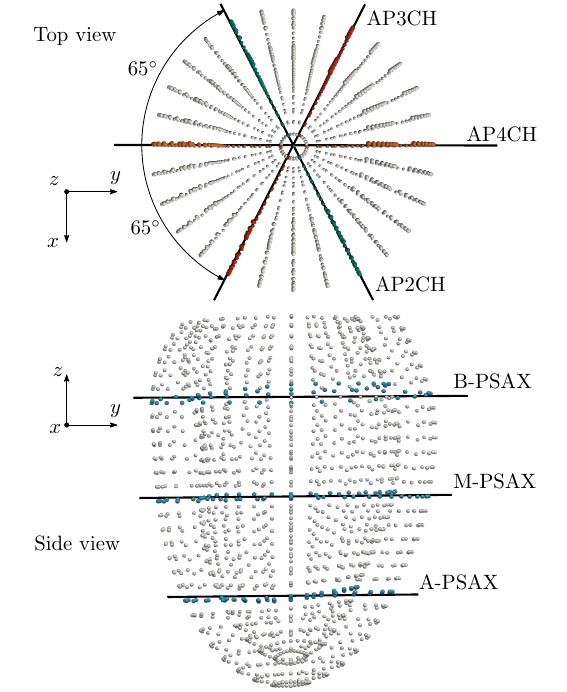}
         \caption{}
         \label{fig:Atlas_example_c}
     \end{subfigure}
        \caption{Dense point cloud data of a left ventricle at end-diastole obtained from the Cardiac Atlas Project \cite{fonseca_cardiac_2011} at \emph{(a)} the epicardium and \emph{(b)} the endocardium with the initial template visualized in the center. \emph{(c)} Point cloud representations of synthetic echocardiograms generated from the dense point cloud in six slices. The longitudinal positions of the PSAX views are approximately based on $25$\%, $50$\%, and $75$\% of the total length of the ventricle. The ellipsoidal template used as the initial condition for the fitting algorithm is shown at \emph{(a)} the epicardium and \emph{(b)} the endocardium.}
        \label{fig:Atlas_example}
\end{figure*}

To benchmark the developed patient-specific isogeometric analysis workflow (Figure~\ref{fig:workflow}), in this section we consider its application to a standard population-averaged data set (Cardiac Atlas Project \cite{fonseca_cardiac_2011}). This data set, based on cardiac MRI, provides a dense point cloud representation of the left ventricle at end-diastole (end filling and prior to contraction). We use the dense point cloud data as a reference case, and use it to construct point cloud representations of six synthetic echocardiogram slices: the apical 2- 3- and 4-chamber views (AP2CH, AP3CH, and AP4CH) and the apical-, mid- and basal-parasternal-short-axis views (A-PSAX, M-PSAX, and B-PSAX) \cite{nihoyannopoulos_echocardiography_2018}. The choice of synthetic slices is limited to and based on the available patient data that will be discussed in Chapter~\ref{sec:Sec6}. However, additional views can be considered, based on a specific clinical echocardiogram protocol of interest. The various point sets are shown in Figure~\ref{fig:Atlas_example}. The workflow is applied to the dense reference data (9744 points, labeled "dense"), to the sparse synthetic echocardiogram data using all six slices (575 points, labeled "spr.6"), and to the sparse synthetic echocardiogram data using only the AP2CH, AP4CH, and M-PSAX slices (344 points, labeled "spr.3"), which are the slices considered in the next chapter. We aim to quantify the effects of data sparsity by comparing the echocardiogram fitting results (Section~\ref{sec:fitting}) and the cardiac mechanics results (Section~\ref{sec:cardres}) to the results obtained using the dense reference case.

\subsection{Fitting results}\label{sec:fitting}
The ellipsoidal multi-patch NURBS template discussed in Section~\ref{sec:Sec2} is used as the initial condition of the fitting algorithm for all cases. The position and dimensions of the template are based on the stress-free geometry as discussed in Willems~\emph{et al.}~\cite{willems_isogeometric_2023} and based on Bovendeerd~\emph{et al.}~\cite{bovendeerd_dependence_1992}. The template is visualized in Figure~\ref{fig:Atlas_example} at both the epicardium and the endocardium surfaces. The developed fitting algorithm is applied to the epicardium and endocardium separately, after which a volumetric multi-patch NURBS parametrization is obtained by through-the-thickness interpolation; see Section~\ref{sec:multipatch}. The control points at the base, $\partial\Gamma^{\mathrm{base}}$, are only allowed to displace in a predefined plane. This plane corresponds to the $XY$-plane containing the average spatial position of the data points that are located at the base. Data points positioned above this plane are neglected as the fitting error cannot be minimized for these points due to the constraint. We use the same settings for the fitting algorithm as for the example case considered in Section~\ref{sec:example}, with the exception that $t_{\mathrm{max}}=500$ iterations are considered, the allowed gradient of the displacement error is set to $\Delta\varepsilon_{\mathcal{D}}=5\cdot10^{-3}$, and the last 20 iterations are only used to correct for continuity. We note that the settings of the algorithm have not been optimized for the case under consideration and that the algorithm is robust in the sense that the results do not change erratically when varying the settings. Optimization of the settings, which is beyond the scope of the current work, can yield further improvements, but will not fundamentally alter the results.

\begin{figure*}
     \centering
     \begin{subfigure}[b]{0.45\textwidth}
         \centering
             \includegraphics[width=\textwidth]{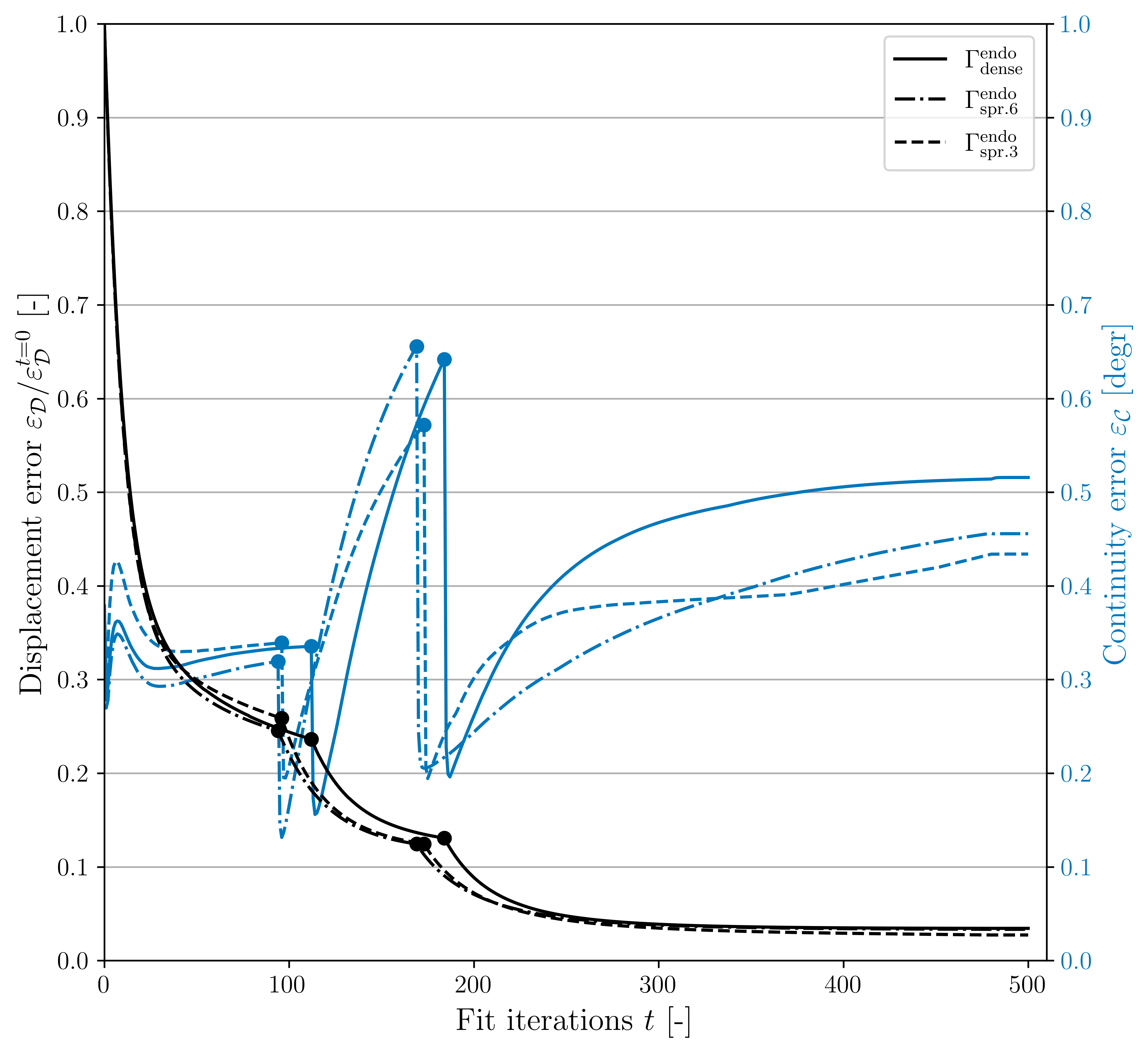}
         \caption{}
         \label{fig:Atlas_error_trace_endo} 
     \end{subfigure}
     \hfill
     \begin{subfigure}[b]{0.45\textwidth}
         \centering
              \includegraphics[width=\textwidth]{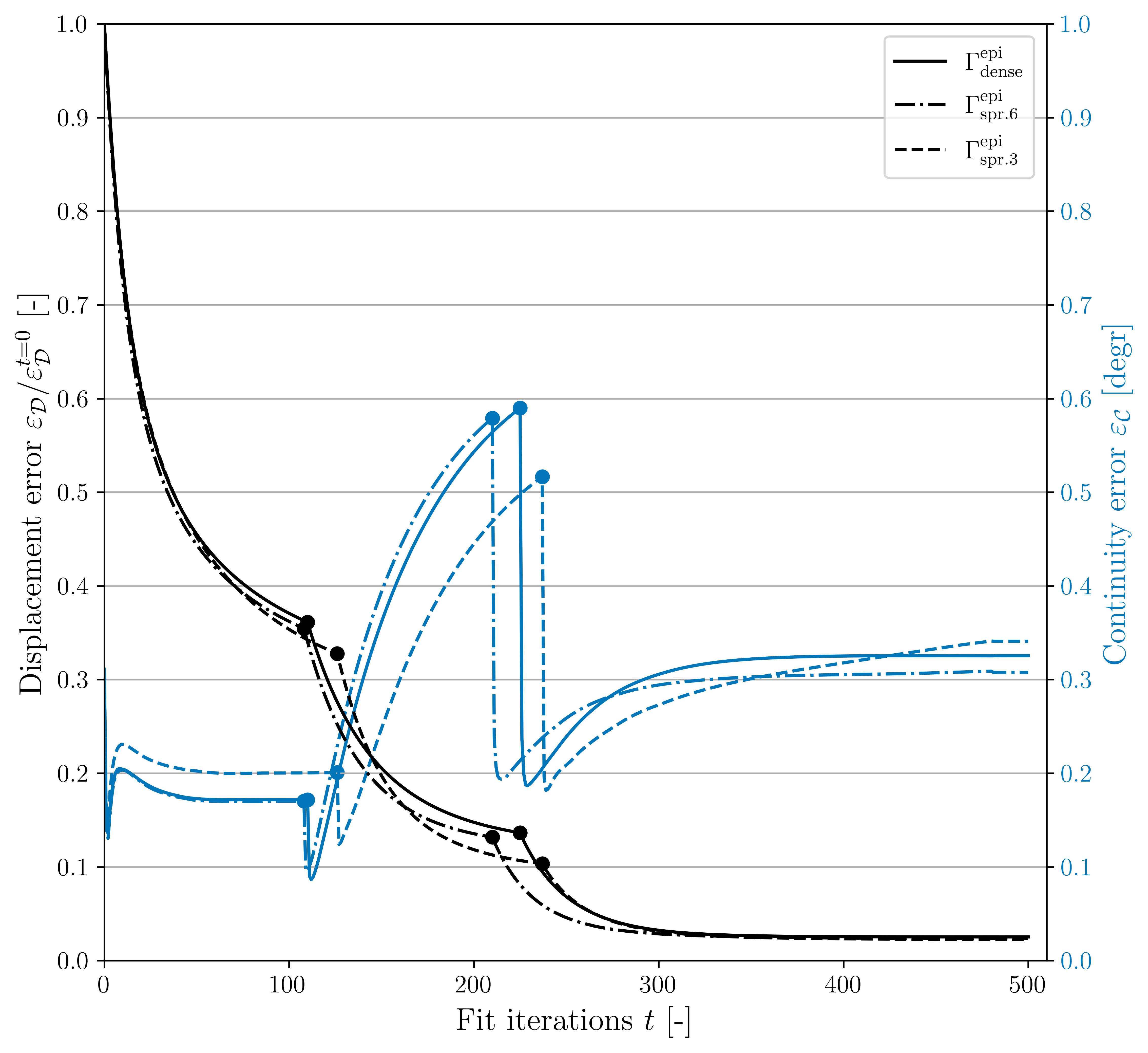}
         \caption{}
         \label{fig:Atlas_error_trace_epi}
     \end{subfigure}
     \caption{Convergence of the displacement error (left axis, black lines) and continuity error (right axis, blue lines) for \emph{(a)} the endocardium and \emph{(b)} the epicardium. The results obtained using different point sets based on the Cardiac Atlas Project are compared.}
        \label{fig:Atlas_error_trace}
\end{figure*}

\begin{figure*}[!t]
     \centering
     \begin{subfigure}[b]{0.45\textwidth}
         \centering
             \includegraphics[width=\textwidth]{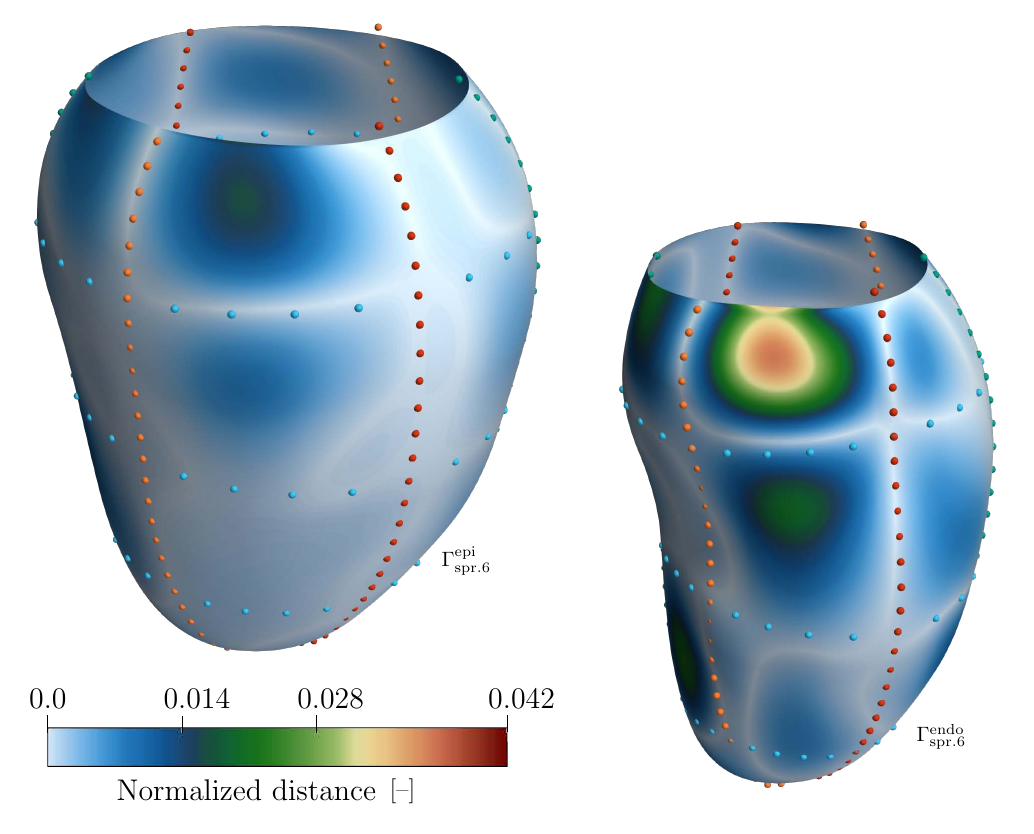}
         \caption{}
         \label{fig:Atlas_error_a} 
     \end{subfigure}
     \hfill
     \begin{subfigure}[b]{0.45\textwidth}
         \centering
              \includegraphics[width=\textwidth]{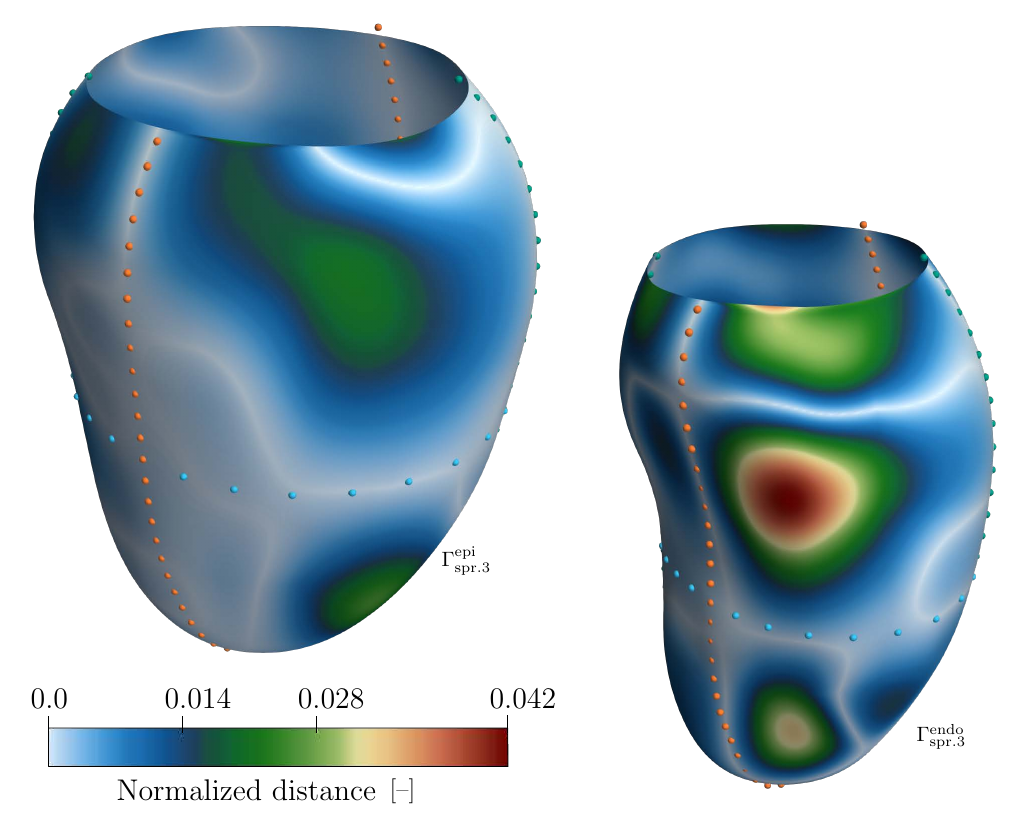}
         \caption{}
         \label{fig:Atlas_error_b}
     \end{subfigure}
     \caption{Distance to the dense point cloud reference result using \emph{(a)} six synthetic echocardiagram slices and \emph{(b)} three synthetic echocardiagram slices. The distance is normalized with respect to the average diameter of $7.18$\,[cm].}
        \label{fig:Atlas_error}
\end{figure*}

In Figure~\ref{fig:Atlas_error_trace} we trace the normalized displacement error (left axis, black lines) and continuity error $\varepsilon_{\mathcal{C}}$ (right axis, blue lines) for the endocardium (Figure~\ref{fig:Atlas_error_a}) and the epicardium (Figure~\ref{fig:Atlas_error_b}). We adopt the displacement and continuity errors following Equations~\eqref{eq:error_displ} and \eqref{eq:error_cont}, where we normalize the displacement error by the initial displacement error at the first iteration, \emph{i.e.}, $\varepsilon_{\mathcal{D}}/\varepsilon^{t=0}_{\mathcal{D}}$. For all three cases, the error behaves similarly as for the example case considered in Section~\ref{sec:example}. The normalized displacement error decreases gradually, reaching a plateau when the best possible fit to the data is obtained. Upon refinement of the multi-patch NURBS (solid circle markers), the error decreases further. Note that because the refinement criterion is dependent on the gradient of the displacement error, refinements occur at different moments for the different cases. After 500 iterations and two refinement steps, the normalized displacement error has reduced to less than 3.5\% percent of the initial template fit error for all cases. The remaining error is a result of the refined template not being able to capture the strongly curved geometry at the apex. For all cases, the continuity error (\emph{i.e.}, the average angle between interfaces) is within 0.55 [deg]. On account of the larger curvature near the apex of the endocardium compared to the epicardium, the continuity error is observed to be larger there.

When comparing the errors for the different cases, the normalized displacement error is observed to decrease with the number of data points. The reason is two-fold: First, a reduction in data points reduces the fitting constraints, which ultimately improves the domain interpolation of the data points. Second, the initial displacement error, $\varepsilon^{t=0}_{\mathcal{D}}$, of the template is relatively higher compared to cases with a greater number of data points. This difference arises from the initial template and point cloud distribution. Provided that the initial template is already close to the data points (in regions away from the apex), a relatively lower initial error is computed for cases with an increased number of data points, which results from the averaging of the displacement error with respect to the number of data points, \emph{cf.} Equation~\eqref{eq:error_displ}. Nonetheless, all three cases show a converged normalized displacement error of which the relative difference is negligible. Unlike the normalized displacement error, the continuity error is not directly influenced by the point cloud distribution. Consequently, there is a minimal difference in its convergence behavior across the various cases.

\begin{figure*}
     \centering
     \begin{subfigure}[b]{0.3\textwidth}
         \centering
             \includegraphics[width=\textwidth]{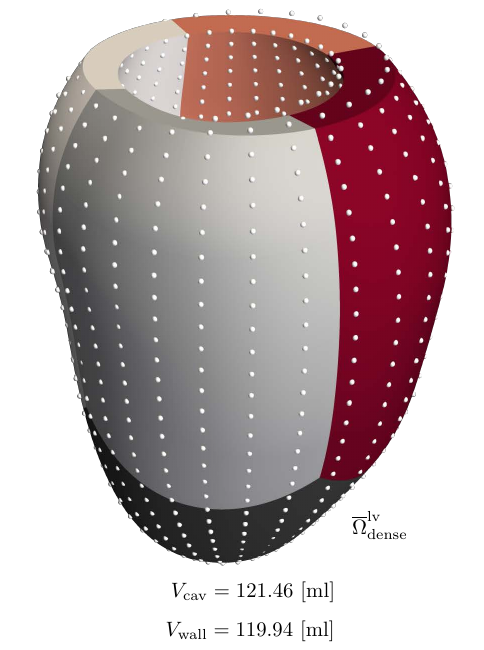}
         \caption{}
         \label{fig:Atlas_fit_a} 
     \end{subfigure}
     \hfill
     \begin{subfigure}[b]{0.3\textwidth}
         \centering
              \includegraphics[width=\textwidth]{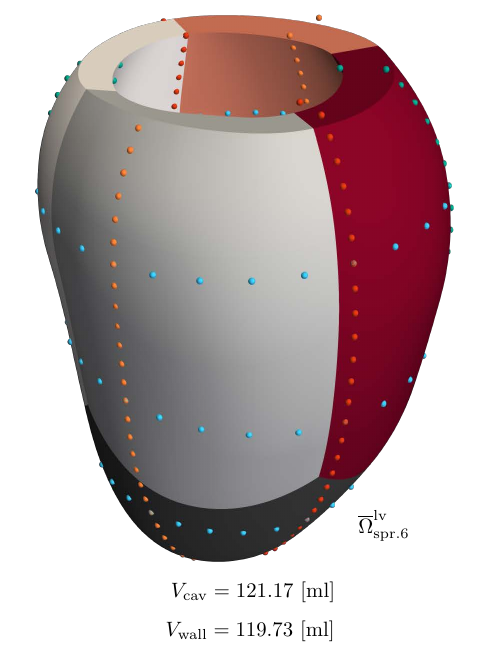}
         \caption{}
         \label{fig:Atlas_fit_b}
     \end{subfigure}
     \hfill
     \begin{subfigure}[b]{0.3\textwidth}
         \centering
              \includegraphics[width=\textwidth]{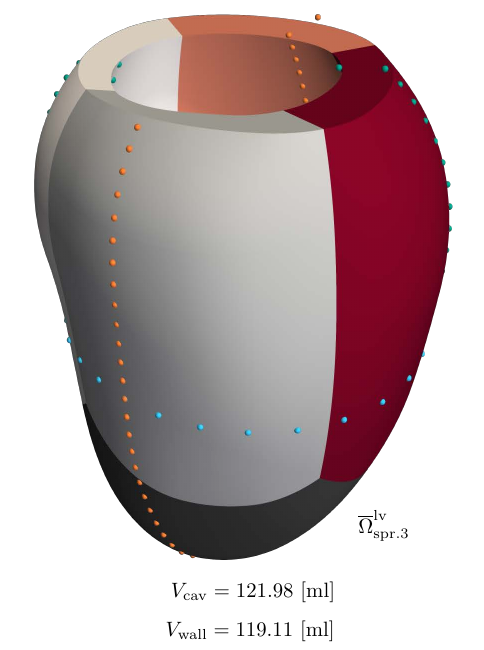}
         \caption{}
         \label{fig:Atlas_fit_c}
     \end{subfigure}
        \caption{Final fitting results for the Cardiac Atlas ventricle using \emph{(a)} the dense point set denoted by $\overline{\Omega}^{\mathrm{lv}}_{\mathrm{dense}}$ (reference result), \emph{(b)} six synthetic echocardiogram slices, denoted by $\overline{\Omega}^{\mathrm{lv}}_{\mathrm{spr.6}}$, and \emph{(c)} three synthetic echocardiogram slices, denoted by $\overline{\Omega}^{\mathrm{lv}}_{\mathrm{spr.3}}$.}
        \label{fig:Atlas_fit}
\end{figure*}

Figure~\ref{fig:Atlas_fit} compares the final volumetric results -- where the endocardial and epicardial boundaries are combined as explained in Section~\ref{sec:multipatch} -- of the fitting algorithm for the three point-set cases. Visually, the results for the various cases are indiscernible. When considering the cavity and wall volumes, it is observed that for the six-slice case (Figure~\ref{fig:Atlas_fit_b}) these volumes deviate from the dense reference case by 0.2\%. For the three-slice case (Figure~\ref{fig:Atlas_fit_c}) the cavity and wall volume deviate by 0.4\% and 0.7\%, respectively. This increase in error upon the removal of data points is expected, as the multi-patch NURBS object needs to interpolate larger data scant regions. This effect is illustrated in more detail in Figure~\ref{fig:Atlas_error}, which shows the distance to the dense point cloud reference result for the six-slice case (Figure~\ref{fig:Atlas_error_a}) and the three-slice case (Figure~\ref{fig:Atlas_error_b}). The distance is normalized with respect to the characteristic diameter of the dense point cloud -- defined as twice the average distance between the data points and the centroid -- which is equal to $7.18$ [cm]. As observed, the distance is largest in the data scant regions, with maximal offsets up to approximately 4\%. For the three-slice case, the area in which the distance errors become relatively large is increased compared to the six-slice case, which in turn results in larger cavity and wall volume errors. Overall, Figures~\ref{fig:Atlas_fit} and \ref{fig:Atlas_error} convey that while a reduction of the number of data points increases the fitting errors, for both the six-slice and the three-slice case the geometry is still fitted with a high degree of accuracy. While for the considered synthetic data cases the fitting errors can be precisely quantified, in general one should view these errors in relation to the uncertainties related to the scanning and segmentation process. In this light, for the considered setting we do not expect the errors related to the fitting procedure to be detrimental to the analysis workflow.

\subsection{Cardiac mechanics results}\label{sec:cardres}
Using the fitted image geometries obtained for the three cases we now evaluate the cardiac response using the model of Section~\ref{sec:cardiacmodel}. For all cases, we adopt parameter settings nearly identical to those employed in the model validation study by Willems \emph{et al.}~\cite{willems_isogeometric_2023}, with the only deviations being in the reference active stress, $T^0$, and the reference volumes of the circulatory model, $V_{\mathrm{ref0}}^{\mathrm{art}}$, $V_{\mathrm{ref0}}^{\mathrm{ven}}$. These parameter values were slightly altered such that a physiological hemodynamic response was obtained. An overview is given in Table~\ref{tab:params}. The reader is referred to Ref.~\cite{willems_isogeometric_2023} for details regarding the choice of parameters. The final refinement of the fitting procedure, with two uniform refinement steps relative to the initial template, is directly used for the cardiac analysis. The IGA cardiac model is constructed using the Nutils Python toolbox~\cite{van_zwieten_nutils_2022}, where we use maximum regularity cubic B-splines for the displacement and contractile length field, resulting in 3885 and 1351 degrees of freedom respectively. Based on the mesh convergence study presented in Ref.~\cite{willems_isogeometric_2023} this discretization setting is expected to yield accurate results. This expectation will be confirmed by consideration of a further uniform refinement at the end of this section.

\begin{table}
\caption{Parameters of the constitutive model components and the circulatory system model, taken from Willems \emph{et al.}~\cite{willems_isogeometric_2023}.}\label{tab:params} 
\centering
\begin{subtable}[t]{0.49\textwidth}
\centering
\begin{tabular}[t]{llSl}
\hline
Passive           & $C$                          & 0.4                       & kPa                     \\
component         & $\kappa$                          & 110.0                       & kPa                     \\
                  & $a_1$                          & 3.0                         & -                       \\
                  & $a_2$                          & 6.0                         & -                       \\
                  & $a_3$                          & 3.0                        & -                       \\ \hline
Active            & $T^{\mathrm{0}}$                   & 160.0                       & kPa                     \\
component         & $E^{\mathrm{a}}$                   & 20.0                        & $\mu\text{m}^{-1}$      \\
                  & $a^{l}$                   & 2.0                         & $\mu\text{m}^{-1}$      \\
                  & $l^{\mathrm{c0}}$                  & 1.5                       & $\mu\text{m}$           \\
                  & $l^{\mathrm{s0}}$                  & 1.9                       & $\mu\text{m}$           \\
                  & $l^{\mathrm{d}}$                   & -1.0                      & $\mu\text{m}$           \\
                  & $v^{\mathrm{0}}$                & 7.5                        & $\mu\text{m} \ \text{s}^{-1}$                    \\
                  & $b$                            & 160.0                       & $\text{ms} \ \mu\text{m}^{-1}$ \\
                  & $T^{\mathrm{act}}$                &  4.0                        & ms                    \\
                  & $\tau^{\mathrm{r}}$                & 75.0                        & ms                      \\
                  & $\tau^{\mathrm{d}}$                & 150.0                       & ms                      \\ \hline
\end{tabular}
\end{subtable}
\begin{subtable}[t]{0.49\textwidth}
\centering
\begin{tabular}[t]{llll}
\hline
Circulatory                & $R^{\mathrm{art}}$                 & $1.00\times 10^{-2}$ & kPa s ml$^{-1}$ \\
system                     & $R^{\mathrm{per}}$                 & $1.20\times 10^{-1}$ & kPa s ml$^{-1}$ \\
                           & $R^{\mathrm{ven}}$                 & $2.00\times 10^{-3}$ & kPa s ml$^{-1}$ \\
                           & $C^{\mathrm{art}}$                 & $2.50\times 10^{1}$  & ml kPa$^{-1}$   \\
                           & $C^{\mathrm{ven}}$                 & $6.00\times 10^{2}$  & ml kPa$^{-1}$   \\
                           & $V_{\mathrm{ref0}}^{\mathrm{art}}$ & $5.30\times 10^{2}$  & ml              \\
                           & $V_{\mathrm{ref0}}^{\mathrm{ven}}$ & $3.10\times 10^{3}$  & ml              \\ 
                           & $V^{\mathrm{tot}}$                 & $5.00\times 10^{3}$  & ml             \\
\hline
\end{tabular}
\end{subtable}
\end{table}

\begin{figure*}[!t]
     \centering
     \begin{subfigure}[b]{0.45\textwidth}
         \centering
             \includegraphics[width=\textwidth]{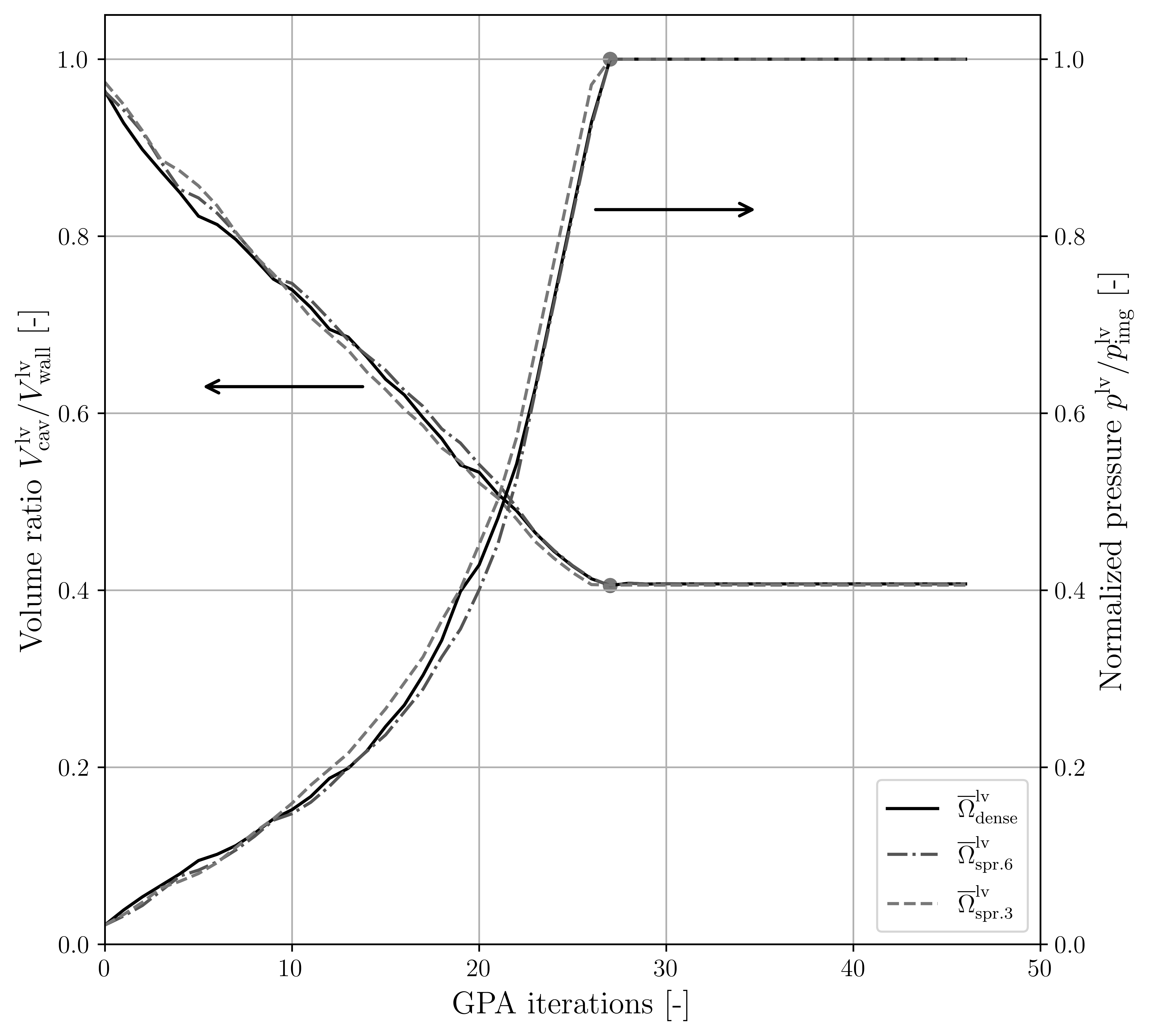}
         \caption{}
         \label{fig:GPA_atlas_a} 
     \end{subfigure}
     \hfill
     \begin{subfigure}[b]{0.4295\textwidth}
         \centering
              \includegraphics[width=\textwidth]{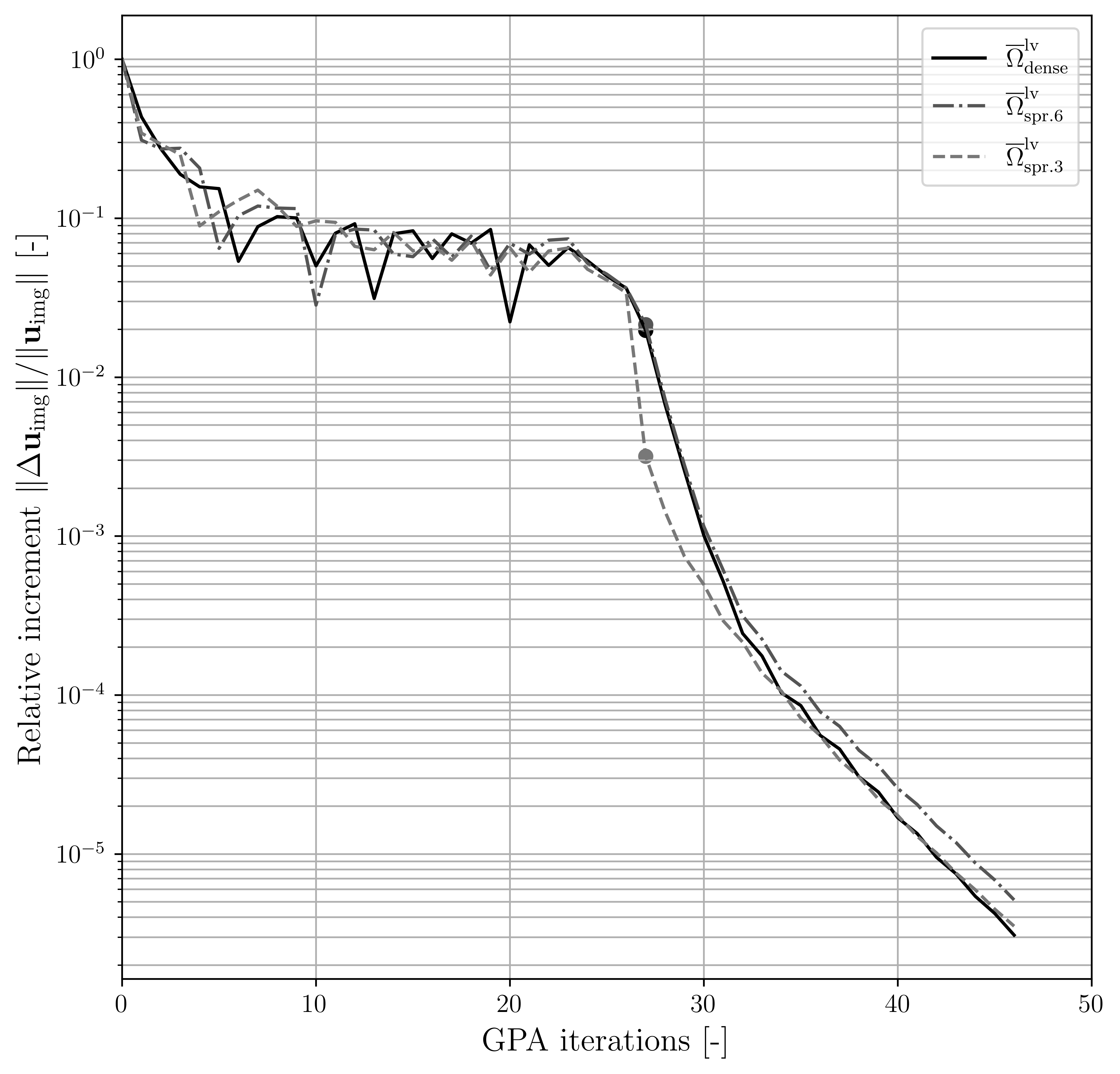}
         \caption{}
         \label{fig:GPA_atlas_b}
     \end{subfigure}
     \caption{Convergence plots of the isogeometric general pre-stressing algorithm: \emph{(a)} the cavity-to-wall volume ratio, and \emph{(b)} the relative displacement increments. The results obtained using different point sets based on the Cardiac Atlas Project are compared.}
        \label{fig:GPA_atlas}
\end{figure*}

\begin{figure*}[!t]
     \centering
     \begin{subfigure}[b]{0.33\textwidth}
         \centering
              \includegraphics[width=\textwidth]{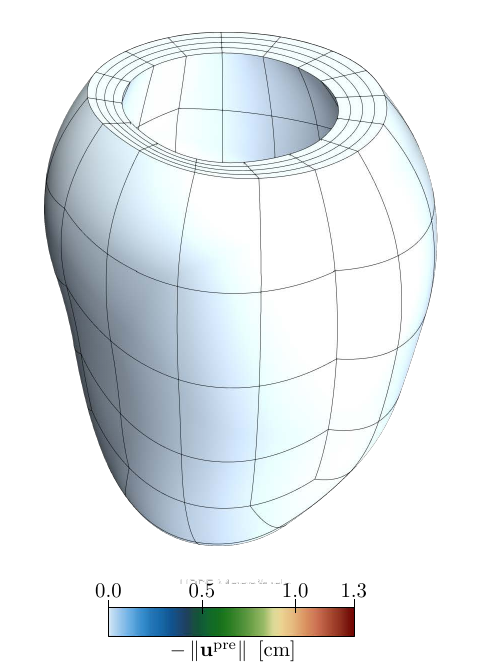}
         \caption{}
         \label{fig:GPA_atlas_deflated_a} 
     \end{subfigure}
     \hfill
     \begin{subfigure}[b]{0.33\textwidth}
         \centering
               \includegraphics[width=\textwidth]{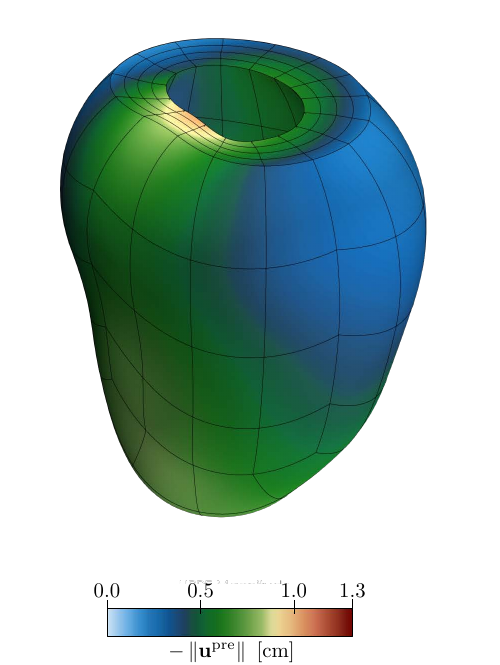}
         \caption{}
         \label{fig:GPA_atlas_deflated_b}
     \end{subfigure}
     \begin{subfigure}[b]{0.33\textwidth}
         \centering
               \includegraphics[width=\textwidth]{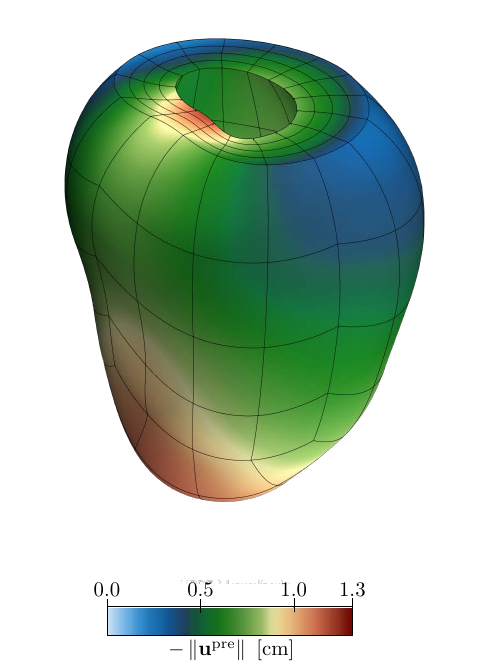}
         \caption{}
         \label{fig:GPA_atlas_deflated_c}
     \end{subfigure}
     \caption{Deflated Cardiac Atlas Project ventricle using the dense point cloud domain, $\overline{\Omega}^{\mathrm{lv}}_{\mathrm{dense}}$, at \emph{(a)} $p^{\rm lv}=0$\,[Pa] (image configuration), \emph{(b)} $p^{\rm lv}=770$\,[Pa] (22 iterations) and \emph{(c)} $p^{\rm lv}=p^{\rm lv}_{\rm img}=1600$\,[Pa] (46 iterations) (unloaded configuration). Note that the unloaded configuration is not used in the analysis and is only included here for illustration purposes.}
        \label{fig:GPA_atlas_deflated}
\end{figure*}

In the first stage of the cardiac analysis, the unloaded configuration is determined using the algorithm of Section~\ref{sec:gpa}. The end-diastolic image pressure is assumed as $1600$ [Pa] or $12$ [mmHg]. An initial pressure increment of $\Delta p^0 = 35$\,[Pa] is used for the isogeometric GPA algorithm. The pressure step is scaled using $n^{\mathrm{Newton}}_{\rm target}=7$ and the iterations are terminated after $n_{\rm fixed}=20$ load-increment-free iterations. The convergence plots for the GPA are shown in Figure~\ref{fig:GPA_atlas}. For all cases, the adaptive pressure load stepper -- characterized by the nonlinear slope of the normalized pressure curve in Figure~\ref{fig:GPA_atlas_a} -- reaches the image pressure after 27 iterations, marked by the solid spheres. The cavity-to-wall volume ratio, shown in Figure~\ref{fig:GPA_atlas_a}, decreases from approximately 95\% in the image configuration, to approximately 40\% in the unloaded configuration. After the pressure is fixed at the image pressure, this ratio is virtually constant. From the relative displacement increment shown in Figure~\ref{fig:GPA_atlas_b}, it is observed that the error decreases linearly while the pressure is fixed at the image pressure, showing the same behavior as for the GPA benchmark problems considered in \ref{app:igpabenchmarks}. The deflation procedure is further illustrated in Figure~\ref{fig:GPA_atlas_deflated}, which shows the deflation process for the dense point cloud case. Although the deflated geometry as shown in this figure is not directly used in the analysis, \emph{i.e.}, all operations are performed on the image geometry resulting from the fitting procedure, this plot clearly illustrates how the ventricle is deflated. It is particularly noteworthy to observe the torsional rotation that takes place during deflation on account of the fiber distribution.

\begin{figure}
    \centering
     \begin{subfigure}[b]{0.48\textwidth}
         \centering
             \includegraphics[width=\textwidth]{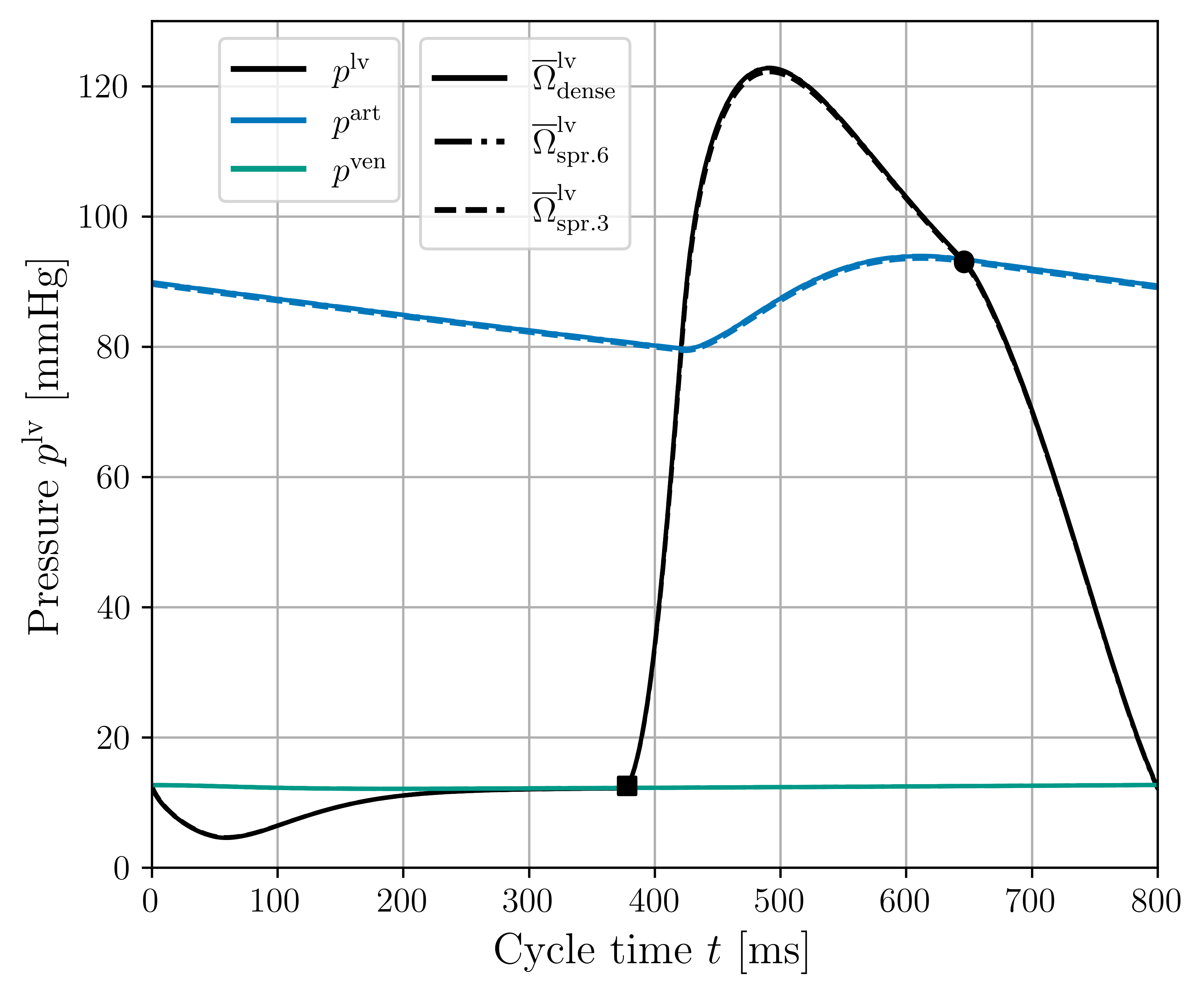}
         \caption{}
         \label{fig:hemodynamics_atlas_a}
     \end{subfigure}
     \hfill
     \begin{subfigure}[b]{0.48\textwidth}
         \centering
              \includegraphics[width=\textwidth]{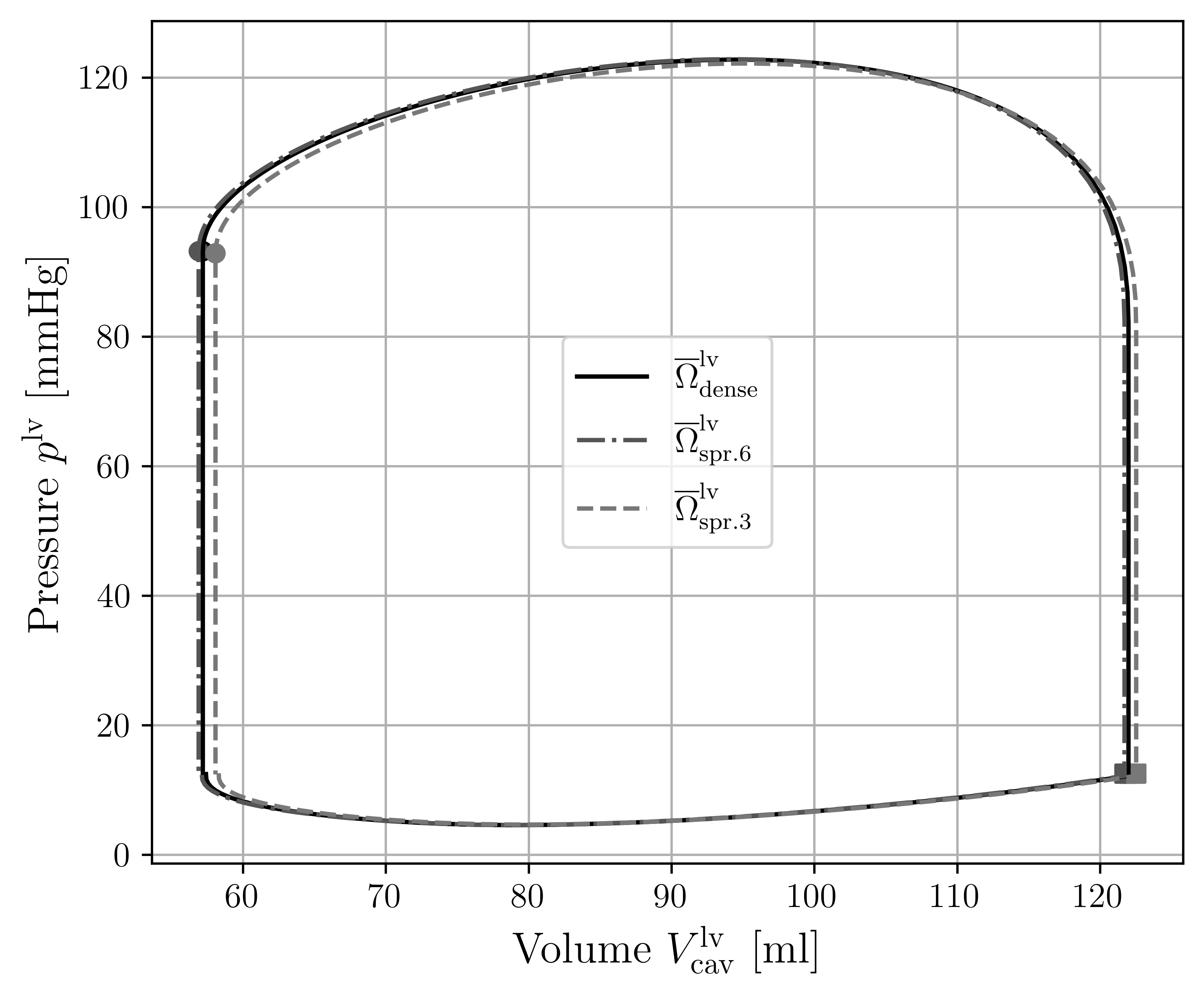}
         \caption{}
         \label{fig:hemodynamics_atlas_b}
     \end{subfigure}
    \caption{Hemodynamic response of the left ventricle with the parameters as in Table~\ref{tab:params}: \emph{(a)} left-ventricle, arterial and venous pressure signals, and \emph{(b)} left-ventricle cavity pressure-volume loop. The results obtained using different point sets based on the Cardiac Atlas Project are compared.}
    \label{fig:hemodynamics_atlas}
\end{figure}

\begin{figure*}
     \centering
     \begin{subfigure}[b]{0.48\textwidth}
         \centering
             \includegraphics[width=\textwidth]{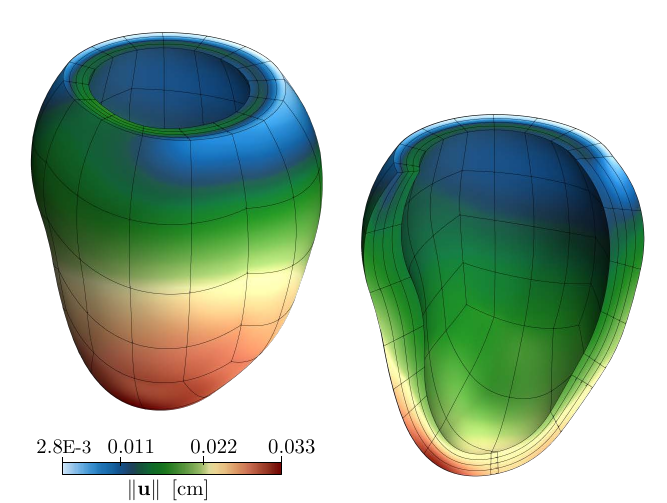}
         \caption{}
         \label{fig:Sim_result_enddiastole} 
     \end{subfigure}
     \hfill
     \begin{subfigure}[b]{0.48\textwidth}
         \centering
              \includegraphics[width=\textwidth]{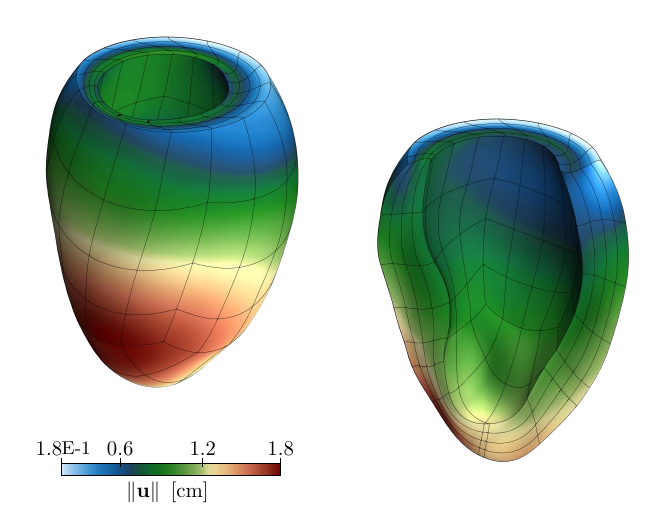}
         \caption{}
         \label{fig:Sim_result_endsystole}
     \end{subfigure}
     \caption{Deformation of the dense point cloud ventricle domain, $\overline{\Omega}^{\mathrm{lv}}_{\mathrm{dense}}$, at \emph{(a)} end-diastole (square marker in Figure~\ref{fig:hemodynamics_atlas}) and \emph{(b)} end-systole (circle marker in Figure~\ref{fig:hemodynamics_atlas}).}
        \label{fig:Atlas_sim_result}
\end{figure*}

With the image deformation state determined through the isogeometric GPA, the time-dependent cardiac model can be evaluated. We use an initial time step size of $\Delta t = 2$\,[ms], which is scaled by a factor of $\frac{1}{2}$ in the case of a non-convergence step. The time step is restored to its original value upon two consecutively converged steps. The simulation is initiated ($t=0$~[ms]) at the end-diastolic image pressure at which contraction starts at $t=T^{\mathrm{act}}=4$~[ms]. The cardiac-cycle duration is fixed at $800$~[ms] of which five consecutive cycles are computed to achieve a cyclic hemodynamic steady state. The hemodynamic response of the fifth cycle is shown in Figure~\ref{fig:hemodynamics_atlas}, showing very similar results for all cases. The minor differences between the cases can be explained directly from the differences in the fitting results. The maximum cavity volume loop is directly related to the cavity volume of the image state, which, as indicated in Figure~\ref{fig:Atlas_fit}, is smallest for the six-slice case, largest for the three-slice case, and with the dense point cloud reference case in between. The deformation field obtained for the dense point cloud case is illustrated in Figure~\ref{fig:Atlas_sim_result} at end-diastole and at end-systole (corresponding to the markers in Figure~\ref{fig:hemodynamics_atlas}). The torsional motion of the ventricle is clearly observable from the mesh lines. The displacement results for the six- and three-slice cases are visually indiscernible from the dense case and have therefore not been included here.

\begin{figure}
    \centering
     \begin{subfigure}[b]{0.48\textwidth}
         \centering
             \includegraphics[width=\textwidth]{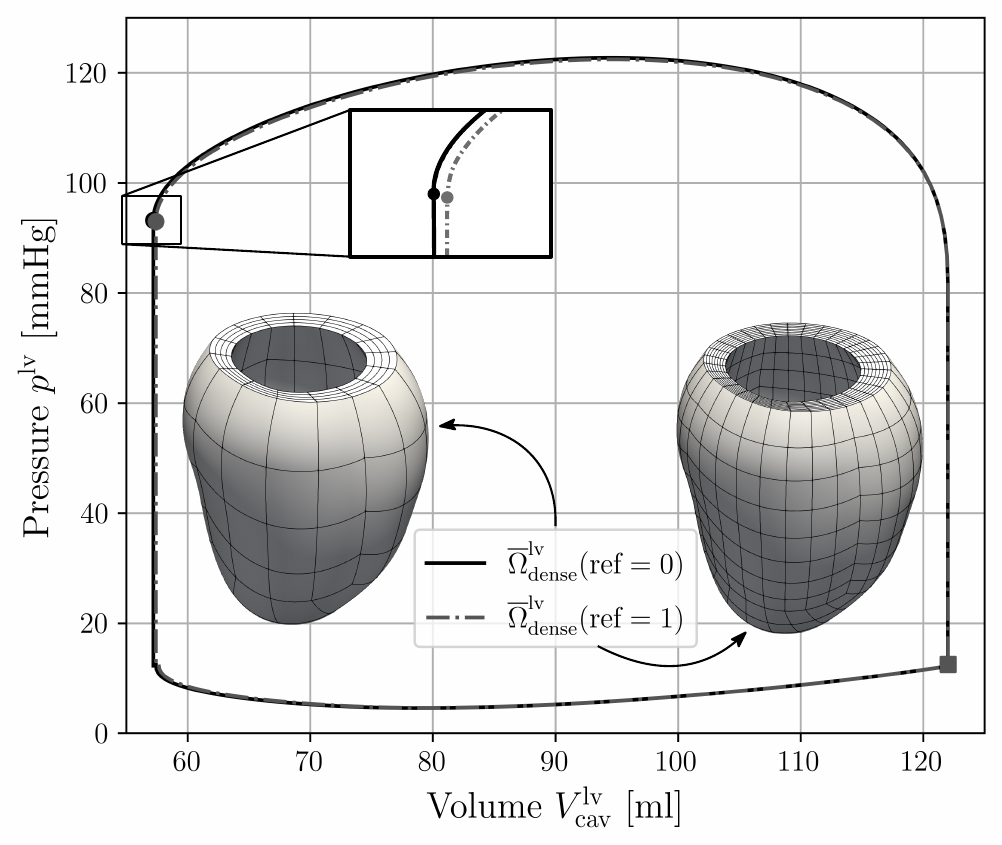}
         \caption{}
         \label{fig:hemodynamics_atlasconvergence_a}
     \end{subfigure}
     \hfill
     \begin{subfigure}[b]{0.48\textwidth}
         \centering
              \includegraphics[width=\textwidth]{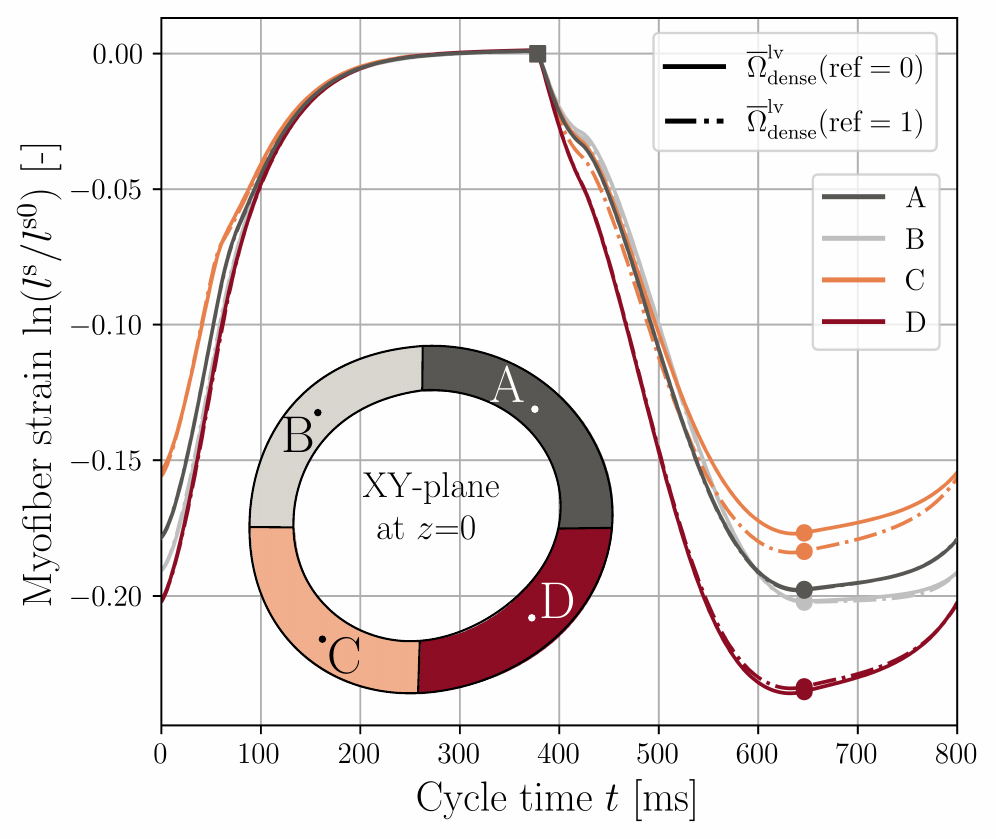}
         \caption{}
         \label{fig:hemodynamics_atlas_convergence_b}
     \end{subfigure}
    \caption{Comparison of \emph{(a)} the pressure-volume loop and \emph{(b)} the myocardium strain computed in four points at the equatorial plane ($z=0$ in Figure~\ref{fig:Atlas_example_c}), using a discretization $\overline{\Omega}^{\mathrm{lv}}_{\rm{dense}}(\mathrm{ref=0})$ with 5241 degrees of freedom and a uniform refinement, $\overline{\Omega}^{\mathrm{lv}}_{\rm{dense}}(\mathrm{ref=1})$, with 22489 degrees of freedom.}
    \label{fig:hemodynamics_atlas_convergence}
\end{figure}

To assess the influence of the mesh size, in Figure~\ref{fig:hemodynamics_atlas_convergence} the results obtained on the mesh considered above, with 5241 degrees of freedom in total, is compared to results with an additional uniform mesh refinement, which results in 22489 degrees of freedom. The obtained pressure-volume loops for the two considered meshes show very close resemblance. When considering local quantities, such as the myofiber strain in specific material points, minor differences are observed. These observations regarding the mesh convergence behavior are in agreement with the detailed mesh study reported in Ref.~\cite{willems_isogeometric_2023}, to which the reader is referred for details.

As for the fitting results discussed above, the global cardiac response results convey that the data sparsity only marginally affects the results when compared to the dense point cloud reference case. With the settings of this benchmark study being chosen to be representative of real-world cases, including the way in which the synthetic echocardiogram data is being generated, the obtained results give confidence that the proposed workflow can yield reliable results in sparse data scenarios. We note that the observations presented here based on the global hemodynamic response do not necessarily carry over to other, possibly more local, quantities of interest.

\newpage

\section{Application to echocardiagram data}\label{sec:Sec6}
The patient-specific isogeometric analysis workflow -- which has been benchmarked based on the Cardiac Atlas Project \cite{fonseca_cardiac_2011} -- is designed to be applicable to real patient data. To demonstrate this applicability, in this section, we perform a cardiac analysis on the echocardiogram data of a clinical patient.

\begin{figure}
    \centering
     \begin{subfigure}[b]{0.48\textwidth}
         \centering
             \includegraphics[width=\textwidth]{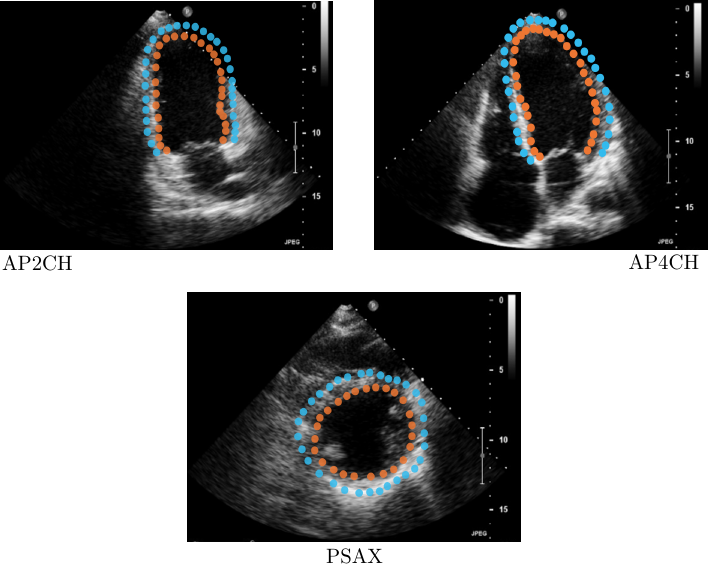}
         \caption{}
         \label{fig:Echo_results_a}
     \end{subfigure}
     \hfill
     \begin{subfigure}[b]{0.48\textwidth}
         \centering
              \includegraphics[width=\textwidth]{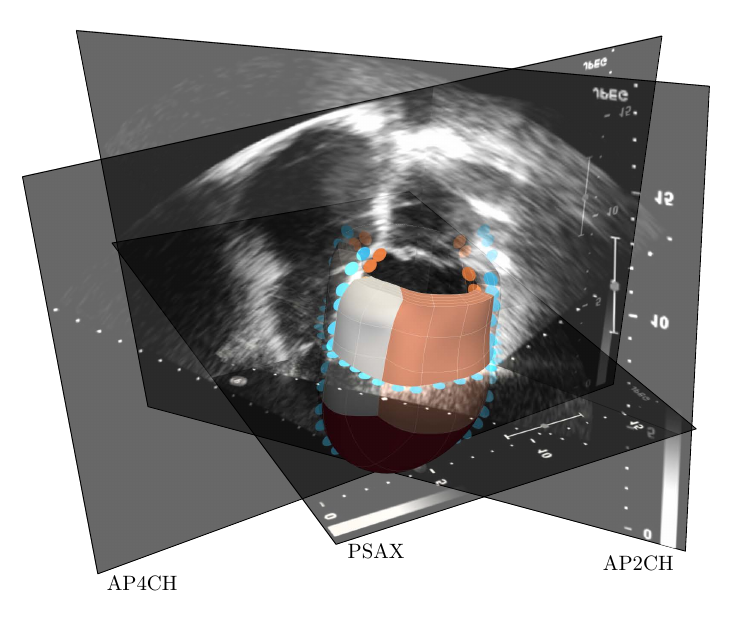}
         \caption{}
         \label{fig:Echo_results_b}
     \end{subfigure}\\
          \begin{subfigure}[b]{0.48\textwidth}
         \centering
             \includegraphics[width=\textwidth]{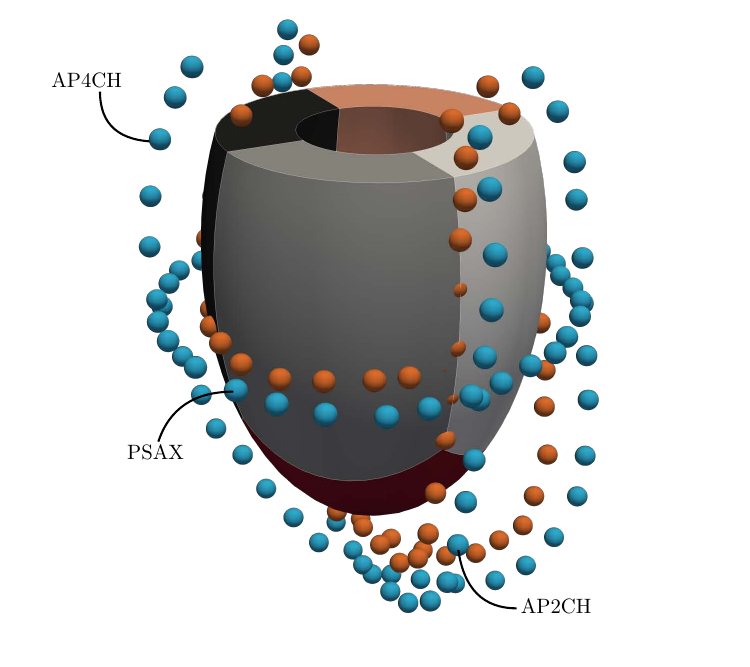}
         \caption{}
         \label{fig:Echo_results_c}
     \end{subfigure}
     \hfill
     \begin{subfigure}[b]{0.48\textwidth}
         \centering
              \includegraphics[width=\textwidth]{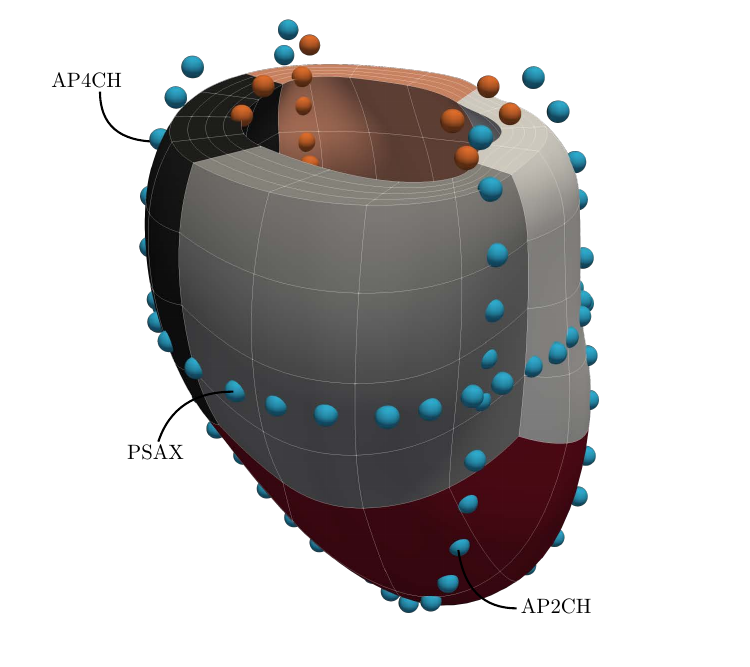}
         \caption{}
         \label{fig:Echo_results_d}
     \end{subfigure}\\
     \begin{subfigure}[b]{0.48\textwidth}
         \centering
             \includegraphics[width=\textwidth]{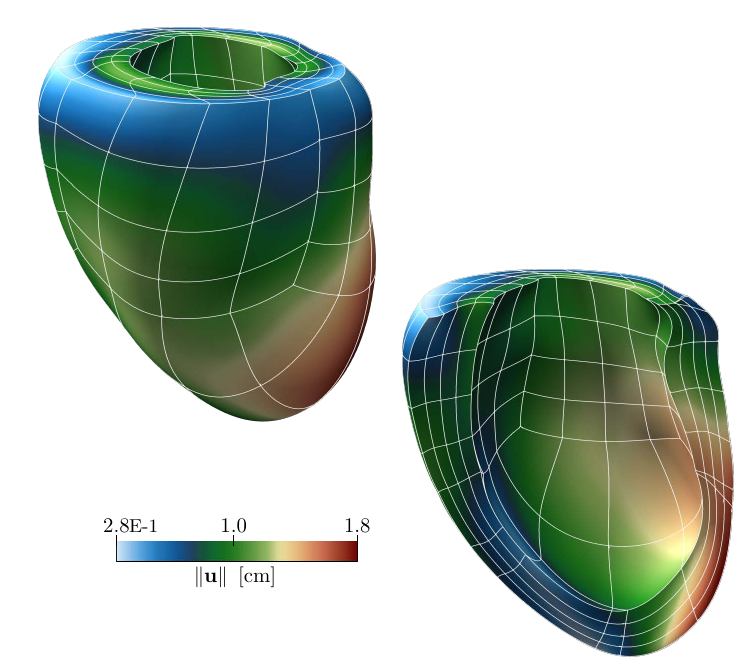}
         \caption{}
         \label{fig:Echo_results_e}
     \end{subfigure}
     \hfill
     \begin{subfigure}[b]{0.48\textwidth}
         \centering
              \includegraphics[width=\textwidth]{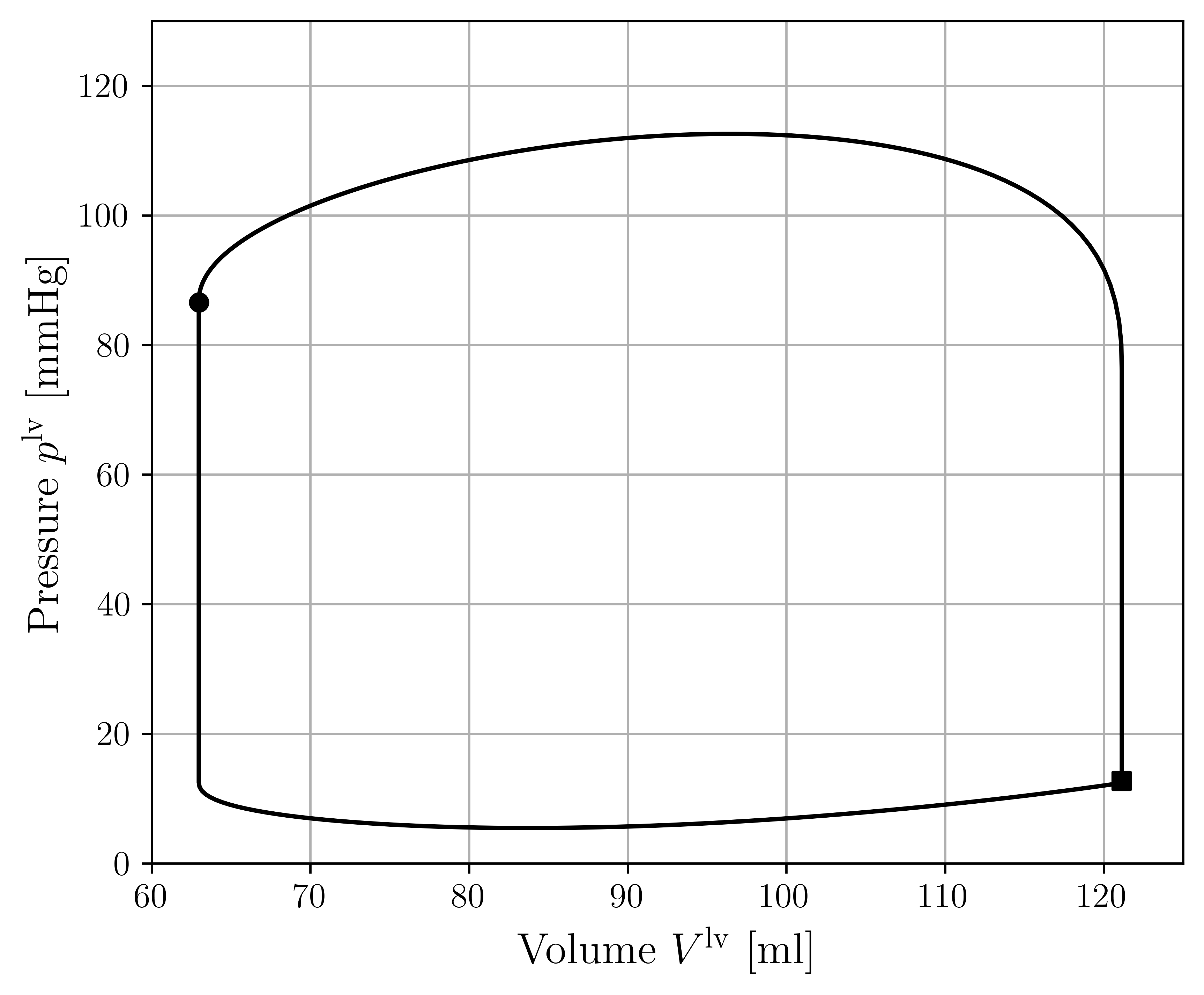}
         \caption{}
         \label{fig:Echo_results_f}
     \end{subfigure}
    \caption{Overview of the cardiac IGA workflow applied to real patient data. (a) Three views of the 2D echocardiogram data, which is segmented for both the endo- (orange) and epicardium (blue) at end-diastole. (b) Combined echocardiogram slices in 3D, resulting in a sparse point cloud (fitted left ventricle already visualized). (c) Sparse point cloud resulting from the slice combination with the initial left ventricle template positioned in the center and slightly below the mitral valve (base). (d) Result of the fitting algorithm after 3 uniform refinements showing indiscernible visual agreement with the data points. (e) Displacement magnitude field resulting from the IGA cardiac simulation at end-systole (noticing the torsional movement). (f) The hemodynamic response corresponding to the cardiac simulation, indicating the end-diastole (square marker) and end-systole (circular marker) cardiac instances.}
    \label{fig:Echo_results}
\end{figure}

The scan data is obtained using a Philips iE33 ultrasound machine, with a spatial resolution of $640\times480$ and a temporal resolution of $37$ frames. Three echocardiogram slices have been obtained, \emph{viz.} apical 2- and 4-chamber views (AP2CH, AP4CH), and the parasternal-short-axis view (PSAX). The scanning results for each slice are stored in a DICOM \cite{noauthor_nema_nodate} file, from which the image showing the end-diastolic configuration is selected based on the closure of the mitral valve. The endo- and epicardium are then segmented manually, by selecting points on these surfaces, which was validated by a cardiologist (Figure~\ref{fig:Echo_results_a}). The echocardiogram slices are combined in 3D space by positioning the AP2CH and AP4CH center at the base and with an angle difference of $60$~[deg]~\cite{nihoyannopoulos_echocardiography_2018}. The PSAX view is then fitted based on these apical views in height and rotation, resulting in a three-dimensional point cloud, as shown in Figure~\ref{fig:Echo_results_b}. 

To initialize the fitting procedure, a truncated ellipsoid template with identical dimensions as mentioned in Section~\ref{sec:fitting} was scaled by a factor of 0.8 and positioned below the mitral valve (Figure~\ref{fig:Echo_results_c}). The fitting algorithm is then executed with the same parameter settings as considered above but with $t_{\mathrm{max}}=220$ fit iterations and $20$ $G^1$-continuity correction iterations for each fit iteration. These changes in the settings are necessary to ensure that the fitting algorithm did not over-correct for $G^1$-continuity, due to the strong curvature in the provided data point set. The resulting fit -- in which the point distance error is $\approx7$\% compared to the initial template and the continuity error is $0.7$ [deg] for the endocardium and $0.3$ [deg] for the epicardium -- is shown in Figure~\ref{fig:Echo_results_d}.

An isogeometric cardiac analysis is then performed directly on the mesh obtained from the fitting procedure, using Nutils~\cite{van_zwieten_nutils_2022}, which was refined 2 times compared to the initial template. This multi-patch NURBS geometry results in a total of $3885$ degrees of freedom for the (cubic) displacement field and $1351$ degrees of freedom for the (cubic) contractile length field. For the analysis, the same parameters are used as for these considered in Section~\ref{sec:Sec5}, with the exception that the general pre-stressing algorithm iterations are terminated after 6 load-increment-free iterations to determine the pre-stresses in the image configuration.

The displacement field obtained from the cardiac analysis is illustrated in the deformed configuration at end-systole in Figure~\ref{fig:Echo_results_e}. As for the benchmark cases considered above, the torsional deformation mode is clearly observed. Compared to the Cardiac Atlas displacement results, visualized in Figure~\ref{fig:Sim_result_endsystole}, this patient-specific case exhibits a similar deformation field and deformation magnitude. From the isogeometric analysis we also obtain the hemodynamic response, which is shown in Figure~\ref{fig:Echo_results_f}, with the end diastole and end systole instance marked by a square and circle respectively. The deformation field and pressure-volume relation are here presented as typical results generated by the isogeometric analysis, but evidently, the full time-dependent mechanical and hemodynamic response is available for analysis.

\newpage

\section{Concluding remarks}\label{sec:ConcRec}
In this work, we have presented a patient-specific cardiac modeling workflow based on the isogeometric analysis (IGA) paradigm. The proposed workflow has two distinctive features, \emph{viz.}: \emph{(i)} Patient-specific geometries can be constructed from sparse and irregular data, for example obtained from echocardiogram slices; \emph{(ii)} The geometry construction operations and the physical analysis are seamlessly integrated, not requiring the use of intermediate geometry clean-up and meshing operations.

The proposed workflow builds on the IGA-based cardiac-mechanics model of Willems~\emph{et al.}~\cite{willems_isogeometric_2023}, from which it also derives its distinctive features. When fitted to sparse data points, the multi-patch NURBS representation of the left ventricle is able to smoothly interpolate the geometry in the regions where data is scant. This same multi-patch NURBS can then be used directly as a basis for a Galerkin method, in this case discretizing the mechanical cardiac response of the left ventricle. To realize this functionality, in this work, we have presented two methodological innovations, \emph{viz.}: \emph{(i)} A multi-patch NURBS fitting algorithm for sparse data; \emph{(ii)} An extension of the computational model of Ref.~\cite{willems_isogeometric_2023} enabling it to accommodate \emph{in vivo} images.

The developed fitting algorithm is capable of dealing with sparse data by assigning data point weights to the various control points. In doing so, the data is closely fitted by the splines, while the splines interpolate the geometry in regions where data is absent. The considered multi-patch NURBS template in principle allows for the occurrence of kinks across patch interfaces, which often (and certainly also in this work) is undesirable. The fitting algorithm minimizes the occurrence of such kinks by amending the global error representing the distance of the geometry to the data points with a term that penalizes non-smooth patch interfaces. In the algorithm, these two error terms are then minimized sequentially through control net manipulations.

Objects observed in \emph{in vivo} images are typically not in an unloaded state. To accommodate the loaded image state in our analysis workflow, the model of Ref.~\cite{willems_isogeometric_2023} has been amended with a multiplicative decomposition of the deformation gradient. In this decomposition, the image state is encoded in the deformation gradient field related to the mapping between the (unknown) unloaded configuration and the image configuration. The advantage of this approach is that all computational operations are performed on the image geometry, for which the mesh quality can be assured through the developed fitting algorithm. To determine the component of the deformation gradient associated with the image loading, the general pre-stressing algorithm originally proposed by Weisbecker \emph{et al.}~\cite{weisbecker_generalized_2014} has been adapted to the isogeometric analysis setting. The most prominent alteration to the algorithm pertains to the representation of the image deformation gradient field by an auxiliary displacement field. This representation by means of control point displacements makes the algorithm compatible with data standards in IGA, thereby promoting integration in existing implementations.

A benchmark study based on left-ventricle data from the Cardiac Atlas Project~\cite{fonseca_cardiac_2011} has demonstrated the capabilities of the proposed workflow. In this study, synthetic sparse echocardiogram data is extracted from a dense reference point cloud, allowing us to assess the effects of data sparsity. The fitting algorithm is demonstrated to maintain a high degree of accuracy when considering a realistic selection of echocardiogram slices. When considering the hemodynamic response computed through the IGA model, only minor deviations from the reference solution are observed. These benchmarking results provide confidence in the suitability of the workflow for real data scenarios. To demonstrate the operation of the workflow in such a scenario, the case of a clinical patient is considered. While a detailed analysis of this specific scenario is beyond the scope of the current work, it demonstrates the robustness of the workflow.

Our patient-specific isogeometric analysis workflow has been developed in the context of cardiac modeling but is to a large extent generic. Specifically, the fitting algorithm is suitable to generate analysis-suitable splines for isogeometric analysis based on general point cloud data, both dense and sparse. The developed workflow is deterministic in nature, \emph{i.e.}, all quantities are assumed to be known precisely. The reality is, however, that uncertainties are present in all steps of the workflow. Depending on the case under consideration, the quantification of such uncertainties may be of interest~\cite{mirams_uncertainty_2016, eck_guide_2016, krygier_quantifying_2021}. In order to open the doors to uncertainty quantification, we envision the need for the development of a reduced-order modeling approach, which would, \emph{e.g.}, allow for the application of stochastic sampling procedures.

\section*{Acknowledgement}
This publication is part of the COMBAT-VT project (project no. 17983) of the research program High Tech Systems and Materials which is partly financed by the Dutch Research Council (NWO). Additionally, this work was performed within the IMPULS framework under the Picasso project (reference no. TKI HTSM/20.0022) of the Eindhoven MedTech Innovation Center (e/MTIC, incorporating Eindhoven University of Technology, Philips Research, and Catharina Hospital), including a PPS-supplement from the Dutch Ministry of Economic Affairs and Climate Policy. We also acknowledge the clinical partners at the Catharina Hospital in Eindhoven who have provided feedback on the echocardiogram segmentation. Lastly, we acknowledge the team of Nutils~\cite{van_zwieten_nutils_2022} for their support regarding the numerical implementation of the cardiac model.

\newpage


\newpage
\appendix
\section{Constraining of the rigid body modes}
\label{app:rigidbodyconstraints}
In our cardiac model, the normal component of the displacement vector is constrained at the basal plane \eqref{eq:B11c}. Since the normal to the basal plane coincides with the $x_3$ direction (with $\mathbf{x}=(x,y,z)$) this constraint can be enforced strongly by setting the $u_3$ displacement field component of the basal plane control points to zero. This constraint removes the rigid body motion in the $x_3$ direction and the rigid body rotations around the $x_1$ and $x_2$ axes. A common strategy to remove the remaining rigid body modes is to strongly constrain the in-plane displacements in one point (in the basal plane), making this a point around which the body can rotate. The rotational rigid body mode is then removed by considering another point in the basal plane and constraining the displacement component perpendicular to the vector between this point and the rotation point. In the patient-specific isogeometric analysis setting considered in this work, this method of constraining is not practical. Directly constraining the displacement in a point is most practical in the patch vertices, where the B-splines are interpolatory. To constrain the displacement in an arbitrary point linear constraints can be used, but, depending on the considered software framework, such constraints are not always practical from an implementation point of view. When using the patch vertices as constraining points, these are in general not positioned in such a way that the rotational rigid body mode can be constrained by a single component of the displacement field. Instead of resorting to linear constraints to solve this problem, in this work, we opt to constrain the rigid body modes by enforcing the conditions \ref{eq:B11d}, \ref{eq:B11e} and \ref{eq:B11f} using Lagrange multipliers.

\begin{figure}
         \centering
     \begin{subfigure}[b]{0.49\textwidth}
         \centering
             \includegraphics[width=\textwidth]{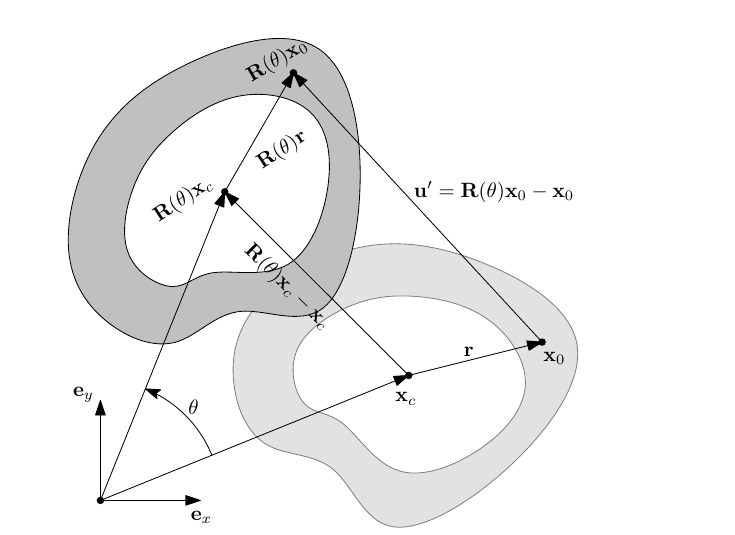}
         \caption{}
         \label{fig:rgbd_rot_trans} 
     \end{subfigure}
     \hfill
     \begin{subfigure}[b]{0.49\textwidth}
         \centering
              \includegraphics[width=\textwidth]{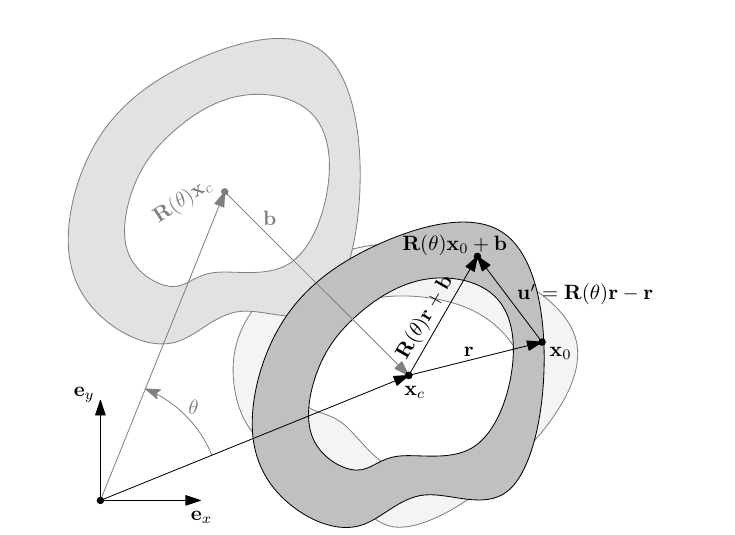}
         \caption{}
         \label{fig:rgbd_rot}
     \end{subfigure}
     \caption{Schematic illustration of an arbitrary domain that undergoes pure rotation around its global origin with an angle $\theta$. (a) The resulting configuration experiences a displacement, $\mathbf{u}^{\prime}$, consisting of a translation, $\mathbf{R}(\theta) \mathbf{x}_c - \mathbf{x}_c$, of the center, $\mathbf{x}_c$, and a rotation, $\mathbf{R}(\theta) \mathbf{x}_0 - \mathbf{x}_0$ around this center, \emph{cf.} Equation~\eqref{eq:rigidbodymotion}. (b) Removing the translational component, \emph{cf.} Equation~\eqref{eq:translation}, ensures that the local rotation around the center is left, which should be removed \emph{cf.} Equation~\eqref{eq:rotation}.}
     \label{fig:rgbd}
\end{figure}

While the average displacement constraints (\ref{eq:B11d} and \ref{eq:B11e}) are standard, the constraint on the average curl of the displacement field \eqref{eq:B11f} is not. To explain how this constraint removes the rigid body rotation, we consider a deformed equilibrium state in the current configuration, $\mathbf{x}$, and apply an arbitrary rigid body transformation to obtain a new configuration (see Figure~\ref{fig:rgbd})
\begin{align}
    \mathbf{x}^{\prime}(b_1,b_2,\theta) &= \mathbf{R}(\theta) \mathbf{x} + \mathbf{b}(b_1,b_2),  &   \left( \begin{array}{c} x_1^\prime \\ x_2^\prime \\ x_3^\prime \end{array} \right) &= \left[ \begin{array}{ccc} \cos{\theta} &  -\sin{\theta} & 0 \\ \sin{\theta} &  \cos{\theta} & 0 \\ 0 & 0 & 1 \end{array} \right] \left( \begin{array}{c} x_1 \\ x_2 \\ x_3 \end{array} \right) + \left( \begin{array}{c} b_1 \\ b_2 \\ 0 \end{array} \right), 
    \label{eq:rigidbodymotion}
\end{align}
where $b_1$ and $b_2$ are arbitrary translations in the $x_1$ and $x_2$ direction, and $\theta$ is an arbitrary angle of rotation around the $x_3$ axis. In the absence of the rigid body constraints (\ref{eq:B11d}--\ref{eq:B11f}), the transformed configuration, $\mathbf{x}^\prime$, satisfies the equilibrium conditions for any choice of the translations and rotation, making the problem ill-posed.

To demonstrate how the rigid body constraints (\ref{eq:B11d}--\ref{eq:B11f}) result in a unique solution, equation \eqref{eq:rigidbodymotion} is expressed in terms of the displacement field as
\begin{align}
     \mathbf{u}^{\prime}(b_1,b_2,\theta) = \mathbf{R}(\theta)\mathbf{u}  + \mathbf{R}(\theta)  \mathbf{x}_0 + \mathbf{b}(b_1,b_2) - \mathbf{x}_0.
     \label{eq:uprime}
\end{align}
Application of the translational constraints, \emph{i.e.}, \ref{eq:B11d} and \ref{eq:B11e}, then yields
\begin{align}
     \mathbf{b}(b_1,b_2)  = \mathbf{x}_c - \mathbf{R}(\theta)  \mathbf{x}_c 
     \label{eq:translation}
\end{align}
where $\mathbf{x}_c = \frac{1}{A^{\rm base}} \mathbf{Q}^{\rm base}$ and
\begin{align}
    A^{\rm base} &= \int_{\Gamma^{\rm base}} \, \mathrm{d}\Gamma^{\rm base}, &  \mathbf{Q}^{\rm base} &= \int_{\Gamma^{\rm base}} \mathbf{x}_0 \, \mathrm{d}\Gamma^{\rm base}.
\end{align}
Back-substitution of the translation vector \eqref{eq:translation} into equation \eqref{eq:uprime} then yields
\begin{align}
    \mathbf{u}^{\prime}(\theta) & = \mathbf{R} (\theta) \left( \mathbf{u} +   \mathbf{r} \right) - \mathbf{r}, &    \left(\begin{array}{c} u_1^\prime \\ u_2^\prime \\ 0 \end{array}\right) &=  \left(\begin{array}{c} (u_1 + r_1) \cos{\theta} - (u_2+r_2) \sin{\theta}  - r_1 \\ (u_1+r_1) \sin{\theta} + (u_2+r_2) \cos{\theta} - r_2 \\ 0 \end{array}\right),\label{eq:rotation}
\end{align}
where $\mathbf{r} = \mathbf{x}_0 - \mathbf{x}_c$. From this expression, the curl of the translated displacement field is obtained as
\begin{align}
    \left( \frac{\partial u_2^\prime}{\partial x_1^\prime} - \frac{\partial u_1^\prime}{\partial x_2^\prime} \right)=\left( \frac{\partial u_2}{\partial x_1}-\frac{\partial u_1}{\partial x_2} \right) + 2 \sin{\theta},
    \label{eq:curl}
\end{align}
which conveys that the rigid body motion \eqref{eq:rigidbodymotion} results in a spatially constant increase of the curl by $2\sin{\theta}$. Note that the well-known relation $\dot{\theta} = \frac{1}{2} \text{curl}(\dot{\mathbf{u}}^\prime)$ \cite{belytschko_nonlinear_2014} directly follows from this expression by taking the time derivative and evaluating the expression at $\theta=0$.

Application of the rotational rigid body constraint \eqref{eq:B11f} to \eqref{eq:curl} gives
\begin{align}
2 A^{\rm base} \sin{\theta} &= 0,
\end{align}
from which it follows that $\theta=2 \pi \cdot i$ or $\theta=2 \pi \cdot i + \pi$ for any integer $i$. The solutions $\theta=2 \pi \cdot i$ result in 
\begin{align}
    \mathbf{u}^{\prime} & = \mathbf{u},
\end{align}
and hence are coincident with the considered configuration before application of the rigid body motion. The solutions $\theta=2 \pi \cdot i + \pi$ correspond to the point reflection of the original configuration in the point $\mathbf{x}_c$. Although this mirrored configuration also satisfies the equilibrium equations, it is not reachable through a continuous path starting from the initial condition. As a result, the only valid rigid body motion \eqref{eq:rigidbodymotion} corresponds to $b_1=b_2=\theta=0$, for which $\mathbf{x}^\prime=\mathbf{x}$. Therefore, application of the rigid body constraints (\ref{eq:B11d}--\ref{eq:B11f}) guarantees a unique solution.
\section{Isogeometric general pre-stress algorithm benchmarks}
\label{app:igpabenchmarks}
\begin{figure}
        \centering
     \begin{subfigure}[b]{0.49\textwidth}
         \centering
             \includegraphics[width=\textwidth]{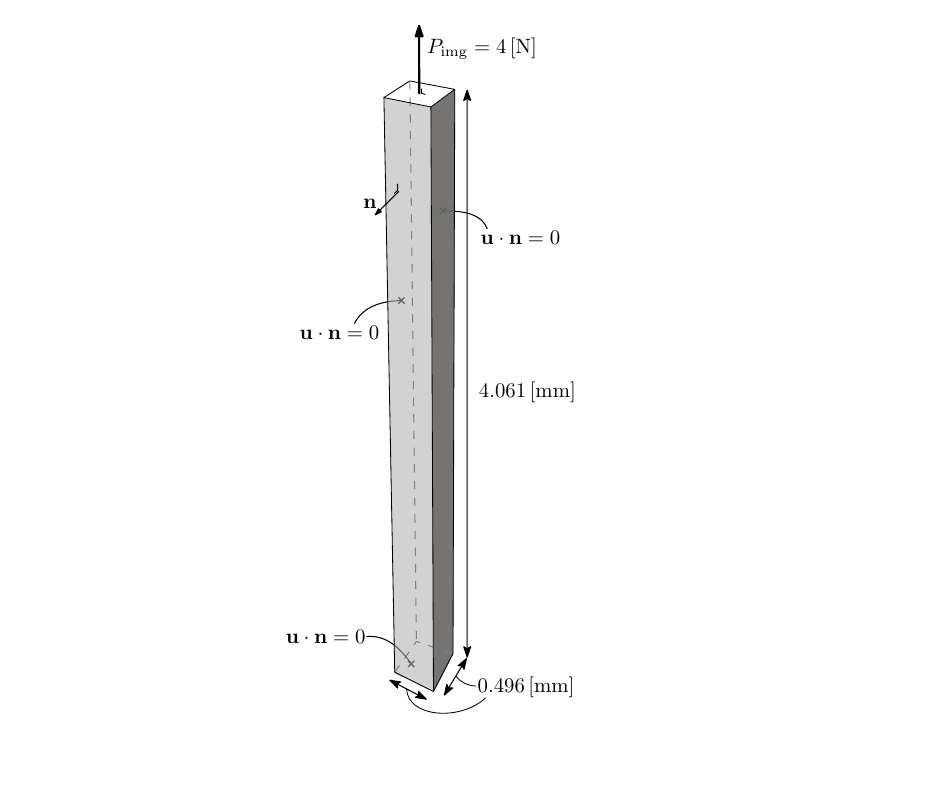}
         \caption{}
         \label{fig:gpabenchmarks_a} 
     \end{subfigure}
     \hfill
     \begin{subfigure}[b]{0.49\textwidth}
         \centering
             \includegraphics[width=\textwidth]{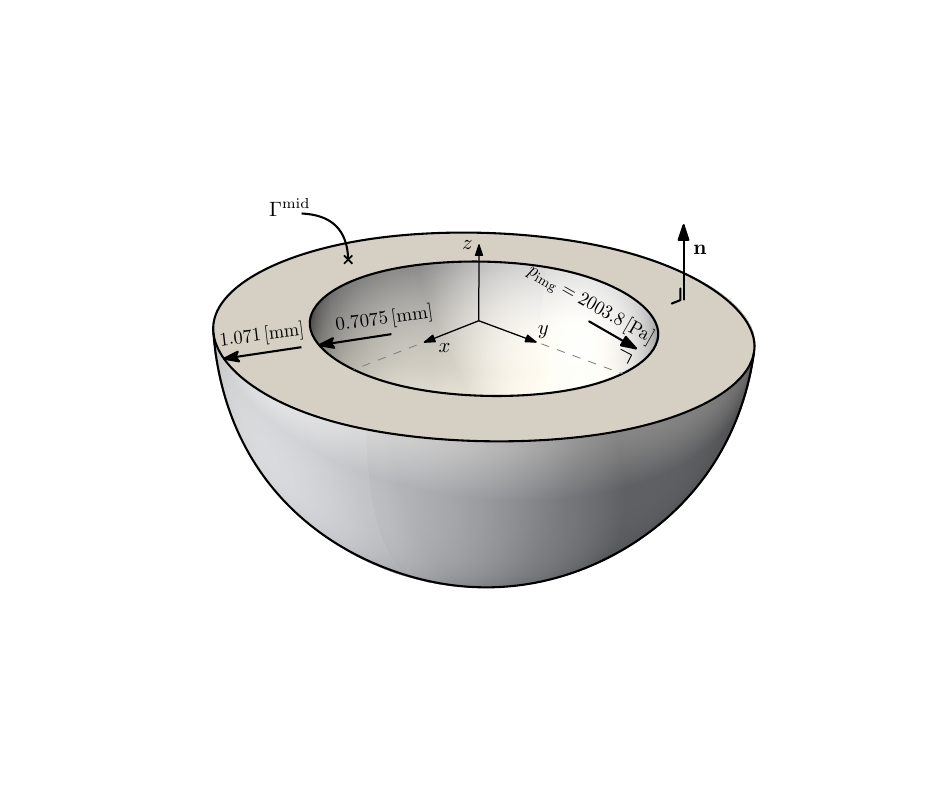}
         \caption{}
         \label{fig:gpabenchmarks_b}
     \end{subfigure}

     \caption{Schematic overview of the benchmark cases considered for the isogeometric GPA. Loaded or \emph{image} configuration of the (a) axially loaded rod and (b) pressure-loaded thick-walled sphere case.}
     \label{fig:gpabenchmarks}
\end{figure}

In this appendix, we present two benchmarks for the isogeometric general pre-stress algorithm (GPA) introduced in Section~\ref{sec:gpa}. In \ref{app:rod} we first consider the axially loaded rod studied in the original GPA paper by Weisbeckker \cite{weisbecker_generalized_2014} (Figure~\ref{fig:gpabenchmarks_a}). In \ref{app:sphere} we revisit the pressure-loaded thick-walled sphere benchmark studied by Willems \emph{et al.} \cite{willems_isogeometric_2023}  (Figure~\ref{fig:gpabenchmarks_b}) to assess the performance of the GPA in a setting that closely resembles the patient-specific cardiac analysis considered in this work.

\subsection{An axially loaded rod}
\label{app:rod}
We consider the axially loaded rod benchmark proposed in Ref.~\cite{weisbecker_generalized_2014} (Figure~\ref{fig:gpabenchmarks_a}). Following Ref.~\cite{weisbecker_generalized_2014}, the material of the rod is modeled as a compressible hyperelastic material \cite{belytschko_nonlinear_2014}, with the shear modulus set to $\mu=1$\,[MPa] and the Poisson ratio to $\nu=0.4999$. We note that although the model is compressible, this parameter setting makes the material nearly incompressible. The rod is loaded by the axial force $P=4$\,[N], which stretches an undeformed unit cube (\emph{i.e.}, $H=W=1$\,[mm]) to a box of height $H=4.061$\,[mm] and width $W=0.4963$\,[mm] as illustrated in Figure~\ref{fig:gpabenchmarks_a}. This deformed configuration is interpreted as being the loaded image configuration.

To test the isogeometric GPA, we discretize the rod in the loaded image configuration using $2 \times 2 \times 2$ elements and second-order ($p=2$) B-splines. The corresponding control net consists of $4 \times 4 \times 4$ control points, for each of which the displacement is discretized in all three directions. The control point displacements on the bottom and the two back surfaces are constrained in accordance with the boundary conditions (Figure~\ref{fig:gpabenchmarks_a}). The single field total Lagrange formulation as considered in Section~\ref{sec:governingequations} is used to solve the problem, which results in a system of equations that is hard to solve on account of the nearly incompressible nature of the material. Using a Newton-Raphson solver with a residual-based line search algorithm allows us to solve the nonlinear system, however. Although this single field discretization is in principle not preferred for nearly incompressible materials, we opt to use it to retain the same setting as in the original benchmark \cite{weisbecker_generalized_2014}.

\begin{figure}
         \centering
     \begin{subfigure}[b]{0.49\textwidth}
         \centering
             \includegraphics[width=\textwidth]{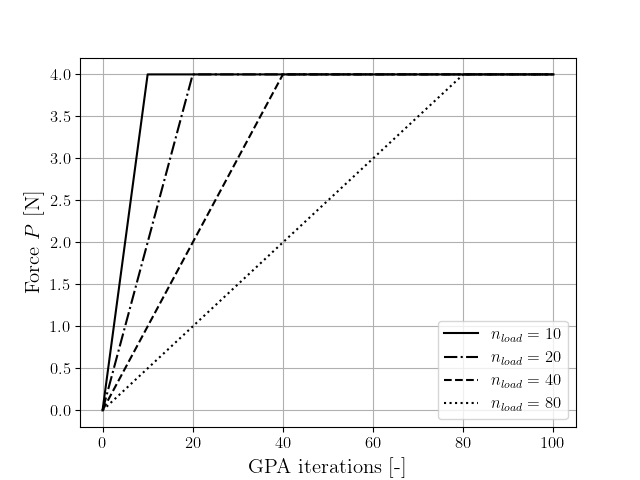}
         \caption{}
         \label{fig:rodresults_a} 
     \end{subfigure}
     \hfill
     \begin{subfigure}[b]{0.49\textwidth}
         \centering
              \includegraphics[width=\textwidth]{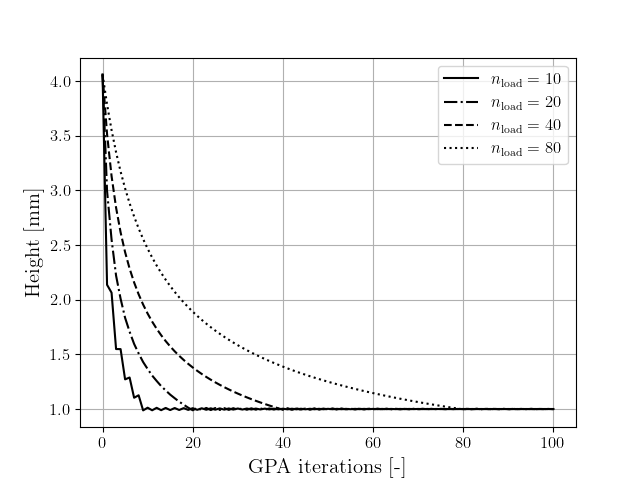}
         \caption{}
         \label{fig:rodresults_b}
     \end{subfigure}

     \caption{GPA iteration results for the axially loaded rod benchmark: (a) applied image force; (b) height of the unloaded geometry. The results for different load application steps are compared.}
     \label{fig:rodresults}
\end{figure}

In Figure~\ref{fig:rodresults} we study the isogeometric GPA by varying the number of steps in which the image load of {$P_{\rm img} = 4$\,[N]} is applied. Figure~\ref{fig:rodresults_a} illustrates the way in which the force is gradually increased up to the image force of 4\,[N]. Since all GPA steps converge, the force steps are constant, resulting in a linear increase of the force. From Figure~\ref{fig:rodresults_b}, it is observed that for all settings of $n_{\rm load}$, the length of the unloaded box converges to the exact value of one. 

\begin{figure}
         \centering
     \begin{subfigure}[b]{0.49\textwidth}
         \centering
              \includegraphics[width=\textwidth]{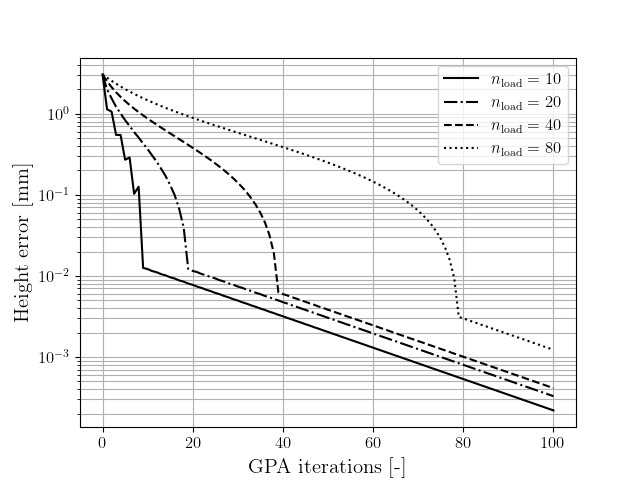}
         \caption{}
         \label{fig:roderrors_a}
     \end{subfigure}
     \hfill
     \begin{subfigure}[b]{0.49\textwidth}
         \centering
              \includegraphics[width=\textwidth]{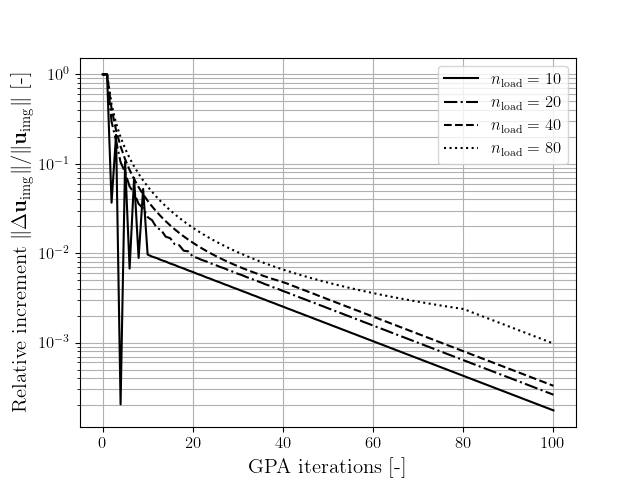}
         \caption{}
         \label{fig:roderrors_b}
     \end{subfigure}
     \caption{Convergence metrics of the GPA iterations for the axially loaded rod benchmark: (a) absolute error in the height of the unloaded geometry; (b) relative increment of the displacement field.  The results for different load application steps are compared.}
     \label{fig:roderrors}
\end{figure}

The convergence of the GPA is studied in more detail in Figure~\ref{fig:roderrors}, which shows the absolute error in the height, and the relative image displacement increment \eqref{eq:uimgincrement} versus the GPA iterations. For both convergence metrics, the loading and fixed point iterations can clearly be distinguished. In the loading stage, the error is governed by the mismatch between the applied force and the desired image force, whereas the error during the fixed point iterations is fully attributed to the non-linearity of the equilibrium condition \eqref{eq:nonlineargpa}. In this second stage, a linear convergence rate is observed. In principle, the results convey that in order to reach a certain error tolerance as fast as possible, the load should be applied in a minimum number of steps. There is, however, a limit to this, as the nonlinear solver may not be able to converge when the load steps are too big. The onset of such convergence problems can be observed for the $n_{\rm load}=10$ case, which is already subject to substantial oscillations in the applied displacement increments. Another important observation from Figure~\ref{fig:roderrors} is that both convergence metrics show very similar behavior. Since for practical cases, the exact unloaded configuration is not known, in such cases only the relative increment of Figure~\ref{fig:roderrors_b} can be monitored. In the case of this benchmark, assessing the convergence based on this increment is also representative of the convergence of the height in Figure~\ref{fig:roderrors_a}.

\subsection{A pressure-loaded thick-walled sphere}
\label{app:sphere}
As a second benchmark, which relates more closely to the cardiac modeling setting considered in this work, we consider the pressure-loaded thick-walled sphere as studied in Ref.~\cite{willems_isogeometric_2023}. The passive material model introduced in Section~\ref{sec:Sec4} is used but with the anisotropic contribution of the fibers omitted (\emph{i.e.}, the elasticity coefficient $a_3$ is set to zero). All other material parameters are taken from Table~\ref{tab:params}. In the image configuration, the sphere is loaded by an internal pressure of $2003.8$\,[Pa]. At this pressure, the sphere has an inner radius of $r_{\rm in}=0.7075$\,[mm] and an outer radius of $r_{\rm out}=1.071$\,[mm], visualized in Figure~\ref{fig:gpabenchmarks_b}. These dimensions correspond to an unloaded state with radii $R_{\rm in}=0.5$\,[mm] and $R_{\rm out}=1.0$\,[mm], following from the exact solution \cite{willems_isogeometric_2023}.

In our analysis, we only consider the bottom part of the domain, prescribing symmetry conditions (\emph{i.e.}, zero normal displacement and zero non-normal traction) at the mid-plane. The domain is discretized using 5 patches with $2 \times 2 \times 2$ elements each. Cubic NURBS are used to parametrize the geometry, resulting in a control net with 135 control points. The boundary conditions are identical to \ref{eq:B11c}, \ref{eq:B11d}, \ref{eq:B11e}, and \ref{eq:B11f}, which are applied to the $\Gamma^{\mathrm{mid}}$. In contrast to the nearly incompressible rod benchmark, this simulation setup does not require the use of a line-search algorithm. 

\begin{figure}
         \centering
     \begin{subfigure}[b]{0.49\textwidth}
         \centering
             \includegraphics[width=\textwidth]{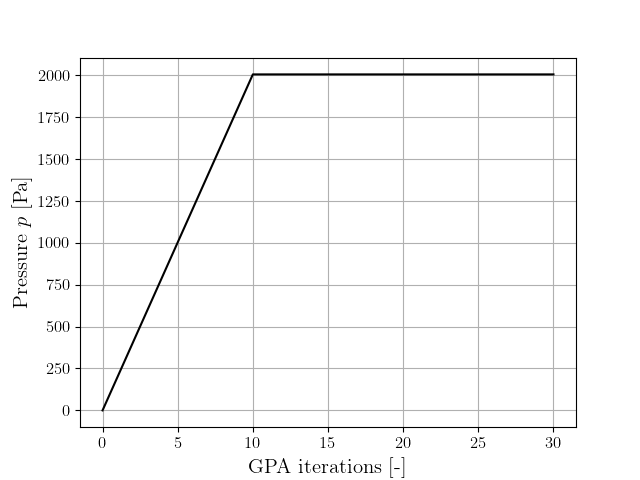}
         \caption{}
         \label{fig:sphereresults_a} 
     \end{subfigure}
     \hfill
     \begin{subfigure}[b]{0.49\textwidth}
         \centering
              \includegraphics[width=\textwidth]{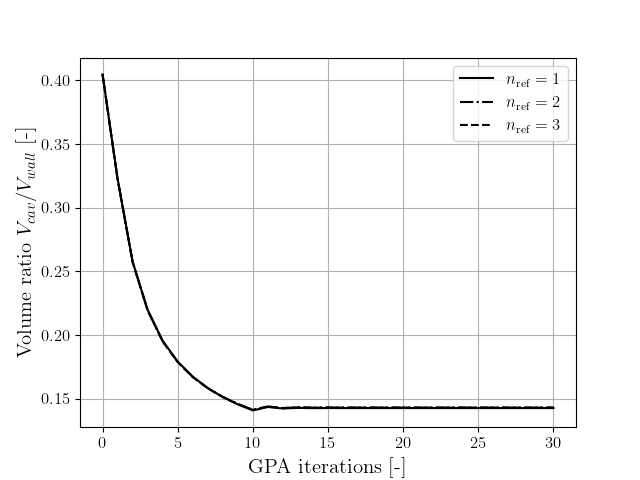}
         \caption{}
         \label{fig:sphereresults_b}
     \end{subfigure}

     \caption{GPA iteration results for the pressure-loaded tick-walled sphere benchmark: (a) applied image pressure; (b) ratio between the cavity volume and the wall volume. The results for different refinement levels are considered. Note that the pressure loading is identical for all cases and that the different cases cannot visually be discerned for the volume ratios.}
     \label{fig:sphereresults}
\end{figure}

Figure~\ref{fig:sphereresults} shows the GPA results for various uniform refinements of the mesh, where the case of $n_{\rm ref}=1$ corresponds to the 135 control points with which the geometry is parametrized. For all refinement cases, the image pressure is applied in 10 steps (Figure~\ref{fig:sphereresults_a}). When considering the ratio between the cavity volume and the wall volume in Figure~\ref{fig:sphereresults_b}, it is observed that it converges to a value that is visually indiscernible from the exact value of $\frac{1}{7}$.

\begin{figure}
         \centering
     \begin{subfigure}[b]{0.49\textwidth}
         \centering
              \includegraphics[width=\textwidth]{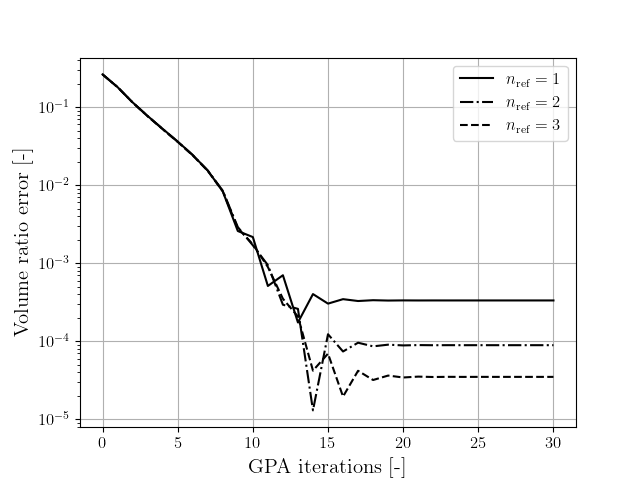}
         \caption{}
         \label{fig:sphereerrors_a}
     \end{subfigure}
     \hfill
     \begin{subfigure}[b]{0.49\textwidth}
         \centering
              \includegraphics[width=\textwidth]{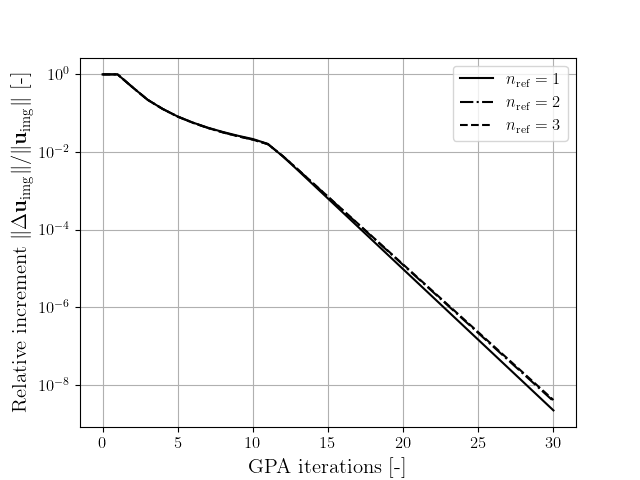}
         \caption{}
         \label{fig:sphereerrors_b}
     \end{subfigure}
     \caption{Convergence metrics of the GPA iterations for the pressure-loaded thick-walled sphere benchmark: (a) absolute error in the cavity-to-wall volume ratio of the unloaded geometry; (b) relative increment of the displacement field. The results for different refinement levels are considered.}
     \label{fig:sphereerrors}
\end{figure}

The error in the volume ratio between the cavity and the wall is inspected closer in Figure~\ref{fig:sphereerrors_a}. It is observed that for all refinement levels, this error decreases under the application of the load and during the first load-free iterations. After approximately 17 steps (7 load-free steps) the errors reach a stable plateau. This plateau corresponds to the discretization error, as the displacement field required to reach the undeformed state cannot be represented exactly by the NURBS basis. Evidently, as the mesh is refined, the exact solution can be represented more accurately. As a consequence, the value at which the error reaches the plateau decreases under mesh refinement. The relative image displacement increment is shown in Figure~\ref{fig:sphereerrors_b}. As for the rod benchmark, the two stages of the GPA algorithm can be observed. In contrast to the volume ratio error, the increments continue to decrease once the discretization accuracy is reached. The reason for this is that the increments are measured relative to the discrete solution, meaning that it will vanish once the discrete solution is attained. As for the rod benchmark, the relative displacement increment provides an adequate metric to assess the convergence of the GPA.

\bibliographystyle{elsarticle-num} 
\bibliography{references}





\end{document}